\documentclass[a4paper,11pt]{article}
\pdfoutput=1 
\usepackage{jheppub} 
\usepackage[T1]{fontenc}
\usepackage{tabulary}
\usepackage[T1]{fontenc}
\usepackage[table]{xcolor}
\usepackage[caption=false]{subfig}
\usepackage{epsfig}
\usepackage{colortbl}
\usepackage[T1]{fontenc}
\usepackage{tikz-feynman}
\usepackage{hepnames}
\usepackage{adjustbox}
\usepackage{textcomp}
\usepackage{slashed}
\usepackage{amsmath}
\usepackage{multirow}
\usepackage[Symbol]{upgreek}
\newcommand{\bea}{\begin{eqnarray}}
\newcommand{\eea}{\end{eqnarray}}
\usepackage{xcolor}

\usepackage[font=small]{caption}
\usepackage[font={small},labelfont={bf}]{caption}
\usepackage{url}
\definecolor{MyDarkBlue}{rgb}{0.1, 0.1, 0.8}
\definecolor{SBlue}{rgb}{0.2, 0.4, 0.7} 
\definecolor{MyLightBlue}{rgb}{0.22,0.51,0.9}
\definecolor{MyGreen}{rgb}{0.0, 0.5, 0.0}
\definecolor{BrickRed}{rgb}{0.8, 0.25, 0.33}
\usepackage{hyperref}
\hypersetup{colorlinks, citecolor=BrickRed,linkcolor=MyDarkBlue, urlcolor=MyGreen}
\title{The Influence of Lepton Portal on the WIMP-pFIMP framework}
\author[a]{Jayita Lahiri, }
\author[b]{Dipankar Pradhan, }
\author[b]{and Abhik Sarkar}
\affiliation[a]{II. Institut f{\"u}r Theoretische Physik, Universit{\"a}t Hamburg, Luruper Chaussee 149, 22761 Hamburg, Germany.}
\affiliation[b]{Department of Physics, Indian Institute of Technology Guwahati,\\	North Guwahati, Assam-781039, India.}
\emailAdd{jayita.lahiri@desy.de}
\emailAdd{d.pradhan@iitg.ac.in}
\emailAdd{sarkar.abhik@iitg.ac.in}
\abstract{
The dynamics and detection possibility of a pseudo-FIMP (pFIMP) dark matter (DM) in the presence of a thermal DM have been studied in different contexts. The pFIMP phenomenology largely depends on the WIMP-like partner DM, as pFIMP interacts with the standard model (SM) particles only via the partner DM loop. Introducing a lepton portal interaction, which connects DM directly to the SM lepton sector, improves its detection prospects. However, such possibilities are constrained strongly by the non-observation of lepton flavor-violating decays. Interestingly, this also makes it possible to probe such models in future low-energy experiments. In this article, we have tried to establish such connections and find parameter space which respects the limits from DM relic density, direct, indirect, and lepton flavor violation (LFV). We also recast the constraints from di-lepton/di-tau plus missing energy signal at the LHC on our model and provide projections for HL-LHC and future lepton colliders. Although the LFV and collider limits mainly concern WIMPs, the parameter space for pFIMPs is also constrained due to its strong connection to WIMPs through DM relic density and detection prospects. }
\begin{document} 
\makeatletter
\gdef\@fpheader{}
\makeatother
\maketitle
\flushbottom
\section{Introduction}
\label{sec1}
From astrophysical and cosmological observations \cite{Clowe:2006eq,Komatsu:2014ioa,Sofue:2000jx,Hayashi:2006kw}, we know that our universe contains a significant amount of `dark' component, known as dark matter (DM) \cite{Zwicky:1933gu, Zwicky:1937zza}, in its total matter content. There has been a tireless effort to observe direct evidence of dark matter in DM-nucleon scattering experiments as well as indirect detection experiments, where DM is searched for via the excess photon/positron/anti-proton from terrestrial sources. On the other hand, dark matter is also searched for at the collider via direct production from the interaction between SM particles. Despite the immense effort and increasing sensitivity of all the experiments mentioned above, we have had null results so far.
Therefore, the current data puts strong constraint on a plethora of dark matter models, especially the ones involving Weakly Interacting Massive Particles (WIMP) \cite{Trimble:1987ee, Griest:1989wd, Jungman:1995df} as DM. Interestingly, the null DM signal can stem from various factors. For example, there are WIMP-dark matter models with extended scalar sector where DM interacts with the SM sector via scalar portal mechanism and mutual cancellation between contributions coming from multiple scalars leads to so-called `blind-spots' in the parameter space~\cite{Badziak:2017uto, Badziak:2015exr,Dey:2019lyr}.
On the other hand, there are scenarios such as co-annihilation~\cite{PhysRevD.43.3191, Edsjo:1997bg}, co-scattering~\cite{DAgnolo:2017dbv, DiazSaez:2024nrq}, which lead to thermal relic via mutual annihilation/scattering within the dark sector. By virtue of such mechanisms, these scenarios can give rise to the observed relic density, even with very small interaction strength with the visible sector.
Alternatively, there is a possibility that the DM is feebly interacting with the SM bath particles and, therefore, contributes to the non-thermal relic density via freeze-in mechanism \cite{Hall:2009bx, Elahi:2014fsa, Blennow:2013jba}. This is the well-known Feebly Interacting Massive Particle (FIMP) scenario, which will also evade all standard methods of DM detection owing to its extremely small interaction strength. However, it's not necessary for the dark sector to consist of only a single type of particle.
It is possible that there are multiple components of DM and all of them contribute to the relic density. It was shown in \cite{Bhattacharya:2022vxm, Bhattacharya:2022dco}, that if the FIMP is associated with a WIMP in a minimal two-component DM scenario, the FIMP not only starts contributing to the thermal relic density but also comes under the domain of sensitivity of future direct and indirect detection experiments via WIMP-loop mediated processes. This scenario was called `pFIMP', to distinguish it from the standard single component FIMP DM.

In this work, we explore a WIMP-pFIMP framework where the WIMP interacts with the SM sector through the lepton portal in addition to the usual scalar-portal interaction. In lepton portal DM models \cite{Bai:2014osa, Kawamura:2020qxo, PandaX:2024pjr, DiazSaez:2022nhp, Asadi:2023csb}, DM couples directly to a charged lepton and mediator. This kind of model has a rich phenomenology at the collider experiments, which will give rise to a unique signature.
Not only that, these interactions also lead to lepton-flavor violating decays, which are absent in the SM. Thus, parameter space of such models can be constrained by low energy experiments, on the other hand, it is also possible to look for signals of this model in future low energy experiments as well.
Consequently, we are interested in exploring the complementarity between LHC constraints and the limits set by lepton flavor violation. At the same time, we also look into the signal at the direct and indirect detection experiments coming from such a scenario.

The plan of our work is as follows. 
In section~\ref{sec2}, we discuss our model. The constraints on our model from lepton flavor violation are taken into consideration in section~\ref{sec3}.
In section~\ref{sec4}, the dark matter phenomenology involving WIMP-pFIMP and all the constraints from the direct and indirect searches are discussed.
Finally, in section~\ref{sec5}, we recast LHC constraints on our model and then calculate projections at HL-LHC as well as future lepton colliders.
We summarize our results in section~\ref{sec6}.
\section{The model}
\label{sec2}
Our main motivation in this work, is to connect the lepton-flavor sector with the dark sector which can be enabled by lepton portal mechanism. In addition, we also want to explore the pFIMP regime, in a lepton portal DM model. Keeping these in mind we write the minimal Lagrangian in Equation~\ref{eq:model}. 

\begin{gather}
\begin{split}
\mathcal{L}_{}\,=\,&\mathcal{L}_{\rm SM}+\mu_H^2H^{\dagger}H-\lambda_H (H^{\dagger}H)^2+\frac{1}{2}|\partial_{\mu}\phi |^2-\frac{1}{2}\mu_{\phi}^2\phi^2-\frac{1}{4\,!}\lambda_{\phi}\phi^4+|\partial_{\mu}\chi |^2-\mu_{\chi}^2|\chi|^2
\\&-\lambda_{\chi}|\chi^*\chi|^2-\frac{1}{2}\mu_{3}\left[\chi^3+\left(\chi^{*}\right)^3\right]-\frac{1}{2}\lambda_{\phi \rm H}\phi^2H^{\dagger}H-\lambda_{\chi \rm H}|\chi|^2H^{\dagger}H-\frac{1}{2}\lambda_{\chi\phi}|\chi|^2\phi^2
\\ & + \overline{\psi}\left[i\gamma^{\mu}\left(\partial_{\mu}+ig^{\prime}{\tt Y} B_{\mu}\right)-m_{\psi}\right]\psi\,-\,\sum_{\ell}\mathtt{y}_{\ell} \overline{\psi}\ell_R\chi+\rm h.c.\,.
\end{split}\label{eq:model}
\end{gather}

In this context, one of the simplest approaches is to define the interaction terms that connect the SM fermions to the DM through charged dark sector particles.
In a simplified scenario, this charged partner is coupled to the right-handed SM fermions ($f_R$) through a renormalizable interaction term, like, $\overline{\Psi} f_R \Phi$ where $\Psi$ and $\Phi$ represent a vector-like Dirac fermion and a scalar, respectively.
The $\Psi$ could serve as a viable DM candidate if $\Phi$ carries SM-like charges, similar to SM fermions.
Alternatively, $\Phi$ could be a DM candidate \cite{Bai:2014osa,Mandal:2018czf}, provided that $\Psi$ is a charged Dirac fermion having SM hypercharge \cite{DiazSaez:2022nhp}. Crucially, the charged particle will ultimately decay into DM and SM fermions in both scenarios.

These kinds of interaction terms can influence a connection between SM flavor anomalies, flavor violating decays and DM phenomenology. Here, we only focused on the LFV decays, while other aspects will be addressed elsewhere.
Introducing a new co-annihilating partner, such as a charged particle, may lead to an under-abundant parameter space, which could be resolved by introducing an additional DM component to account for the remaining relic density.
Different types of multiparticle DM scenarios are possible, depending on the interactions between DM and SM particles, as well as DM-DM interactions.
The most promising scenarios: WIMP-WIMP \cite{Bhattacharya:2013hva,Bhattacharya:2016ysw}, WIMP-FIMP \cite{DuttaBanik:2016jzv,Bhattacharya:2021rwh}, and WIMP-pFIMP \cite{Bhattacharya:2022vxm,Bhattacharya:2022dco,Bhattacharya:2024nla}, among others.
In this work, we focus exclusively on the WIMP-pFIMP scenario, which provides richer phenomenology than the WIMP-WIMP case, particularly in the presence of lepton portal interactions.
The pFIMP does not directly couple to leptons and only interacts via the WIMP loop, which is comparatively suppressed relative to the WIMP.
Nevertheless, a correlation can still be established between the parameter space permitted by LFV decay and the parameter space responsible for pFIMP regime.

Our model consists of two DM fields: a real scalar $\phi$ which transforms under $\mathcal{Z}_2$ and a complex scalar $\chi$ is transformed under $\mathcal{Z}_3$.
The dark sector is further extended by introducing a charged vector-like lepton $\psi$, also transformed under $\mathcal{Z}_3$, that interacts exclusively only with right-handed charged leptons ($\ell_R$).
The charges of dark fields under $\mathcal{Z}_2\otimes \mathcal{Z}_3$ symmetry are shown in Tab\,.~\ref{tab:tab1},
\begin{table}[htb!]
\begin{center}
\begin{tabular}{|c|cc|}\hline
\rowcolor{magenta!15}{\bf Dark Fields} &$\mathcal{Z}_2$&$\mathcal{Z}_3$\\
\rowcolor{cyan!15}  $\phi$&$-1$&$+1$\\
\rowcolor{lime!15}  $\chi$ & $+1$&$\omega/\omega^2$\\
\rowcolor{gray!15}  $\psi$ &$+1$&$\omega/\omega^2$ \\\hline
\end{tabular}
\end{center}
\caption{Dark sector fields and their corresponding quantum numbers while the charge fermion has $U(1)_{\tt{Y}}$ hypercharge, $\tt{Y}=-1$.}
\label{tab:tab1}
\end{table} 
and the SM extended Lagrangian is written as,

\noindent
where $g^{\prime}_{}=(2/v)\sqrt{m_Z^2-m_W^2}$ is the $\rm U(1)_{\tt Y}$ gauge coupling.
$\psi$\footnote{The source of origin of low energy lepton portal renormalizable interaction term, $\overline{\psi}\ell_R\chi$, could be a dimension-5 effective operator $\left(C_{\ell}/\Lambda \right) \overline{\uppsi}H\ell_R\chi$, where $\uppsi=(\psi^0~\psi^-)^{\rm T}$ is a vector like Lepton doublet, and also transform similarly under $\mathcal{Z}_3$ like $\chi$. After EWSB, obtain a term like $\left(C_{\ell}v/\sqrt{2}\Lambda\right)\overline{\psi}\chi \ell_R\equiv \mathtt{y}_{\ell} \overline{\psi}\chi \ell_R$.} is a vector-like lepton (VLL) \cite{Athron:2021iuf, Bai:2014osa, Kawamura:2020qxo, PandaX:2024pjr} with weak hypercharge ${\tt Y} =-1$ and charged under $\mathcal{Z}_3$. For theoretical constraints on the model parameters, see the following \cite{Belanger:2012zr}.
\section{Constraints on model parameters}
\label{sec3}
The observed DM relic density is $\Omega_{\rm DM}h^{2} = 0.1200\pm0.0012$ \cite{Planck:2018vyg}. In this work, we investigate the parameter space of the model consistent with this relic density. The perturbative limit on the lepton portal coupling is $\mathtt{y}_{\ell}<\sqrt{4\pi}$ and $m_{\psi}>m_{\chi}+m_e$ to ensure the stability of DM candidate $\chi$. Furthermore, the charged fermion is strongly constrained by the LEP \cite{L3:2001xsz, ALEPH:2002nwp, OPAL:2003nhx, DELPHI:2003uqw, L3:2003fyi}, $m_{\psi}\gtrsim 103.5$ GeV.
The other theoretical constraints (unitarity, perturbativity, vacuum stability) on the model parameters are available in the following references \cite{Athron:2018ipf, Choi:2021yps}. Higgs invisible decay constraints are measured by ATLAS $\mathcal{B}( h \rm \to inv )<0.107$ \cite{ATLAS:2023tkt} and CMS $\mathcal{B}(h \rm \to inv )<0.15$ \cite{CMS:2023sdw} at 95\% CL, applicable when $m_{\rm DM}\leq m_h/2$.
The observed total decay width of the Higgs boson (based on indirect measurement) is $\Gamma_h=3.2^{+2.4}_{-1.7}\rm~ MeV$ \cite{CMS:2022ley}, while the SM expectation is $4.1\rm ~MeV$ \cite{LHCHiggsCrossSectionWorkingGroup:2016ypw}.
The loop-mediated decay of $Z$ boson to WIMP is constrained by recent $Z$ invisible decay width bound has come from various experiments like \cite{CMS:2021qbc, CMS:2022ett, ALEPH:2005ab},
\bea\begin{split}
\rm\Gamma_{Z\to invisible}
<  \begin{cases}\rm523\pm 16~MeV~~~( CMS)\,,\\\rm 503\pm 16~MeV~~~(LEP)\,,\\\rm498\pm 17~ MeV~~~(L3)\,.\end{cases}
\end{split}\label{Z_invisible_decay}
\eea
\subsection{Lepton flavor constraints}
A stringent bound on the couplings of the DM particle and the heavy VLL appears from the measurements of the anomalous magnetic moments of leptons \cite{Chacko:2001xd, Lindner:2016bgg, Barducci:2018esg, Aoyama:2019ryr, Acaroglu:2022hrm,Leveille:1977rc,Acaroglu:2023cza,DAlise:2022ypp}.These processes are lepton flavor conserving in nature. 
\begin{figure}[htb!]
\centering
\begin{tikzpicture}
\begin{feynman}
\vertex(a){\(\ell_i\)};
\vertex[ right =1.5cm  of a] (a1);
\vertex[ right =3.5cm  of a] (a2);
\vertex[ right =5cm of a] (a3){\(\ell_j\)};
\vertex[ above right =1cm and 1cm of a1] (b);
\vertex[ above right=1cm and 1cm of b] (b1){\(\gamma\)};
\diagram*{
(a) -- [line width=0.25mm, fermion, arrow size=0.7pt,style=black] (a1),
(a2) -- [line width=0.25mm, half left, charged scalar, arrow size=0.7pt,style=gray!75,edge label={\({\color{black}\rm\chi} \)}]  (a1), 
(a2)-- [line width=0.25mm, fermion,arrow size=0.7pt,edge label={\(\rm \)},style=black] (a3),
(a1) -- [line width=0.25mm, fermion,quarter left, arrow size=0.7pt,style=gray!75,edge label={\({\color{black}\rm\psi} \)}] (b),
(b) -- [line width=0.25mm, fermion, quarter left,arrow size=0.7pt,style=gray!75,edge label={\({\color{black}\rm\psi} \)}] (a2),
(b)-- [line width=0.25mm, boson,arrow size=0.7pt,edge label={\({\color{black}\rm} \)},style=black] (b1)};
\node at (a1)[circle,fill,style=gray,inner sep=1pt]{};
\node at (b)[circle,fill,style=gray,inner sep=1pt]{};
\node at (a2)[circle,fill,style=gray,inner sep=1pt]{};
\end{feynman}
\end{tikzpicture}\quad
\begin{tikzpicture}
\begin{feynman}
\vertex(a){h};
\vertex[right = 2cm  of a] (b);
\vertex[above right = 1cm and 1cm  of b] (a1);
\vertex[below right = 1cm and 1cm  of b] (a2);
\vertex[above right = 0.25cm and 1cm of a1] (a3){\(\ell_i\)};
\vertex[below right = 0.25cm and 1cm of a2] (a4){\(\ell_j\)};
\diagram*{
(a) -- [line width=0.25mm,scalar, arrow size=0.7pt,style=black] (b),
(a2) -- [line width=0.25mm,charged scalar, arrow size=0.7pt,style=gray!75,edge label={\({\color{black}\rm\chi} \)}] (b)  -- [line width=0.25mm,charged scalar, arrow size=0.7pt,style=gray!75,edge label={\({\color{black}\rm\chi} \)}] (a1), 
(a3) -- [line width=0.25mm, fermion,arrow size=0.7pt] (a1),
(a2) -- [line width=0.25mm, fermion,arrow size=0.7pt] (a4),
(a1) -- [line width=0.25mm, fermion,arrow size=0.7pt,style=gray!75,edge label={\({\color{black}\rm\psi} \)}] (a2)};
\node at (a1)[circle,fill,style=gray,inner sep=1pt]{};
\node at (b)[circle,fill,style=gray,inner sep=1pt]{};
\node at (a2)[circle,fill,style=gray,inner sep=1pt]{};
\end{feynman}
\end{tikzpicture}
\caption{\textit{Left}: Diagram contributing to $\ell_i\to\ell_j\gamma$. \textit{Right}: Higgs boson decay to lepton pair, $h\to\ell_i\ell_j$.}
\label{fig:clfv}
\end{figure}
Although the anomalous magnetic dipole moment of electron ($\Delta a_e=g_e/2-1$) \cite{Schwartz:2014sze, Peskin:1995ev,pal2014introductory} is known precisely \cite{Fan:2022eto}, SM prediction \cite{Aoyama:2012qma, Aoyama:2019ryr} relies on the measurement of fine structure constant using the recoil velocity/frequency of atoms that absorb a photon. Currently, there is a $5.5 \sigma $ discrepancy between the measurements using Rubidium-87 \cite{Morel:2020dww} and Cesium-133 \cite{Parker:2018vye}.
\bea
\Delta a_{e}\equiv a_{e}^{\rm exp}-a_{e}^{\rm SM}=\begin{cases}
(-8.8\pm 3.6)\times 10^{-13}~(-2.4\sigma) &{\rm(Cs)}\\
(+4.8\pm 3.0)\times 10^{-13}~(+1.6\sigma)& {\rm(Rb)}
\end{cases}\,.
\label{eq:ae}
\eea
The anomalous magnetic moment of muon $a_{\mu}=g_{\mu}/2-1$ has been measured by BNL E821 and FNAL experiments yields a $4.2\sigma$ discrepancy \cite{Muong-2:2021ojo} from the SM prediction,
\bea
\Delta a_{\mu}\equiv a_{\mu}^{\rm exp}-a_{\mu}^{\rm SM}=(2.49\pm 0.48)\times 10^{-9}\,~\text{ \cite{Muong-2:2023cdq, Datta:2023iln}}.
\label{eq:amu}
\eea
\begin{figure}[htb!]
\centering
\subfloat[]{\includegraphics[width=0.45\linewidth]{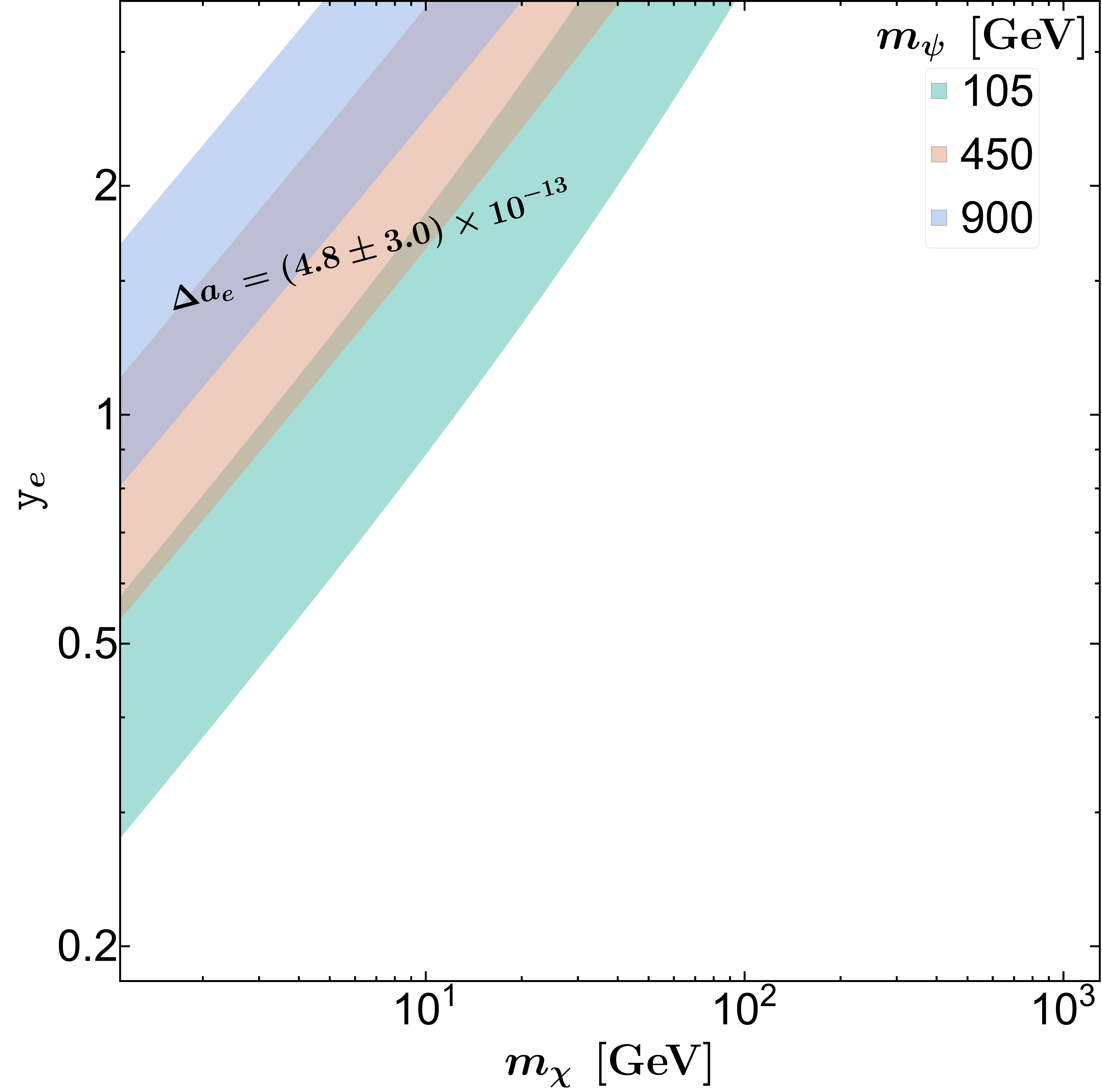}\label{fig:efc}}\quad
\subfloat[]{\includegraphics[width=0.45\linewidth]{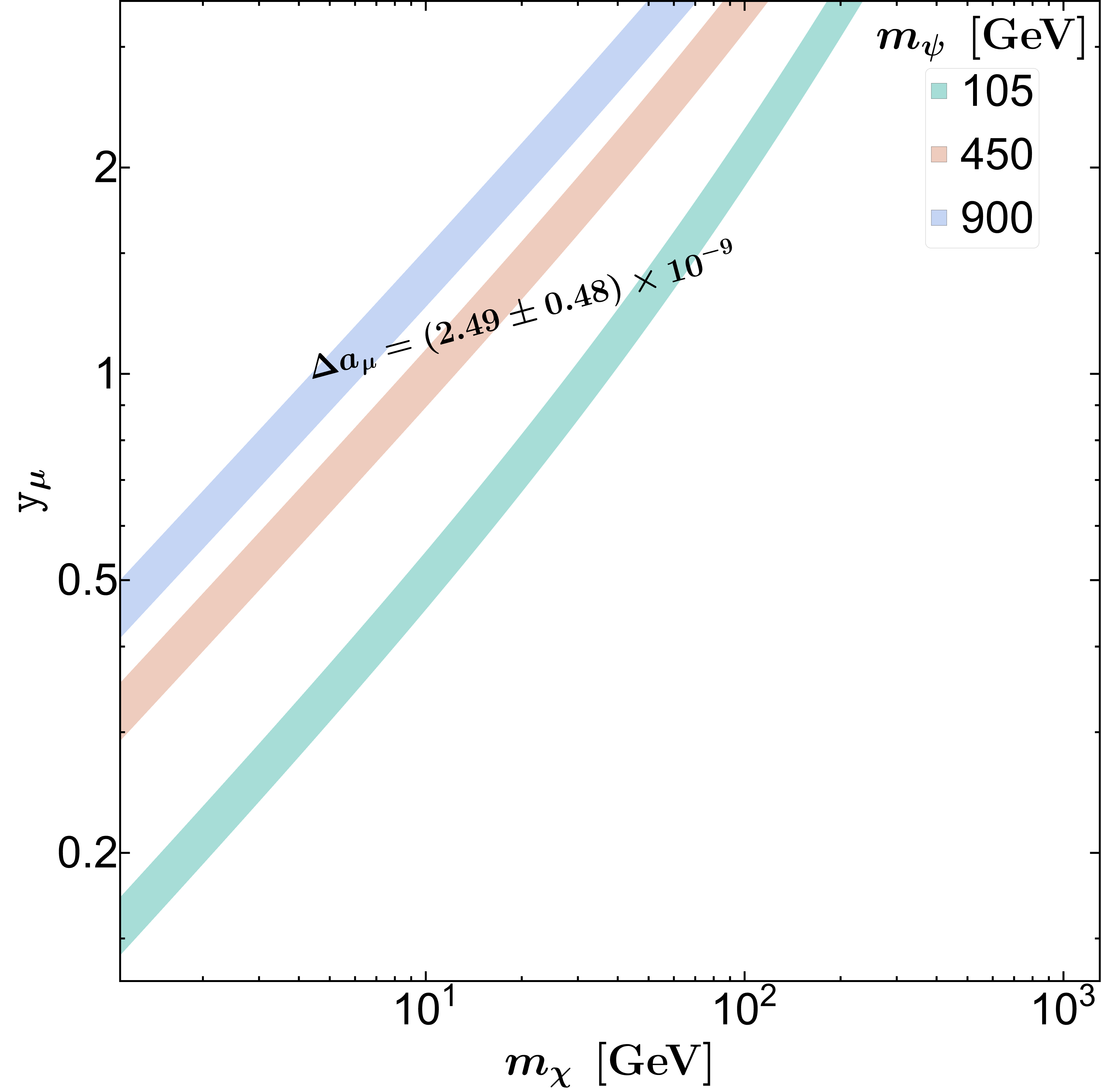}\label{fig:mufc}}
\caption{In Figs\,.~\ref{fig:efc}, and \ref{fig:mufc}, the bands represent the allowed regions on the $m_{\chi}-\mathtt{y}_{\ell} $ parameter space derived from the uncertainties in $\Delta a_{e}$ and $\Delta a_{\mu}$ measurements, respectively.}
\label{fig:lfc}
\end{figure}
Fig\,.~\ref{fig:lfc} illustrates the allowed parameter space in the $m_{\chi}-\mathtt{y}_{\ell}$ plane for the flavor conserving quantities, namely lepton anomalous magnetic moment $(\Delta a_{\ell})$. We mention that, these anomalies and their significant deviation from the SM expectations cannot be addressed in our analysis, where $\mathtt{y}_{\ell} \lesssim 0.1$ due to constraints from lepton-flavor violating decays, which will be discussed in detail. This limitation can be addressed by further extensions of our model in the future. On the other hand, the anomalous magnetic moment of the tau lepton, \(a_\tau = (g_\tau - 2) / 2\), can act as a sensitive indicator of potential new physics \cite{Volkotrub:2888478,Verducci:2023cgx}. However, achieving a precise measurement of \(a_\tau\) is considerably more challenging than for the magnetic moments of electrons and muons.
In addition to the lepton flavor conserving processes, there are potential contributions to lepton flavor violating (LFV) decays, $\ell_i \to\ell_j \gamma$ with $i\neq j$ in our model. The most stringent limits on LFV decays come from $\mathcal{B}_{\mu\to e\gamma}$\footnote{MEG II claims to achieve a future sensitivity for the $\mu\to e\gamma$ branching ratio of $6 \times 10^{-14}$ \cite{MEGII:2018kmf}.} followed by $\mathcal{B}_{\tau\to e\gamma}$ and $\mathcal{B}_{\tau\to \mu\gamma}$.
\bea
\begin{split}
\mathcal{B}_{\mu\to e\gamma} &< 3.1\times 10^{-13}~(90\%\rm~ C.L.)\,~\text{ \cite{MEGII:2023ltw}}, \\
\mathcal{B}_{\tau\to e\gamma} &< 3.3\times 10^{-8}~(90\%\rm~ C.L.)\,~\text{ \cite{BaBar:2009hkt}}, \\
\mathcal{B}_{\tau\to \mu\gamma} &< 4.2\times 10^{-8}~(90\%\rm~ C.L.)\,~\text{ \cite{Belle:2021ysv}}.
\label{eq:mueg}
\end{split}
\eea
\begin{figure}[htb!]
\centering
\subfloat[]{\includegraphics[width=0.32\linewidth]{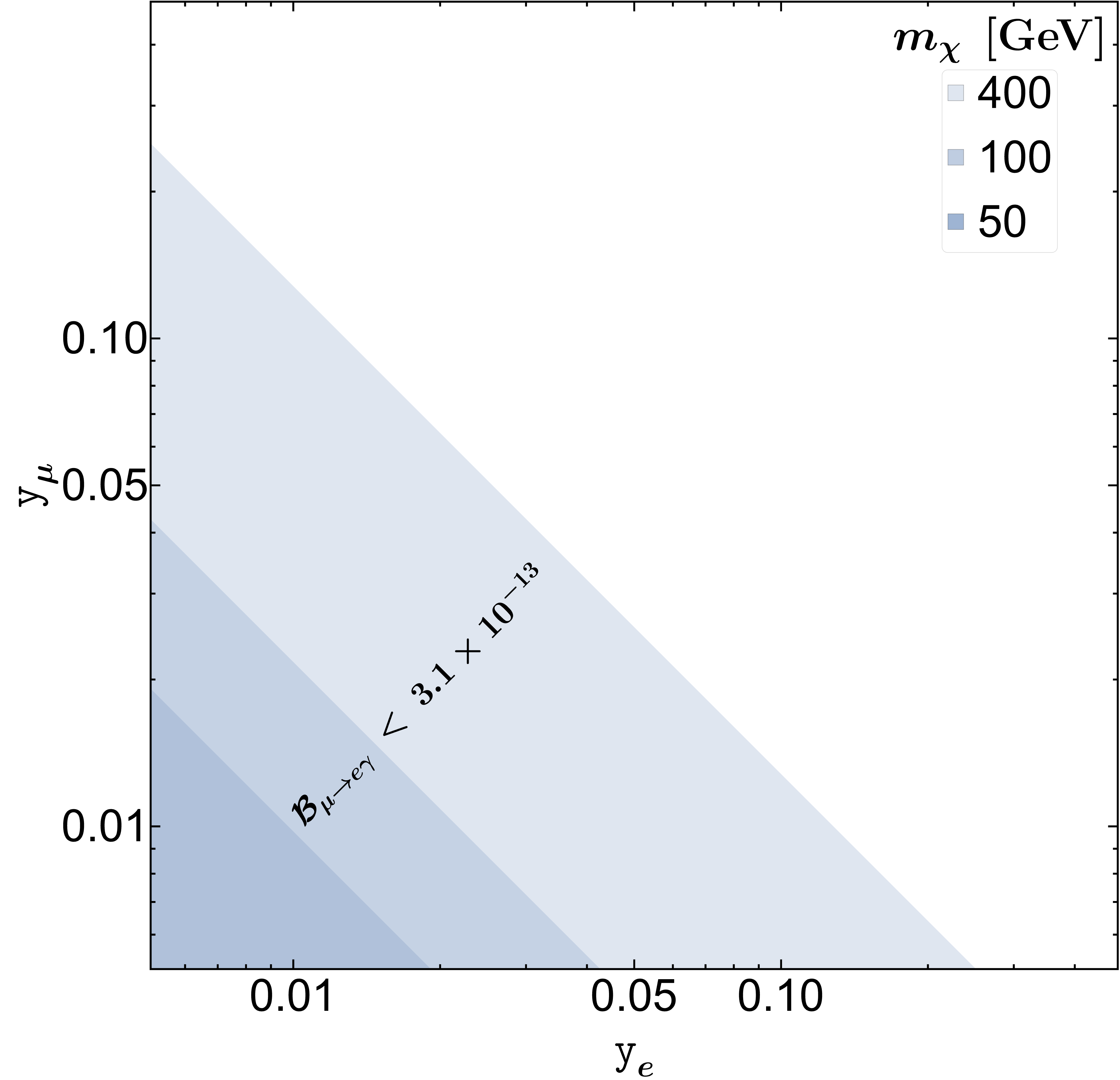}\label{fig:lfv1}}~
\subfloat[]{\includegraphics[width=0.32\linewidth]{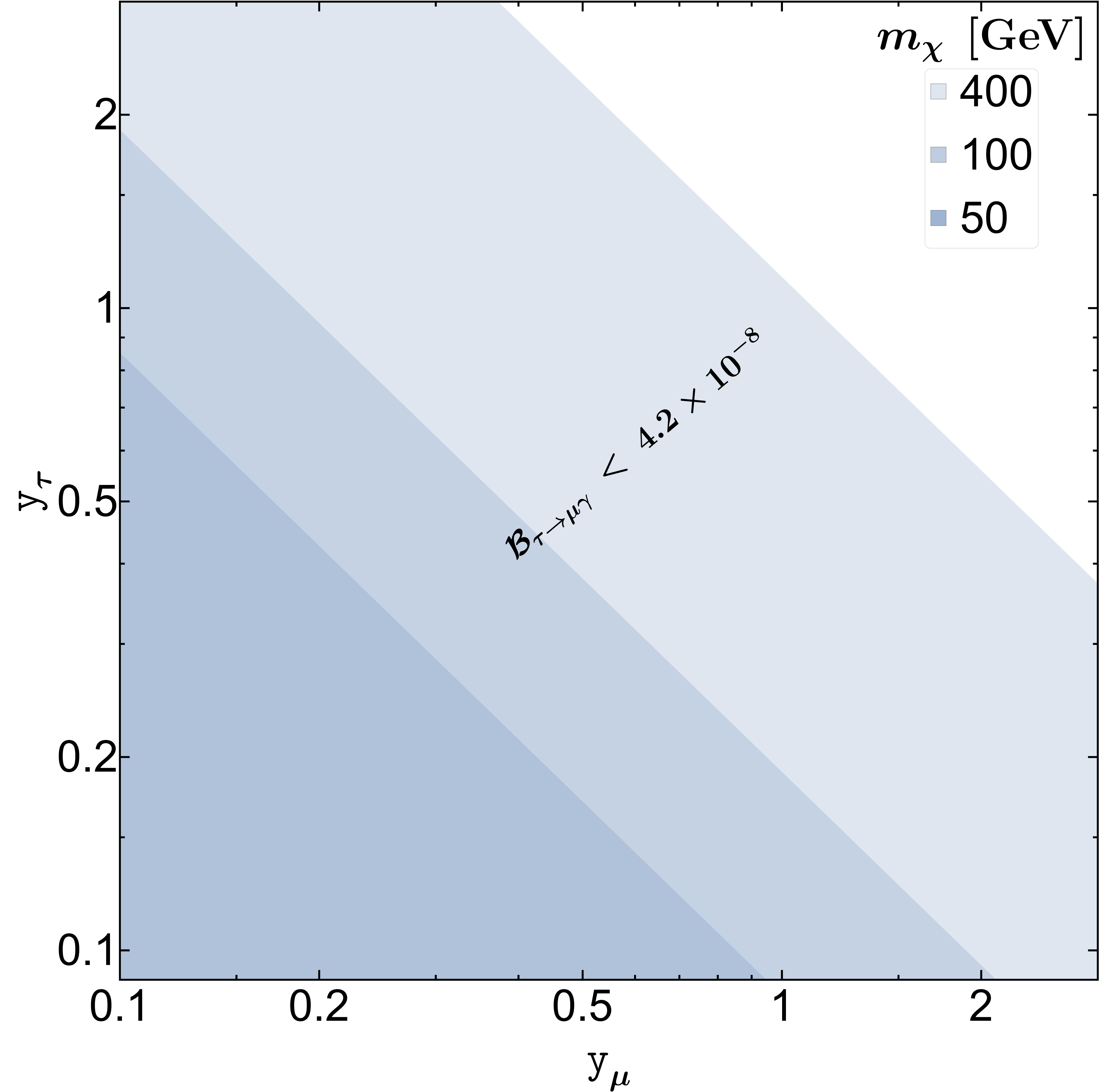}\label{fig:lfv2}}~
\subfloat[]{\includegraphics[width=0.32\linewidth]{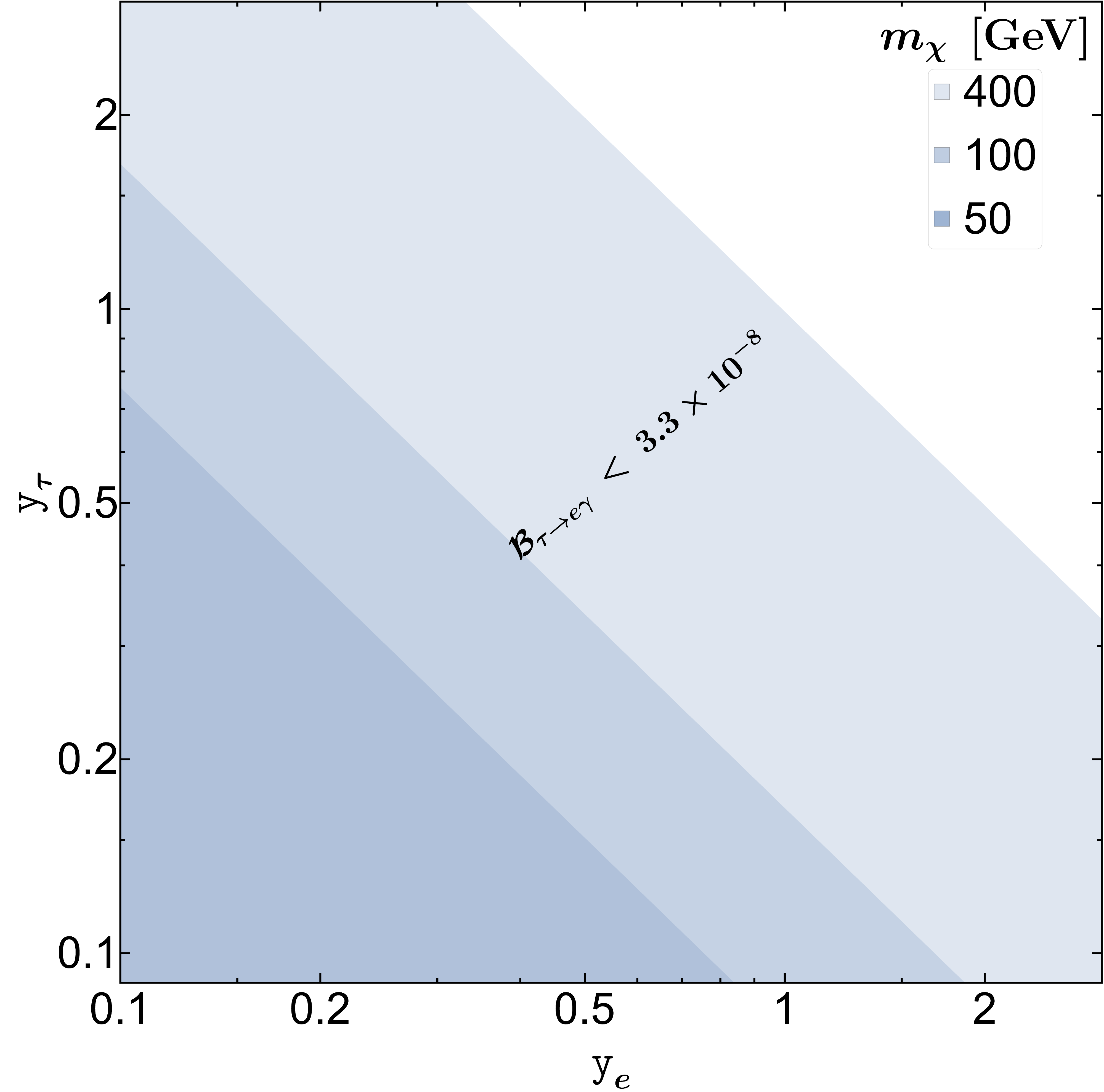}\label{fig:lfv3}}
\caption{In Figs\,.~\ref{fig:lfv1},~\ref{fig:lfv2}, and \ref{fig:lfv3}, the shaded regions represent the allowed parameter spaces for different $m_{\chi}$ from experimental measurements of LFV $\mu \to e \gamma$, $\tau \to  \mu\gamma$ and $\tau \to e \gamma$ decay channels, respectively. The region below each shade is indefinitely allowed for that shade; however, it overlaps with the next shade. In the analysis, we have set the charged fermion mass as $m_{\psi}=450~\rm GeV$.}
\label{fig:lfv}
\end{figure}
Fig\,.~\ref{fig:clfv} show the Feynman diagrams contributing to lepton flavor conserving ($i=j$) and violating ($i\neq j$) processes. Fig\,.~\ref{fig:lfc} illustrates the parameter space that can accommodate the anomalies $\Delta a_e$ and $\Delta a_{\mu}$, represented by different color shades according to mass of the mediator $\psi$.
Fig\,.~\ref{fig:lfv} illustrates the allowed parameter space constrained by the lepton flavor violating processes, represented by different color shades according to the WIMP mass. Figs\,.~\ref{fig:lfv1}, \ref{fig:lfv2} and \ref{fig:lfv3}, illustrate the allowed parameter space for the LFV processes $\mu\to e\gamma$, $\tau\to \mu\gamma$, and $\tau\to e\gamma$, respectively, within the experimental bounds (see Eq. \ref{eq:mueg}) in the $\mathtt{y}_e-\mathtt{y}_{\mu}$, $\mathtt{y}_{\mu}-\mathtt{y}_{\tau}$, and $\mathtt{y}_{e}-\mathtt{y}_{\tau}$ planes. The varying color shades represent different WIMP masses in each case.

\subsection{Higgs decay to lepton pairs}
At the LHC, for a Higgs boson mass of 125 GeV, the observed upper limits on the branching fraction of its decay to lepton pairs, at 95\% C.L. are given below:
\begin{equation}
\begin{split}
& \mathcal{B}_{h\to ee} < 3.0\times 10^{-4}~\quad\text{\cite{CMS:2022urr}}\,, \\
& \mathcal{B}_{h\to \mu\mu} < 2.6^{+1.3}_{-1.3} \times 10^{-4}\quad\text{\cite{ATLAS:2022vkf}}\,, \\
& \mathcal{B}_{h\to \tau\tau} < 6.0^{+0.8}_{-0.7}\times 10^{-2}\quad\text{\cite{ATLAS:2022vkf}}\,, \\
& \mathcal{B}_{h\to e \mu} < 4.4\times 10^{-5}~\quad\text{\cite{CMS:2023pte}}\,, \\
& \mathcal{B}_{h\to \mu \tau} < 1.5\times 10^{-3}~ \quad\text{\cite{CMS:2021rsq}}\,, \\
& \mathcal{B}_{h\to e \tau} < 2.0\times 10^{-3}~\quad\text{\cite{ATLAS:2023mvd}}\,,
\end{split}
\end{equation}
where $\mathcal{B}_{h\to \ell_i\ell_j}= \Gamma_{h\to \ell_i \ell_j}/\Gamma_{h}^{\rm total}$ and $\Gamma^{\rm total}_{h}=4.1~\rm MeV$ \cite{Hong:2020qxc,Fajfer:2021cxa}. However, even the most stringent bound on $h\to e \mu$ is far less restrictive compared to the $\mu \to e \gamma$ constraint, hence they do not constrain our parameter spaces any further.
\section{Dark Matter phenomenology}
\label{sec4}
In the extended SM Lagrangian, as presented in Eq\,.~\eqref{eq:model}, we have introduced three new particles $\psi$, $\chi$, and $\phi$. The real scalar particle $\phi$ is absolutely stable, in the absence of its decay term owing to $\mathcal{Z}_2$ symmetry. Among the complex scalar and vector-like charged fermion, the lightest one will serve as a stable DM candidate, and the responsible symmetry is $\mathcal{Z}_3$ in this case. The corresponding charges for these particles under these symmetries transformations are provided in Tab\,.~\ref{tab:tab1}. Another key point to highlight is that a charged DM \cite{Stebbins:2019xjr,Munoz:2018pzp,DeRujula:1989fe,Agrawal:2016quu,Kadota:2016tqq,Davidson:2000hf,Iles:2024zka,Berlin:2024lwe,Fiorillo:2024upk} is heavily constrained by observations, and hence we do not consider the charged DM possibility in our context. Therefore, we always choose $m_{\psi} > m_{\chi} + m_e$ to ensure that $\chi$ remains a stable DM candidate. Finally, these two DM components ($\phi$ and $\chi$) would contribute to the total DM relic density. We emphasize that, the Higgs-portal coupling of $\phi$, $\lambda_{\phi \rm H}$ is taken $10^{-12}$ to ensure the pFIMP dynamics as discussed in \cite{Bhattacharya:2022vxm, Bhattacharya:2022dco}.
In order to avoid $\phi$ equilibrating with the SM states, $\lambda_{\phi \rm H}$ must be less than $10^{-6}$, yet it cannot be zero in order to allow its production by Higgs. Therefore we choose a value $10^{-12}$, though we could increase or decrease that value by several orders of magnitude with no effect on this analysis.
In the rest of the paper, we exclusively consider this regime. 
\subsection{cBEQ and relic density}
Let us assume CP conservation exists inside the dark sector. The coupled Boltzmann equations (cBEQ) for the two DMs, where $Y_{\rm w}=Y_{\chi}+Y_{\chi^*}+Y_{\psi^{+}}+Y_{\psi^{-}}$ and $Y_{\rm w}^{\rm eq}=Y_{\chi}^{\rm eq}+Y_{\chi^*}^{\rm eq}+Y_{\psi^{+}}^{\rm eq}+Y_{\psi^{-}}^{\rm eq}$, is
\begin{align}
\nonumber\dfrac{dY_{\phi}}{dx}\,=\,&\frac{2~\bf{s}}{x~\mathcal{H}(x)}\left[\frac{1}{\textbf{s}} \left(Y_{h}^{\rm eq }-Y_{h}^{\rm eq}\frac{Y_{\phi}^2}{Y_{\phi}^{\rm eq^2}}\right)\langle\Gamma\rangle_{h\to \phi~\phi}\,+\,\left(Y_{\rm SM}^{\rm eq^2}-Y_{\rm SM}^{\rm eq^2}\frac{Y_{\phi}^2}{Y_{\phi}^{\rm eq^2}}\right)\langle\sigma v\rangle_{\rm{SM~ SM}\to\rm \phi ~\phi}
\right.\\ &\left.+\left(Y_{\rm w}^2-Y_{\rm w}^{\rm eq^2}\frac{Y_{\rm \phi}^2}{Y_{\rm \phi}^{\rm eq^2}}\right)\langle\sigma v\rangle^{\rm eff}_{\rm conv}\right]\,,\\
\nonumber\dfrac{dY_{\rm w}}{dx}\,=\,&-\frac{\bf{s}}{x~\mathcal{H}(x)}\left[\left(Y_{\rm w}^2-Y_{\rm w}^{\rm eq^2}\right)\langle\sigma v\rangle^{\rm eff}_{\rm ann}+\frac{1}{2}\left(Y_{\rm w}^2-Y_{\rm w}Y_{\rm w}^{\rm eq}\right)\langle\sigma v\rangle^{\rm eff}_{\rm semi}+\left(Y_{\rm w}^2-Y_{\rm w}^{\rm eq^2}\frac{Y_{\phi}^2}{Y_{\phi}^{\rm eq^2}}\right)\langle\sigma v\rangle^{\rm eff}_{\rm conv}\right]\,,
\end{align}
where $Y_i=n_i/{\bf s}$, $n_i$ is the number density of $i^{\rm th}$ particle, {\bf s} is the entropy density, and other terms carry the usual meaning.
{\small
\begin{flalign}
\langle\sigma v\rangle^{\rm eff}_{\rm ann}&=\langle\sigma v\rangle_{\chi~\chi^*\to \rm SM~SM} \dfrac{n_{\chi}^{\rm eq^2}}{n_{\rm w}^{\rm eq^2}}+\langle\sigma v\rangle_{\psi^-~\psi^+\to \rm SM~SM} \dfrac{n_{\psi}^{\rm eq^2}}{n_{\rm w}^{\rm eq^2}}+2\langle\sigma v\rangle_{\psi^-\chi^*\to \rm SM~SM} \dfrac{n_{\psi}^{\rm eq}n_{\chi}^{\rm eq}}{n_{\rm w}^{\rm eq^2}}\,,\\
\langle\sigma v\rangle^{\rm eff}_{\rm semi}&=2\langle\sigma v\rangle_{\psi^-~\chi\to \rm \chi^*~SM} \dfrac{n_{\psi}^{\rm eq}n_{\chi}^{\rm eq}}{n_{\rm w}^{\rm eq^2}}+2\left(\langle\sigma v\rangle_{\chi~\chi\to \rm \psi^+~SM}+\langle\sigma v\rangle_{\chi~\chi\to \rm \chi^*~SM}\right) \dfrac{n_{\chi}^{\rm eq^2}}{n_{\rm w}^{\rm eq^2}}\,,\\
\langle\sigma v\rangle^{\rm eff}_{\rm conv}&=\langle\sigma v\rangle_{\chi~\chi^*\to\phi ~\phi} \dfrac{n_{\chi}^{\rm eq^2}}{n_{\rm w}^{\rm eq^2}}\,,
\end{flalign}}
where ${\rm SM}=\{h,\rm~W^{\pm},~Z,~leptons,~quarks\}$. The total relic density is given by, in terms of DM yields as the solution of cBEQ,
\bea
{\rm \Omega_{\rm DM}h^2} = 2.744 \times 10^8 \left[m_{\chi}Y_{\rm w}+m_{\phi}Y_{\phi}\right]_{x\to \infty}\,.
\eea
We numerically solved the cBEQ using \texttt{micrOMEGAs} \cite{Alguero:2023zol} after importing the model generated with \texttt{FeynRules} \cite{Christensen:2008py, Alloul:2013bka}. The results of the cBEQ solution and the relic density allowed parameter space are illustrated in Figs\,.~\ref{fig:wimp-dd},~\ref{fig:pfimp-relic},~and~\ref{fig:wimp-id}.
\subsection{Direct detection}
DM-nucleus scattering is one of the crucial methods for detecting dark matter (DM). Experiments such as XENON1T \cite{XENON:2018voc}, XENONnT \cite{XENON:2023cxc}, and LUX-ZEPLIN \cite{LZ:2022lsv} have set an upper limit on the DM-nucleon scattering cross-section, while PandaX-xT \cite{PandaX:2024oxq} and DARWIN/XLZD \cite{Baudis:2023pzu} provides projected limits. In our model, both DMs are weakly coupled with the visible sector and might have a possibility of detection in future direct detection experiments. In Fig\,.~\ref{fig:feyn-dd}, we represent the possible Feynman diagram of DM-nucleon scatterings.
\begin{figure}[htb!]
\centering
\subfloat[]{\begin{tikzpicture}
\begin{feynman}
\vertex (a);
\vertex[left=1cm and 1cm  of a] (a1){\(\chi\)};
\vertex[right=1cm and 1cm  of a] (a2){\(\chi\)}; 
\vertex[below=2cm of a] (b); 
\vertex[left=1cm and 1cm of b] (c1){\(\rm N\)};
\vertex[right=1cm and 1cm of b] (c2){\(\rm N\)};
\diagram*{
(a1) -- [line width=0.25mm,charged scalar, arrow size=0.7pt, style=black] (a),
(a) -- [line width=0.25mm,charged scalar, arrow size=0.7pt, style=black] (a2),
(a) -- [line width=0.25mm,scalar, edge label={\(\color{black}{h}\)}, style=gray] (b),
(c1) -- [line width=0.25mm,fermion, arrow size=0.7pt] (b),
(b) -- [line width=0.25mm,fermion, arrow size=0.7pt] (c2)};
\node at (a)[circle,fill,style=gray, inner sep=1pt]{};
\node at (b)[circle,fill,style=gray, inner sep=1pt]{};
\end{feynman}
\end{tikzpicture}
\label{feyn:wimp-dd}}\quad
\subfloat[]{\begin{tikzpicture}
\begin{feynman}
\vertex (a);
\vertex[below=1cm of a] (b);
\vertex[above=2cm of b] (c); 
\vertex[left=1cm  of c] (a1){\(\phi\)};
\vertex[right=1cm of c] (a2){\(\phi\)};  
\vertex[left=1cm of b] (c1){\(\rm N\)};
\vertex[right=1cm of b] (c2){\(\rm N\)};
\diagram*{
(a1) -- [line width=0.25mm,scalar, style=black, arrow size=0.7pt] (c),
(c) -- [line width=0.25mm,scalar, style=black, arrow size=0.7pt] (a2),
(a) -- [line width=0.25mm,charged scalar,half left, edge label={\(\color{black}{\chi}\)}, style= gray, arrow size=0.7pt](c),
(c) -- [line width=0.25mm,charged scalar,half left, edge label={\(\color{black}{\chi}\)}, style= gray, arrow size=0.7pt] (a),
(a) -- [line width=0.25mm,scalar, edge label={\(\color{black}{h}\)}, style=gray] (b),
(c1) -- [line width=0.25mm,fermion, arrow size=0.7pt] (b),
(b) -- [line width=0.25mm,fermion, arrow size=0.7pt] (c2)};
\node at (a)[circle,fill,style=gray,inner sep=1pt]{};
\node at (b)[circle,fill,style=gray,inner sep=1pt]{};
\node at (c)[circle,fill,style=gray,inner sep=1pt]{};
\end{feynman}
\end{tikzpicture}\label{feyn:pfimp-dd}}
\caption{The Feynman diagrams are corresponding to the direct detection of WIMP (\ref{feyn:wimp-dd}) and pFIMP (\ref{feyn:pfimp-dd}).}
\label{fig:feyn-dd}
\end{figure}
\subsubsection*{WIMP}
The complex scalar WIMP $\chi$ interacts with the target nucleus through a Higgs-mediated process shown in Fig\,.~\ref{feyn:wimp-dd}, enabled by the Higgs portal interaction. In the WIMP-pFIMP scenario, we use the effective WIMP-nucleon cross-section, which is, $\rm \sigma_{\chi N}^{\rm eff}=(\Omega_{\chi}h^2/\Omega_{\rm DM}h^2)\sigma_{\chi \rm N}^{\rm SI}$ where $\sigma_{\chi N}^{\rm SI}$ is the spin-independent (SI) WIMP-nucleon scattering cross-section.
\begin{figure}[htb!]
\centering
\subfloat[]{\includegraphics[width=0.475\linewidth]{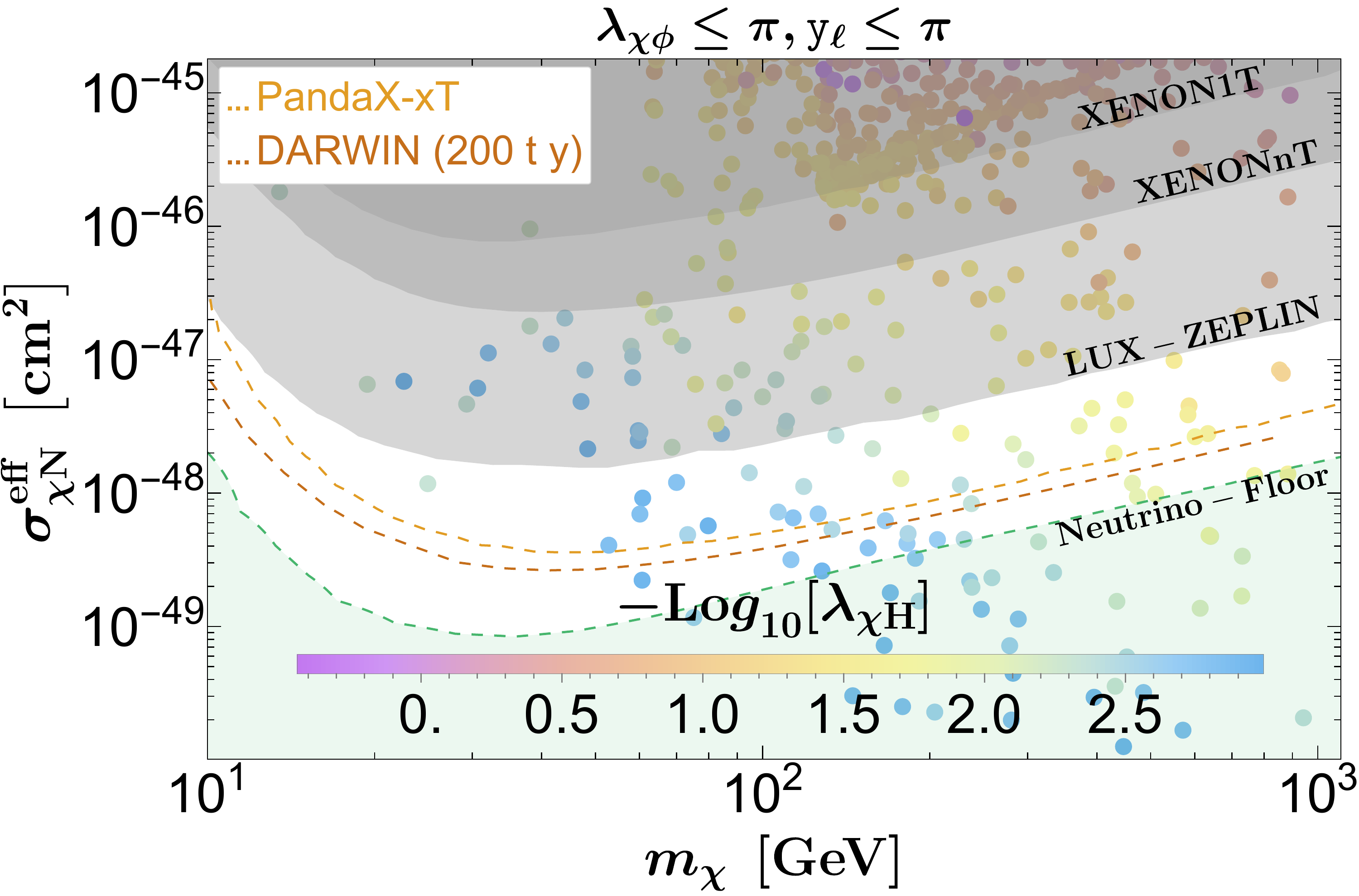}\label{figa}}\quad
\subfloat[]{\includegraphics[width=0.475\linewidth]{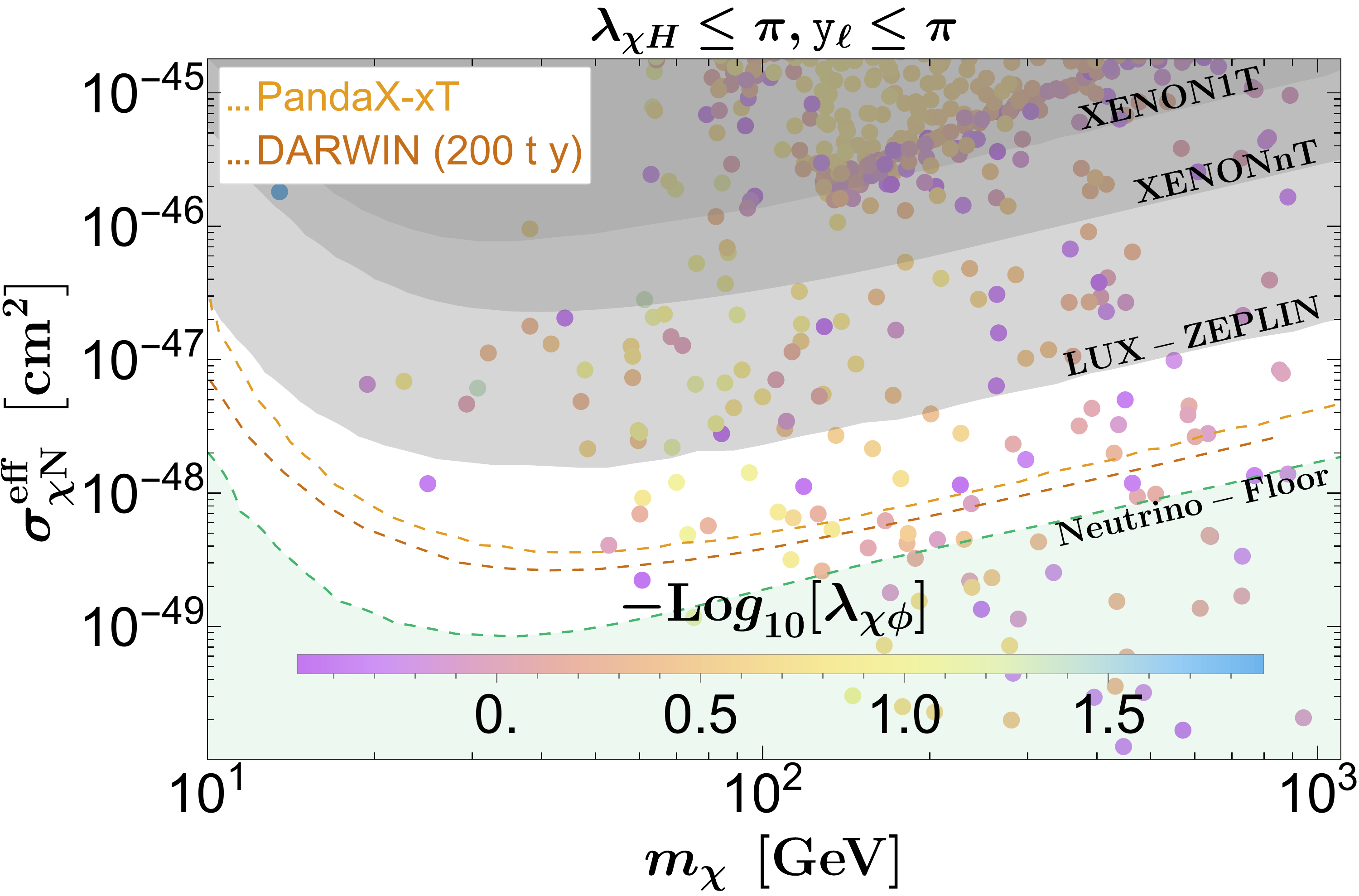}\label{figb}}

\subfloat[]{\includegraphics[width=0.475\linewidth]{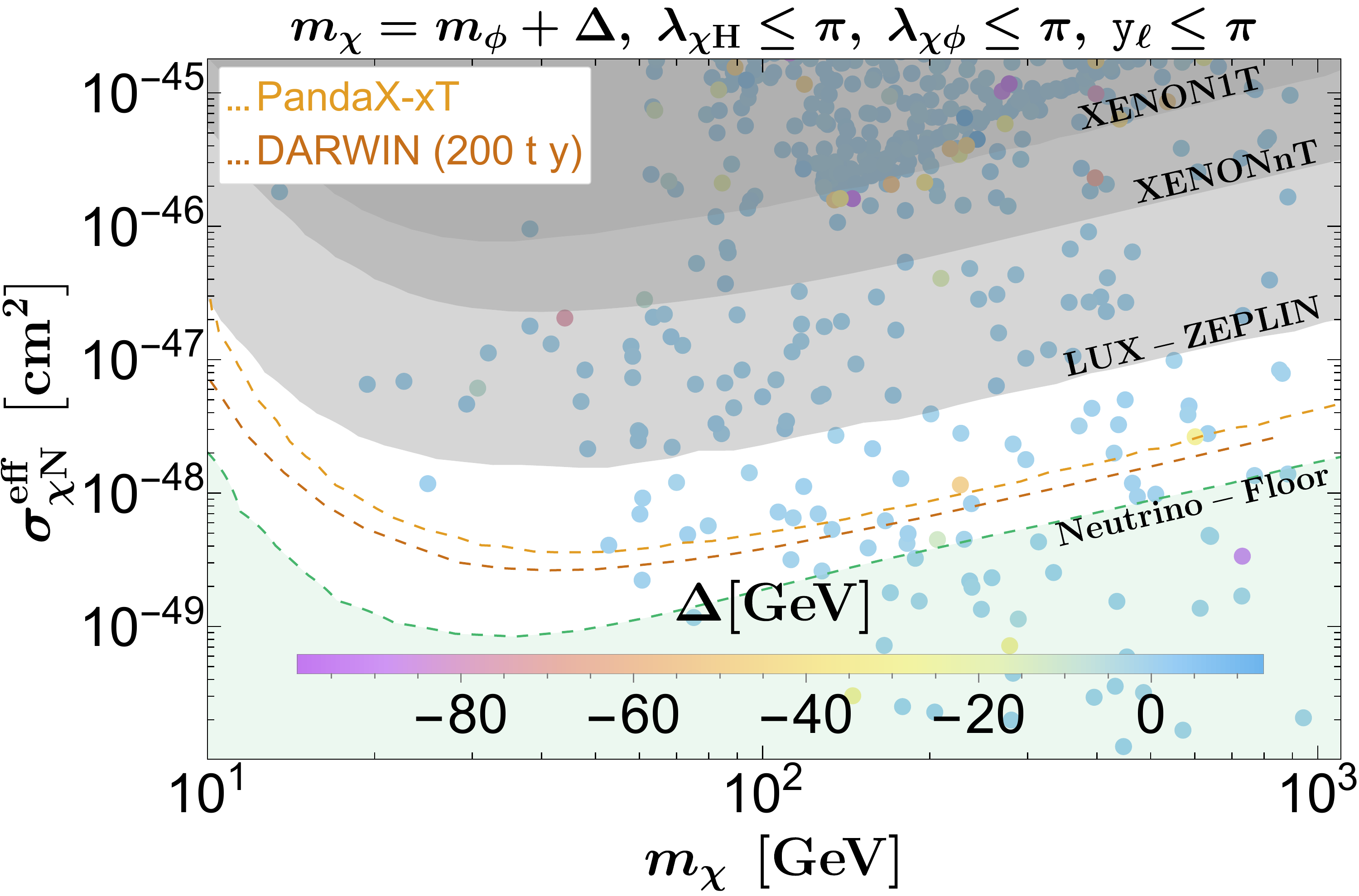}\label{figc}}\quad
\subfloat[]{\includegraphics[width=0.475\linewidth]{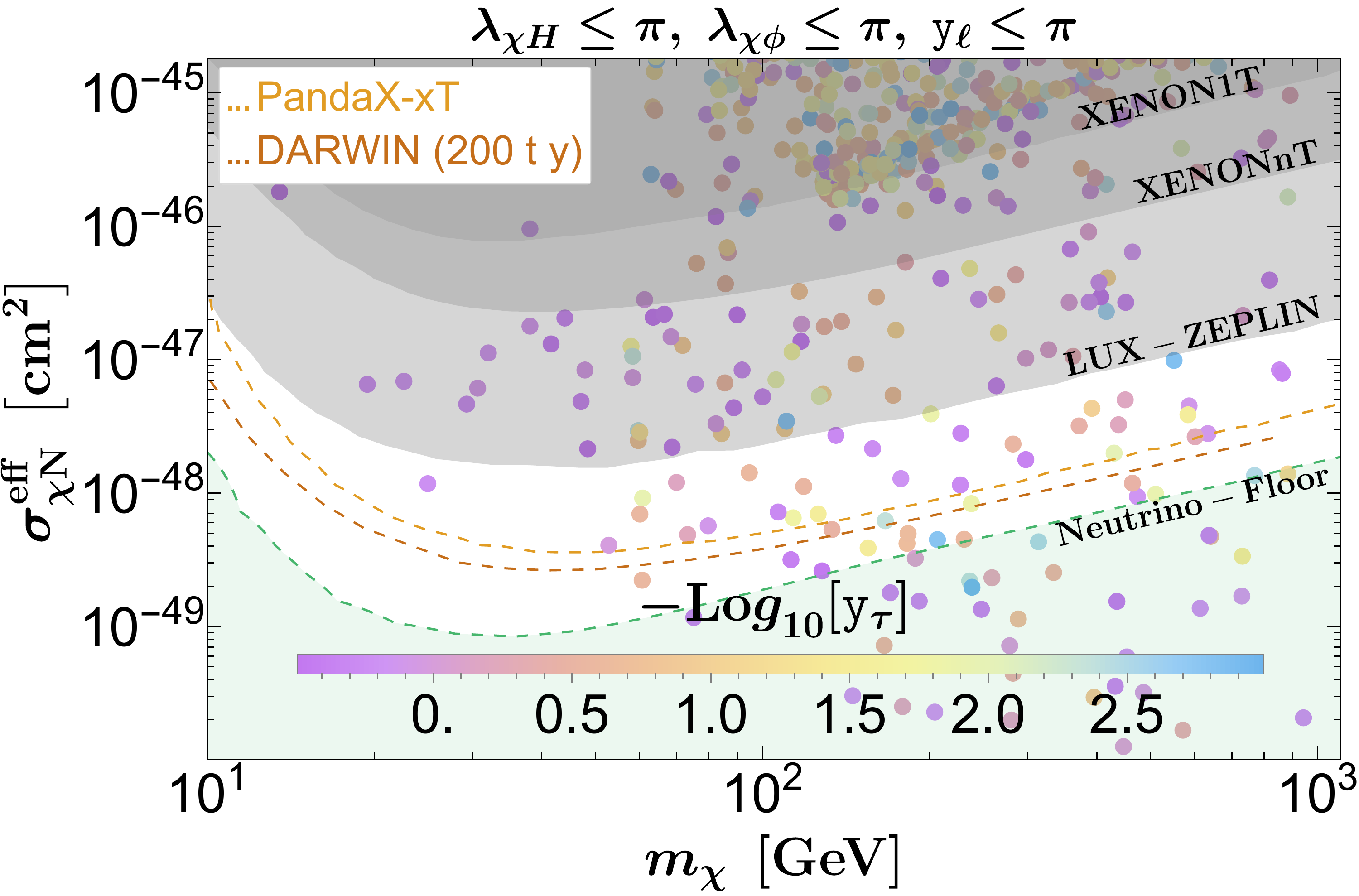}\label{figd}}

\subfloat[]{\includegraphics[width=0.475\linewidth]{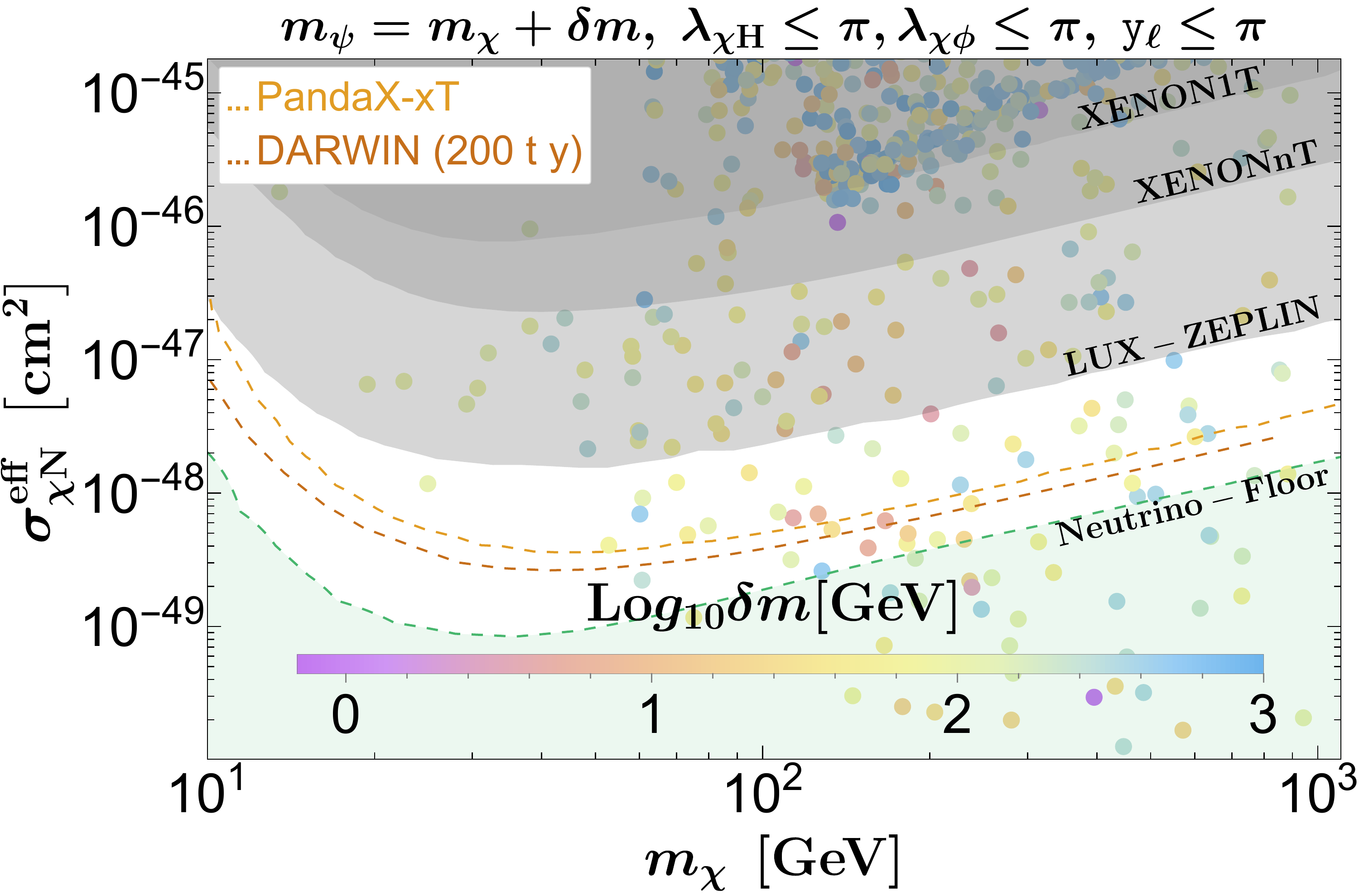}\label{fige}}\quad
\subfloat[]{\includegraphics[width=0.475\linewidth]{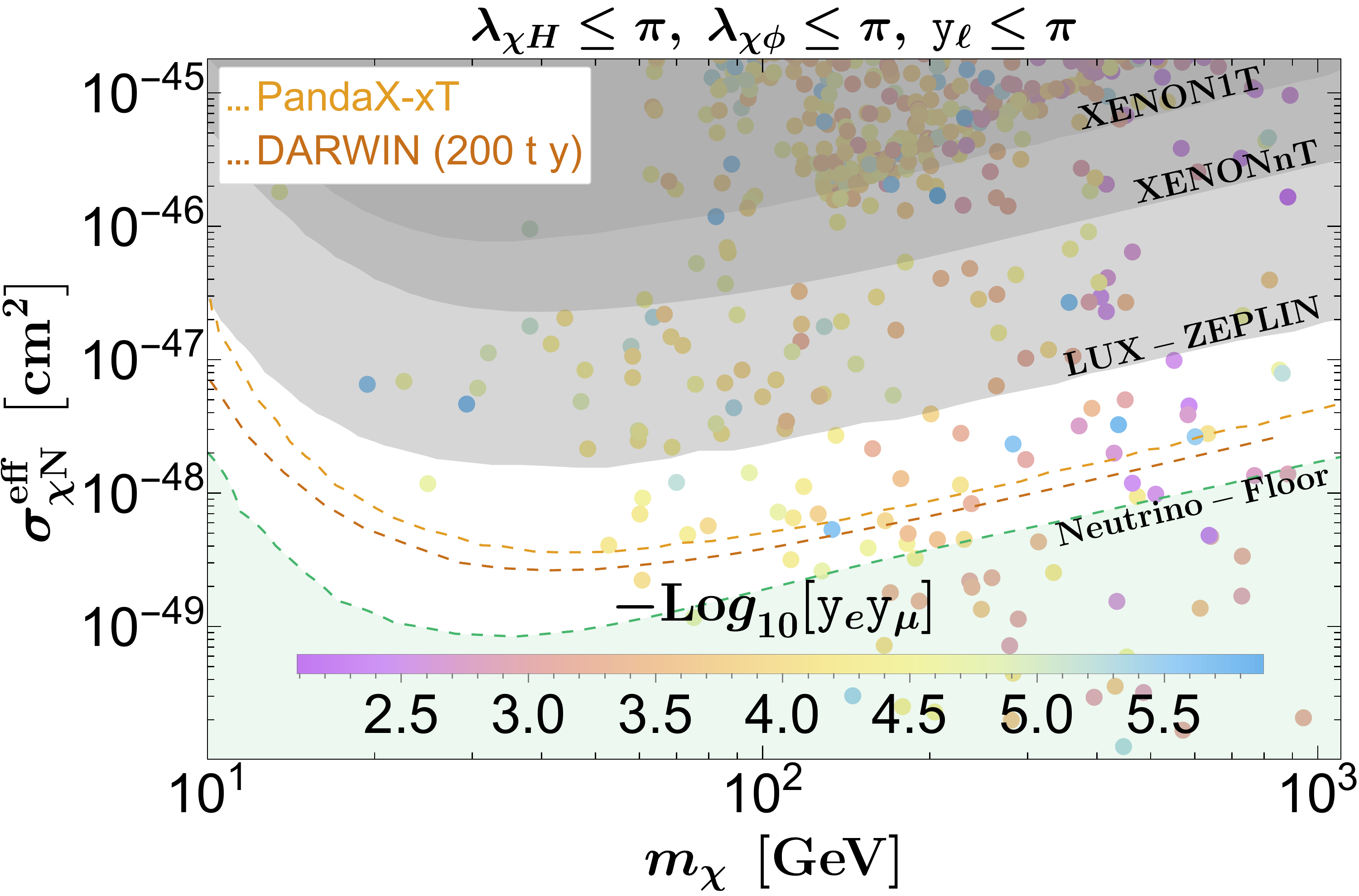}\label{figf}}
\caption{In Figs\,.~\ref{figa} to \ref{figf}, the relic density allowed parameter space for WIMP are shown in the $ m_{\chi} - \sigma_{\chi\rm N}^{\rm eff}$ plane. The grey-shaded regions are excluded by the observed limits from XENON1T, XENONnT, and LUX-ZEPLIN, while the projected limits from PandaX-xT (orange) and DARWIN/XLZD (brown) are represented by dashed lines. In all plots, we fix the parameters as, $\lambda_{\phi \rm H} = 10^{-12}$, $\lambda_{\chi} = 1$, and $\mu_{3} = m_{\chi}$. The lepton portal couplings, $\mathtt{y}_{\ell}$, are randomly varied below the value of $\pi$. All points shown are consistent with the LFV constraints from $\mu^+ \to e^+ \gamma$, $\tau^+ \to e^+ \gamma$, and $\tau^+ \to \mu^+ \gamma$.}
\label{fig:wimp-dd}
\end{figure}
In Fig\,.~\ref{fig:wimp-dd}, we illustrate the relic density allowed parameter space in the $m_{\chi}-\sigma^{\rm eff}_{\chi \rm N}$ plane, with the color bar representing the variation of the parameters as described above the color bar, while the remaining parameters are detailed in the figure's inset. All the points in the plotted parameter space are allowed by the LFV constraints from the processes $\mu^+ \to e^+ \gamma$, $\tau^+ \to e^+ \gamma$, and $\tau^+ \to \mu^+ \gamma$.

In Figs\,.~\ref{figa},~\ref{figb},~\ref{figc},~\ref{figd},~\ref{fige}~and~\ref{figf}, we show the dependency of $\lambda_{\rm \chi H}$, $\lambda_{\rm \chi \phi}$, $\Delta$, $\mathtt{y}_{\tau}$, $\delta m$, and $\mathtt{y}_{e}\mathtt{y}_{\mu}$, respectively, on the relic density and DD, illustrated through the pastel color bar. Here, we have defined, $\Delta = m_{\chi} - m_{\phi}$ and $\delta m = m_{\psi} - m_{\chi}$. The DM masses, $m_{\chi}$ and $m_{\phi}$, are varied from $10~\rm GeV$ to $1000~\rm GeV$. The VLL mass, $m_{\psi}$, is varied from $105~\rm GeV$ (LEP bound) to $1000~\rm GeV$. We further ensure that $m_{\psi} > m_{\chi} + m_{e}$ to allow on-shell decay of the VLL. The parameters $\lambda_{\chi\phi}$, $\lambda_{\rm \chi H}$ and $\mathtt{y}_{\ell}$ are scanned from $10^{-3}$ to $\pi$, whereas, $\lambda_{\chi}$, $\lambda_{\phi H}$, and $\mu_{3}$ are fixed at $1.0$, $10^{-12}$ and $m_{\chi}$, respectively. Below the Higgs resonance, most of the region is excluded by the LFV limit, although it remains viable under the DM relic density constraint due to the large lepton portal coupling, and is allowed by the SI DD constraint for small values of $\lambda_{\chi \rm H}$, regardless of $\lambda_{\chi \phi}$.
In the low mass regime, i.e\,.~below Higgs mass, the DM relic density is adjusted by the other processes that are not involved with $h\chi\chi^*$ vertex, which is reflected in Fig\,.~\ref{figa}.
Above the Higgs mass, $\psi$ mass also gradually increases following $m_{\psi}=m_{\chi} + m_{e}$, hence, the LFV bound is much more relaxed.
In this mass regime, $\lambda_{\chi\phi}$ coupling plays a crucial role in DM relic density, as we see in Fig\,.~\ref{figb}.
However, the DM-DM conversion plays an important role in the DM relic density for both DM mass hierarchical scenarios, mostly through the four-point interaction with vertex factor $\lambda_{\chi\phi}$.
In Fig\,.~\ref{figc} we have shown the effect of DM mass hierarchy through $\Delta=m_{\chi}-m_{\phi}$. If $m_{\chi} > m_{\phi}$, then $\chi\to\phi$ conversion is more efficient compared to the opposite. For positive $\Delta$, around the Higgs mass, $\chi \to \phi$ conversion makes the $\chi$ under abundant and $\phi$ mostly contributes to the total DM relic density. This is also true for the higher mass regimes, but all kinds of contributions are feasible.
In Fig\,.~\ref{figd} we have shown the variation of the $\tau$ lepton portal coupling ($\mathtt{y}_{\tau}$). This coupling is more relevant for a minute Higgs portal coupling to avoid the DD bound.
In Fig\,.~\ref{fige}, we show the variation of the mass difference between WIMP and charged fermion by $\delta m$ and represent it through the pastel color bar.
Fig\,.~\ref{figf} illustrates the variation of the product of lepton portal couplings, $\mathtt{y}_{e}\mathtt{y}_{\mu}$, as shown in the color bar. This coupling product determines the most stringent constraint from LFV, originating from the $\mu \to e \gamma$ channel. This parameter space will be useful for collider analysis.
\subsubsection*{pFIMP}
The real scalar pFIMP ($\phi$) interacts with the target nucleus through the WIMP loop and Higgs-mediated process shown in Fig\,.~\ref{feyn:pfimp-dd}, enabled by the WIMP-Higgs portal and WIMP-pFIMP interactions. Here, we are using the SI effective pFIMP-nucleon cross-section, which is $\rm \sigma_{\phi\rm  N}^{\rm eff}=(\Omega_{\phi}h^2/\Omega_{\rm DM}h^2)\sigma_{\phi\rm N}^{\rm SI}$.
\begin{figure}[htb!]
\centering
\subfloat[]{\includegraphics[width=0.475\linewidth]{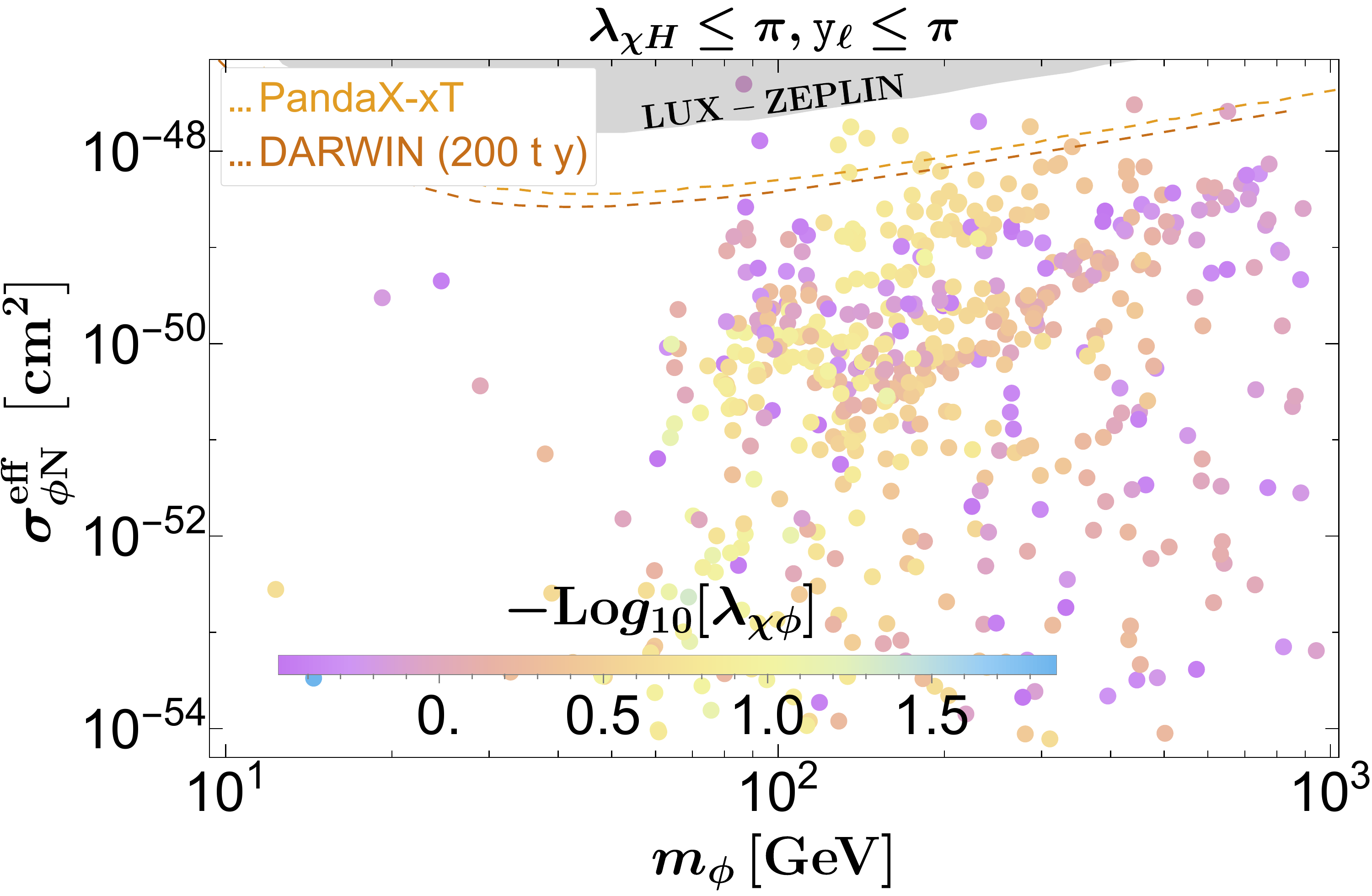}\label{fig:pfimp-dd}}~~
\subfloat[]{\includegraphics[width=0.475\linewidth]{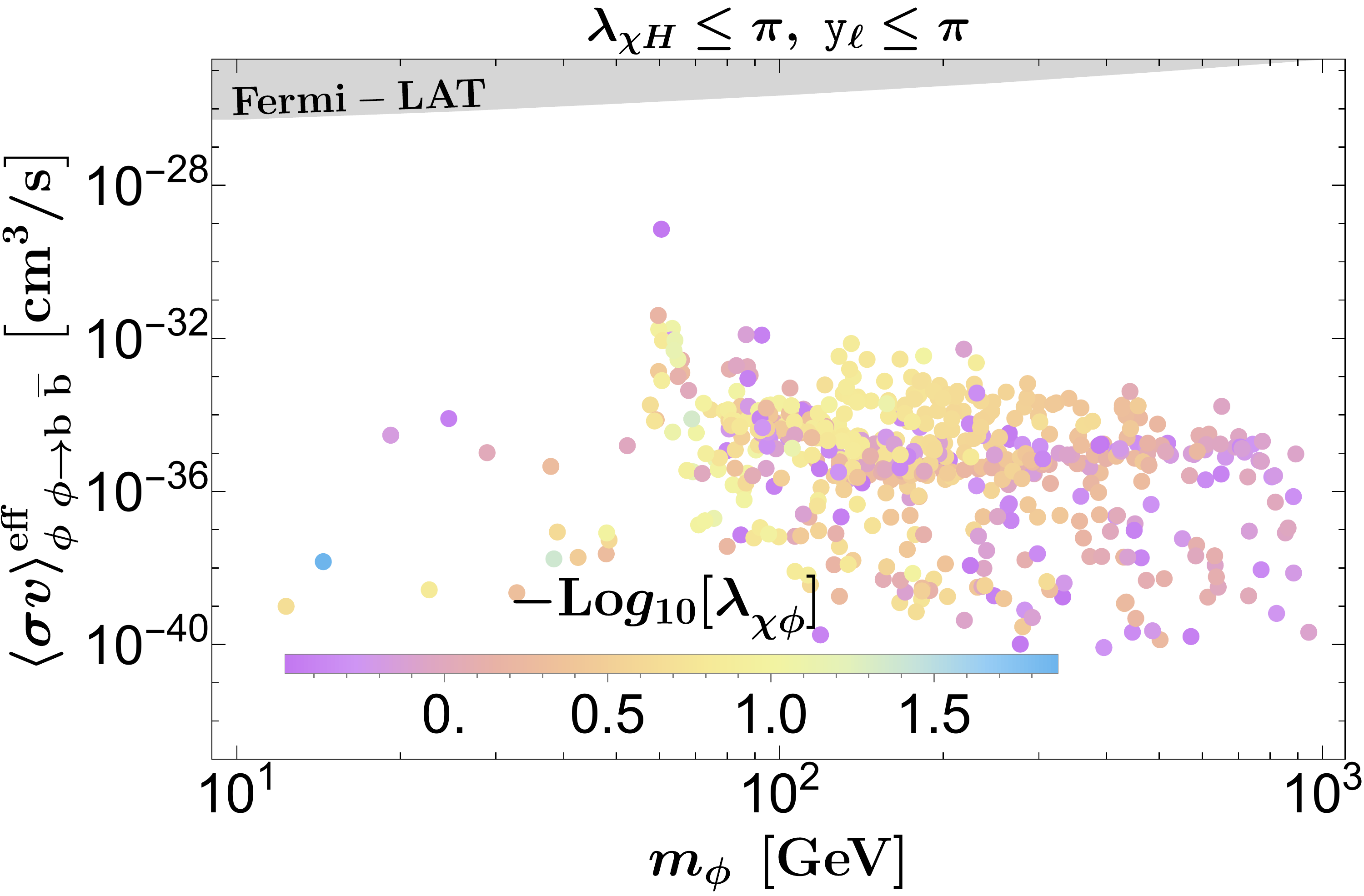}\label{fig:pfimp-id}}
\caption{The pFIMP relic density allowed parameter space are shown in $ m_{\phi}-\sigma_{\phi\rm N}^{\rm eff}$ and $m_{\phi}-\langle\sigma v\rangle^{\rm eff}_{\phi\phi\to b\overline{b}}$ plane. In all plots, we have fixed the parameters: $\lambda_{\phi \rm H} = 10^{-12},~\lambda_{\chi}=1$ and $\mu_{3} = m_{\chi}$. The lepton portal couplings, $\mathtt{y}_{\ell}$, are randomly varied below the value of $\pi$. All the points are allowed from the LFV ($\mu^+ \to e^+ \gamma$, $\tau^+ \to e^+ \gamma$ and $\tau^+ \to \mu^+ \gamma$) constraints, see Eq\,.~\eqref{eq:mueg}.}
\label{fig:pfimp-relic}
\end{figure}
The pFIMP-nucleon scattering is only possible via a 1-loop mediated process, see Fig\,.~\ref{feyn:pfimp-dd}. This $\sigma_{\phi\rm N}^{\rm SI}$ cross-section predominantly depends on $\lambda_{\chi\rm H}$ and $\lambda_{\chi\phi}$, while the effect of the DM mass in the loop is minimal as it only contributes through logarithms. We already have discussed the $\lambda_{\chi\phi}$ effect in DM relic density in Fig\,.~\ref{figb}, where we have shown the correlation between $\lambda_{\chi\phi}$ and $\lambda_{\chi \rm H}$, which is inversely correlated to each other to adjust the DM relic density and respect the observed DD limit. The use of a small $\lambda_{\chi\rm H}$ value allows the WIMP to stay within the current DD limit. However, this also increases the relic density of the WIMP, which is then adjusted by the enhancement of $\lambda_{\chi\phi}$. However, this choice does not impact $\rm \sigma_{\phi\rm N}^{SI}$, and this effect is more noticeable at higher mass ranges. In Figure \ref{fig:pfimp-dd}, the relic density allowed parameter space is depicted in the $m_{\phi}-\sigma_{\phi\rm  N}^{\rm eff}$ plane. The parameters are varied over the same ranges as in Fig\,.~\ref{fig:wimp-dd}. As these points depend on the effective relic density contribution of pFIMP, the scanning behaviour relies on the characteristics of WIMP. However, around the Higgs resonance, some points fall within the direct detection constraint, while above it, the parameter space is mostly open for future detection.

\subsection{Indirect detection}
The AMS-02 experimental data \cite{Ibarra:2013zia, Weng:2020zmr} on cosmic ray positrons shows that at high energies, the primary source of positrons is either dark matter annihilation or other astrophysical sources. Therefore, using the AMS-02 data on positron flux sets an upper limit on the rate of dark matter annihilation to electron-positron pairs with a branching ratio of 100\%.
The gamma-ray observations of Milky Way dSphs from six years of Fermi Large Area Telescope (Fermi-LAT) data \cite{Essig:2013goa, Fermi-LAT:2015kyq, Fermi-LAT:2016uux} reported no significant detections. They presented upper limits on the DM self-annihilation cross-section for 15 dwarf spheroidal satellite galaxies (dSphs) and projected sensitivity for 45 dSphs with 15 years of observation \cite{Meurer:2009ir}.
The DM semi-annihilation cross-section is constrained by gamma-ray observations from Fermi-LAT \cite{Fermi-LAT:2015att}, as well as the projected limits from H.E.S.S \cite{HESS:2016mib} and CTA \cite{Silverwood:2014yza}.
\begin{figure}[htb!]
\centering
\subfloat[]{\begin{tikzpicture}
\begin{feynman}
\vertex (a);
\vertex[above left=1cm and 1cm  of a] (a1){\(\chi\)};
\vertex[below left=1cm and 1cm  of a] (a2){\(\chi\)}; 
\vertex[right=1cm of a] (b); 
\vertex[above right=1cm and 1cm of b] (c1){\(\rm {b}\)};
\vertex[below right=1cm and 1cm of b] (c2){\(\rm b\)};
\diagram*{
(a1) -- [line width=0.25mm, charged scalar, arrow size=0.7pt,style=black] (a),
(a) -- [line width=0.25mm,charged scalar, arrow size=0.7pt,style=black] (a2),
(a) -- [line width=0.25mm,scalar, edge label={\(\color{black}{h}\)}, style=gray] (b),
(c1) -- [line width=0.25mm,fermion, arrow size=0.7pt] (b),
(b) -- [line width=0.25mm,fermion, arrow size=0.7pt] (c2)};
\node at (a)[circle,fill,style=gray,inner sep=1pt]{};
\node at (b)[circle,fill,style=gray,inner sep=1pt]{};
\end{feynman}
\end{tikzpicture}
\label{feyn:wimp-id}}
\subfloat[]{\begin{tikzpicture}
\begin{feynman}
\vertex (a);
\vertex[ left=0.75cm and 0.75cm  of a] (a1){\(\chi\)};
\vertex[ right=0.75cm and 0.75cm  of a] (a2){\(\chi\)}; 
\vertex[ below=2.5cm of a] (b); 
\vertex[ left=0.75cm and 0.75cm of b] (c1){\(\rm \chi\)};
\vertex[ right=0.75cm and 0.75cm of b] (c2){\(h\)};
\diagram*{
(a1) -- [line width=0.25mm,charged scalar, arrow size=0.7pt,style=black] (a),
(a2) -- [line width=0.25mm,charged scalar, arrow size=0.7pt,style=black] (a),
(b) -- [line width=0.25mm,charged scalar, arrow size=0.7pt, edge label'={\(\color{black}{\chi}\)}, style=gray] (a) ,
(c1) -- [line width=0.25mm,charged scalar, arrow size=0.7pt] (b),
(b) -- [line width=0.25mm,scalar, arrow size=0.7pt] (c2)};
\node at (a)[circle,fill,style=gray,inner sep=1pt]{};
\node at (b)[circle,fill,style=gray,inner sep=1pt]{};
\end{feynman}
\end{tikzpicture}
\label{feyn:wimp-semi}}
\subfloat[]{\begin{tikzpicture}
\begin{feynman}
\vertex (a);
\vertex[above left=1cm and 1cm  of a] (a1){\(\phi\)};
\vertex[below left=1cm and 1cm  of a] (a2){\(\phi\)}; 
\vertex[right=0.75cm of a] (c); 
\vertex[right=0.75cm of c] (b); 
\vertex[above right=1cm and 1cm of b] (c1){\(\rm b\)};
\vertex[below right=1cm and 1cm of b] (c2){\(\rm b\)};
\diagram*{
(a1) -- [line width=0.25mm,charged scalar, arrow size=0.7pt,style=black] (a),
(a) -- [line width=0.25mm,charged scalar, arrow size=0.7pt,style=black] (a2),
(c) -- [line width=0.25mm,scalar, edge label={\(\color{black}{h}\)}, style=gray] (b),
(c1) -- [line width=0.25mm,fermion, arrow size=0.7pt] (b),
(b)  -- [line width=0.25mm,fermion, arrow size=0.7pt] (c2),
(a) -- [line width=0.25mm,charged scalar,half left, edge label={\(\color{black}{\chi}\)}, style= gray, arrow size=0.7pt](c),
(c) -- [line width=0.25mm,charged scalar,half left, edge label={\(\color{black}{\chi}\)}, style= gray, arrow size=0.7pt] (a)};
\node at (a)[circle,fill,style=gray,inner sep=1pt]{};
\node at (b)[circle,fill,style=gray,inner sep=1pt]{};
\node at (c)[circle,fill,style=gray,inner sep=1pt]{};
\end{feynman}
\end{tikzpicture}
\label{feyn:pfimp-id}}
\subfloat[]{\begin{tikzpicture}
\begin{feynman}
\vertex(a);
\vertex[above left=0.5cm and 1cm of a] (a1){\(\chi\)};
\vertex[right=1cm of a] (b);
\vertex[above right=0.5cm and 1cm of b] (b1){\(\gamma\)};
\vertex[below=1cm of a] (c);
\vertex[below left=0.5cm and 1cm of c] (c1){\(\chi\)};
\vertex[right=1cm of c] (d);
\vertex[below right=0.5cm and 1cm of d] (d1){\(\gamma\)};
\diagram*{
(a1) -- [line width=0.25mm,charged scalar, arrow size=0.7pt,edge label={\(\rm \)},style=black] (a),
(c) -- [line width=0.25mm,charged scalar, arrow size=0.7pt,edge label={\(\rm \)},style=black] (c1),
(b) -- [line width=0.25mm,boson, arrow size=0.7pt,edge label={\(\rm \)},style=black] (b1),
(d1) -- [line width=0.25mm,boson, arrow size=0.7pt,edge label={\(\rm \)},style=black] (d),
(b) -- [line width=0.25mm,fermion, arrow size=0.7pt,style=gray!75,edge label' = {\(\rm \color{black}{\psi/e} \)}] (a),
(a) -- [line width=0.25mm,fermion, arrow size=0.7pt,style=gray!75,edge label' = {\(\rm \color{black}{\psi/e} \)}] (c),
(c) -- [line width=0.25mm,fermion, arrow size=0.7pt,style=gray!75,edge label' = {\(\rm \color{black}{\psi/e} \)}] (d),
(d) -- [line width=0.25mm,fermion, arrow size=0.7pt,style=gray!75,edge label' = {\(\rm \color{black}{\psi/e} \)}] (b)};
\node at (a)[circle,fill,style=gray,inner sep=1pt]{};
\node at (b)[circle,fill,style=gray,inner sep=1pt]{};
\node at (c)[circle,fill,style=gray,inner sep=1pt]{};
\node at (d)[circle,fill,style=gray,inner sep=1pt]{};
\end{feynman}
\end{tikzpicture}\label{fig:id-gamma-simp1}}
\caption{The Feynman diagrams are related to the indirect detection of WIMP and pFIMP.}
\label{fig:feynman-id}
\end{figure}

In this work, the key processes depicted in Figs\,.~\ref{fig:feynman-id} play a crucial role in establishing the limit on the self-annihilation cross-section of WIMPs and pFIMPs, as well as on the semi-annihilation of WIMPs. In Figs\,.~\ref{fig:wimp-id} and \ref{fig:pfimp-id}, we have represented the relic density allowed parameter space. The parameters are varied over the same ranges as in Fig\,.~\ref{fig:wimp-dd}. Some points are excluded by indirect limits on DM self and semi-annihilation cross-sections from different indirect observations.
Here, we use the effective DM annihilation cross-section, which is the DM annihilation cross-section multiplied by the normalized effective DM number density.
In Figs\,.~\ref{fig:chiachiee}, \ref{fig:chiachibb}, \ref{fig:chiachiaa}, and \ref{fig:chichiachih}, we have represented the relic density allowed parameter space in 
$m_{\chi}-\langle\sigma v\rangle^{\rm eff}_{\chi\chi^*\to e^-e^+}$, 
$m_{\chi}-\langle\sigma v\rangle^{\rm eff}_{\chi\chi^*\to b\overline{b}}$, 
$m_{\chi}-\langle\sigma v\rangle^{\rm eff}_{\chi\chi^*\to \gamma\gamma}$, and 
$m_{\chi}-\langle\sigma v\rangle^{\rm eff}_{\chi\chi\to \chi^* \rm h}$ plane, respectively.

\subsubsection*{WIMP}
\begin{figure}[htb!]
\centering
\subfloat[]{\includegraphics[width=0.475\linewidth]{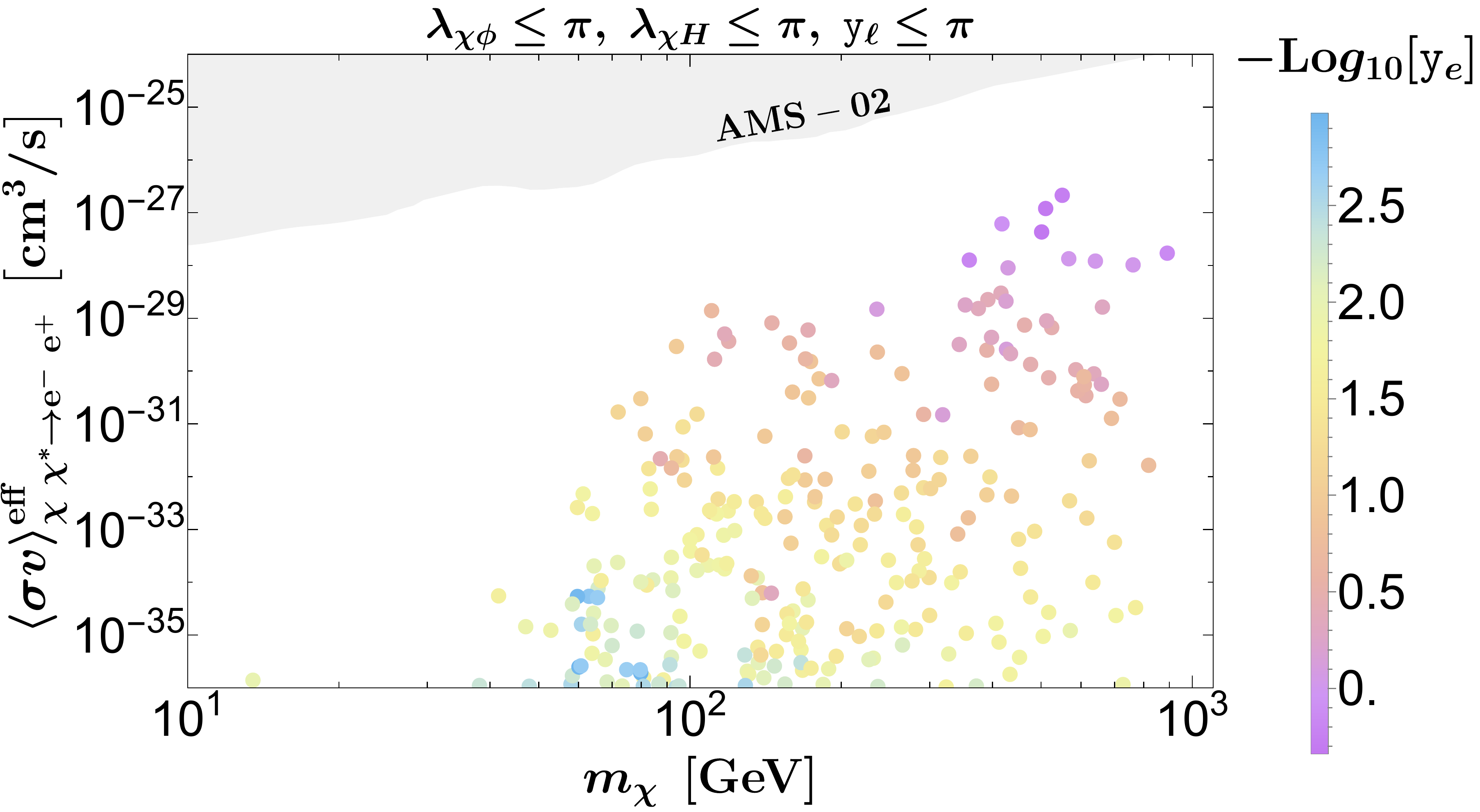}\label{fig:chiachiee}}\quad
\subfloat[]{\includegraphics[width=0.475\linewidth]{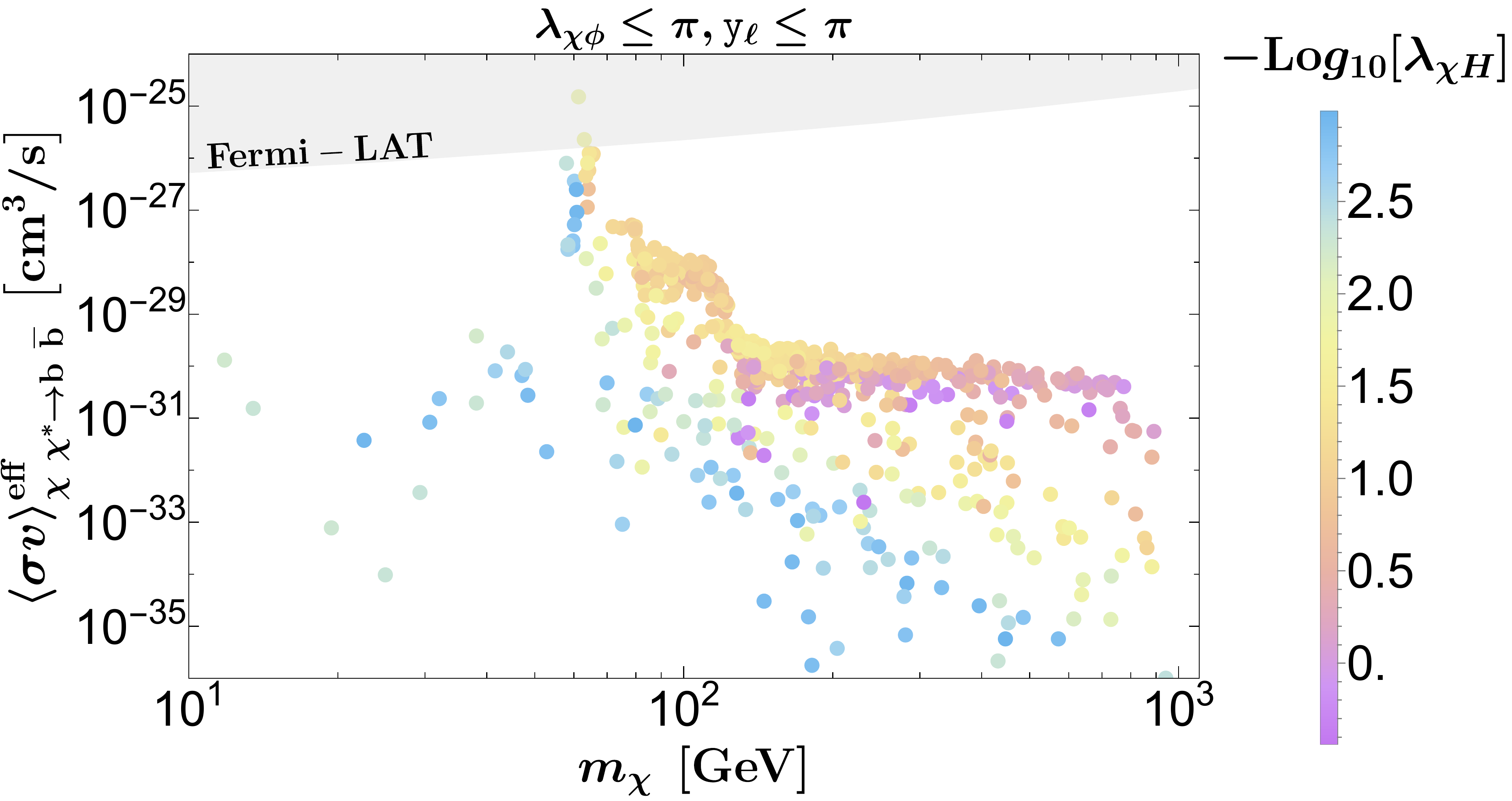}\label{fig:chiachibb}}

\subfloat[]{\includegraphics[width=0.475\linewidth]{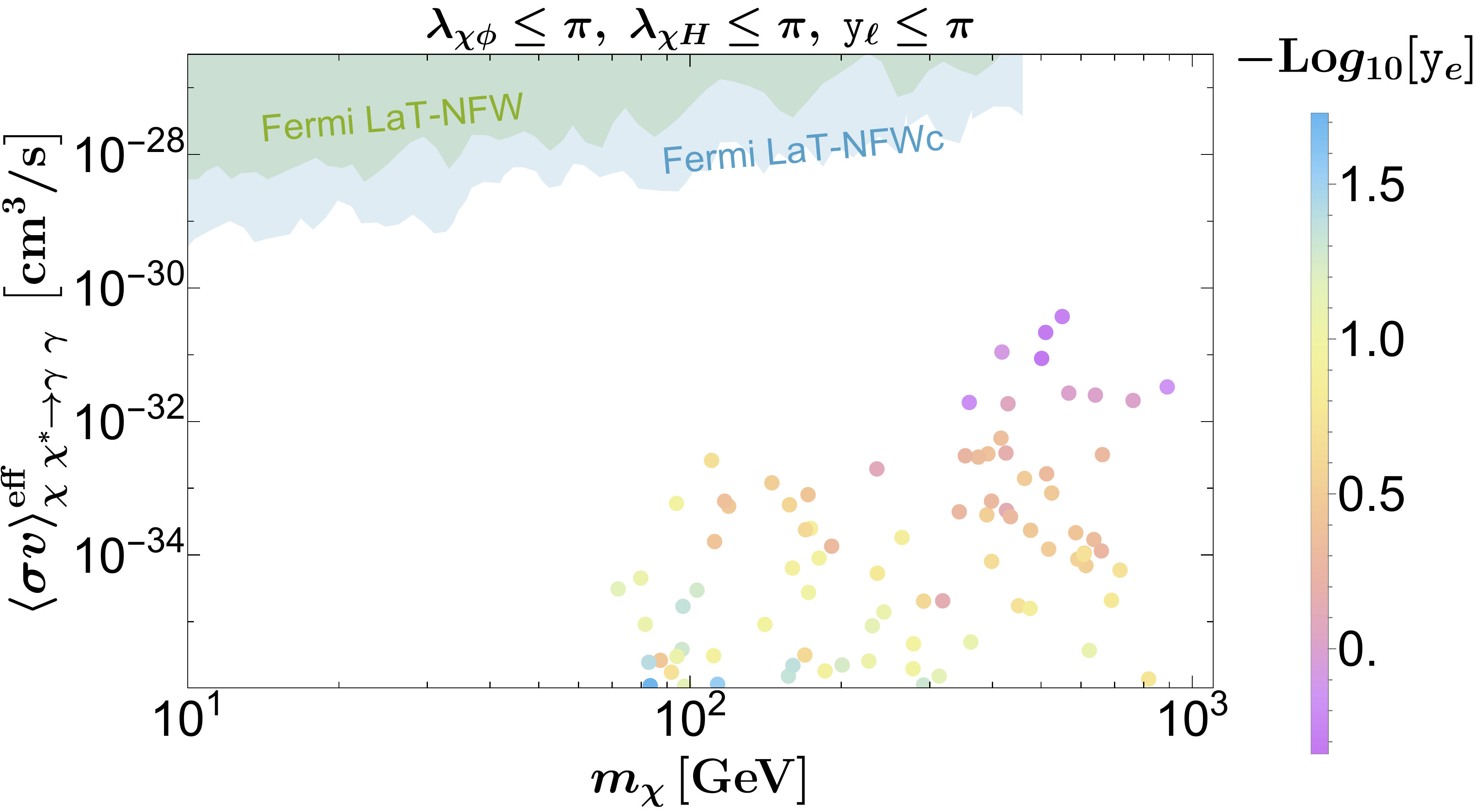}\label{fig:chiachiaa}}\quad
\subfloat[]{\includegraphics[width=0.475\linewidth]{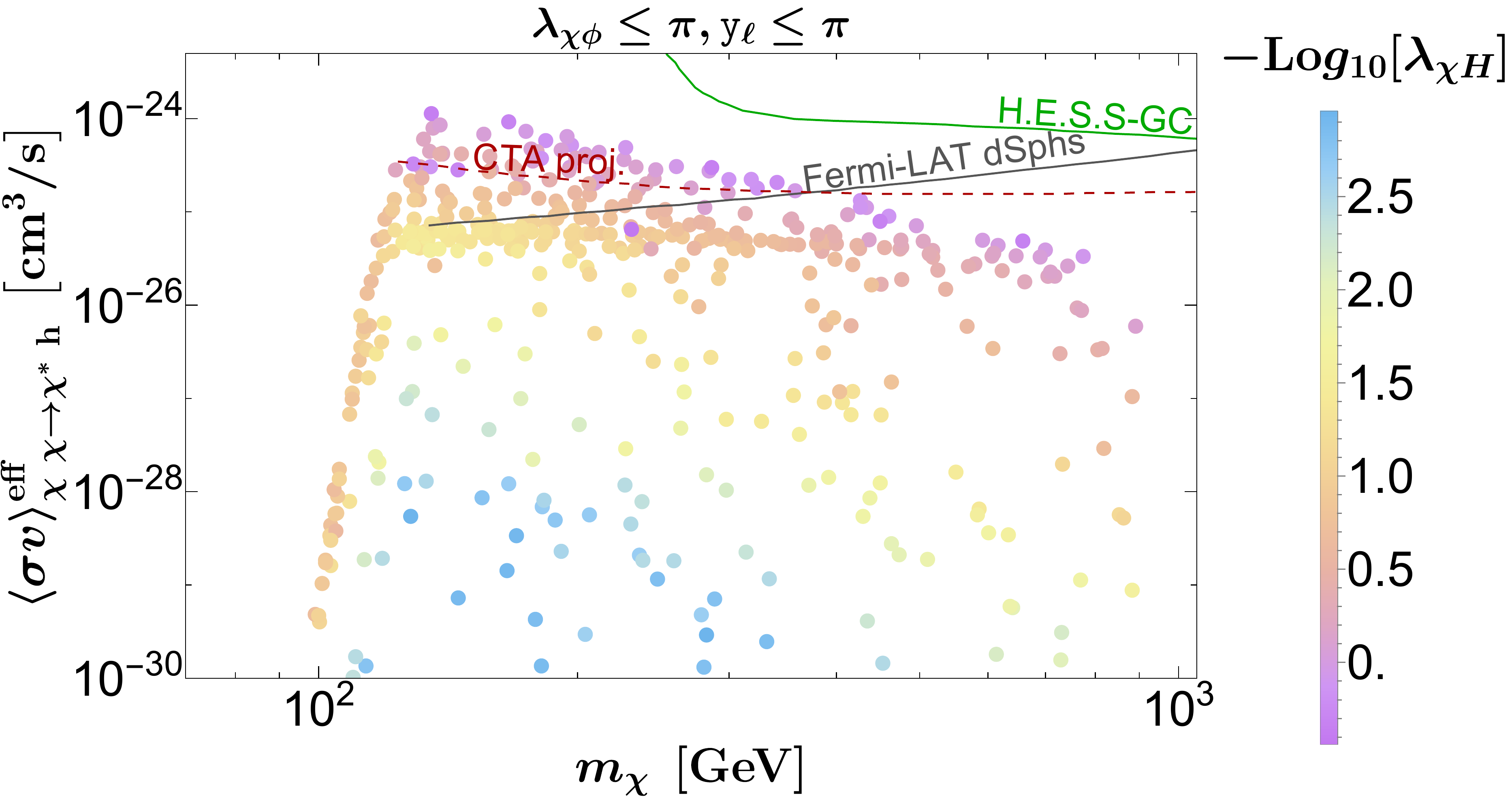}\label{fig:chichiachih}}
\caption{In this figure, we represent the relic density allowed parameter space. Figs\,.~\ref{fig:chiachiee} and \ref{fig:chiachibb} show the exclusion region from AMS-02 and Fermi-LAT limit on WIMP annihilation to electron and bottom pair, respectively. Fig\,.~\ref{fig:chiachiaa} represents the exclusion region using data from Fermi-LAT and Planck, indicated by different color shades. We're narrowing down our parameter space for WIMP semi-annihilation in Fig\,.~\ref{fig:chichiachih} using H.E.S.S, Fermi-LAT, and CTA data. In all plots, we have fixed the parameters: $\lambda_{\phi \rm H} = 10^{-12},~\lambda_{\chi}=1$ and $\mu_{3} = m_{\chi}$. The lepton portal couplings, $\mathtt{y}_{\ell}$, are randomly varied below the value of $\pi$. All the points are allowed from the LFV ($\mu^+ \to e^+ \gamma$, $\tau^+ \to e^+ \gamma$ and $\tau^+ \to \mu^+ \gamma$) constraints, as shown in Eq\,.~\eqref{eq:mueg}.}
\label{fig:wimp-id}
\end{figure}
In Fig\,.~\ref{fig:chiachiee}, the indirect bound on the DM annihilation to electron-positron pair, using AMS-02 data, excludes some of the relic density allowed parameter space in below Higgs mass regime. This is because the larger lepton portal coupling is required to achieve the correct relic density. In this regime, the Higgs portal processes are always suppressed. A similar explanation is also applicable to other processes, as shown in Fig\,.~\ref{fig:wimp-id}. In Fig\,.~\ref{fig:chiachibb}, the Higgs resonance regime is excluded by the Fermi-LAT data on DM annihilation to the bottom pair.
Additionally, there is a stringent limit from WIMP annihilation into photon pairs based on Fermi-LAT observations. However, this does not constrain the parameter space (Fig\,.~\ref{fig:chiachiaa}), as the process is only possible through 1-loop box diagrams (Fig\,.~\ref{fig:id-gamma-simp1}).
Finally, in Fig\,.~\ref{fig:chichiachih}, some of the WIMP parameter space is also constrained, and near Higgs mass, some points are excluded by the Fermi-LAT, CTA, and H.E.S.S limit on the WIMP semi-annihilation to the Higgs.
\subsubsection*{pFIMP}
The pFIMP annihilation occurs exclusively through a 1-loop mediated process, as shown in Fig\,.~\ref{feyn:pfimp-id}. In the case of pFIMPs, most parameter space remains viable, except near the Higgs resonance region. In this area, certain points are excluded due to a significant resonance enhancement of the cross-section, despite the presence of loop suppression, as illustrated in Fig\,.~\ref{fig:pfimp-id}. Alternative indirect detection constraints on pFIMP annihilation and semi-annihilation become ineffective due to the fact that the associated processes are suppressed by loop-level interactions where the 1-loop processes are $\phi ~\phi \to e^- ~e^+$ and $\rm\phi ~\phi \to \phi ~h$,  and the 2-loop process is $\phi~ \phi \to \gamma ~\gamma$.
\section{Sensitivity at collider experiments}
\label{sec5}
In this section, we study the collider sensitivity of the DM candidates. DM searches at colliders are primarily done via missing energy (at Lepton colliders) or missing transverse energy (primarily at Hadron colliders) signals in association with visible particles. Typical DM searches are done concerning mono-X (X = $\gamma, Z, h, j$) signal where DM is produced in association with a visible species (X, here). However, such signatures are subjected to huge background contamination, especially at hadron colliders where the hadronic activities are indomitable. Even at lepton colliders, the large SM neutrino background hinders the signal extraction to a great extent. In our model, the possibility of WIMP detection is greatly enhanced by the presence of lepton portal interactions. The lepton portal opens up the di-lepton (also, di-tau) signal possibility, which brings into play a wide range of kinematic observables, aiding in better discrimination of the DM signal from the SM background. Due to the absence of tree-level interaction with the visible sector, pFIMP detection at colliders is possible only through WIMP loop-mediated processes and, hence, always suppressed. In the following sections, we explore the features of di-lepton/di-tau searches at the LHC runs and its possible manifestation at proposed future lepton colliders.
\subsection{Recasting the LHC limits}
\begin{figure}[htb!]
\centering
\includegraphics[width=0.6\linewidth]{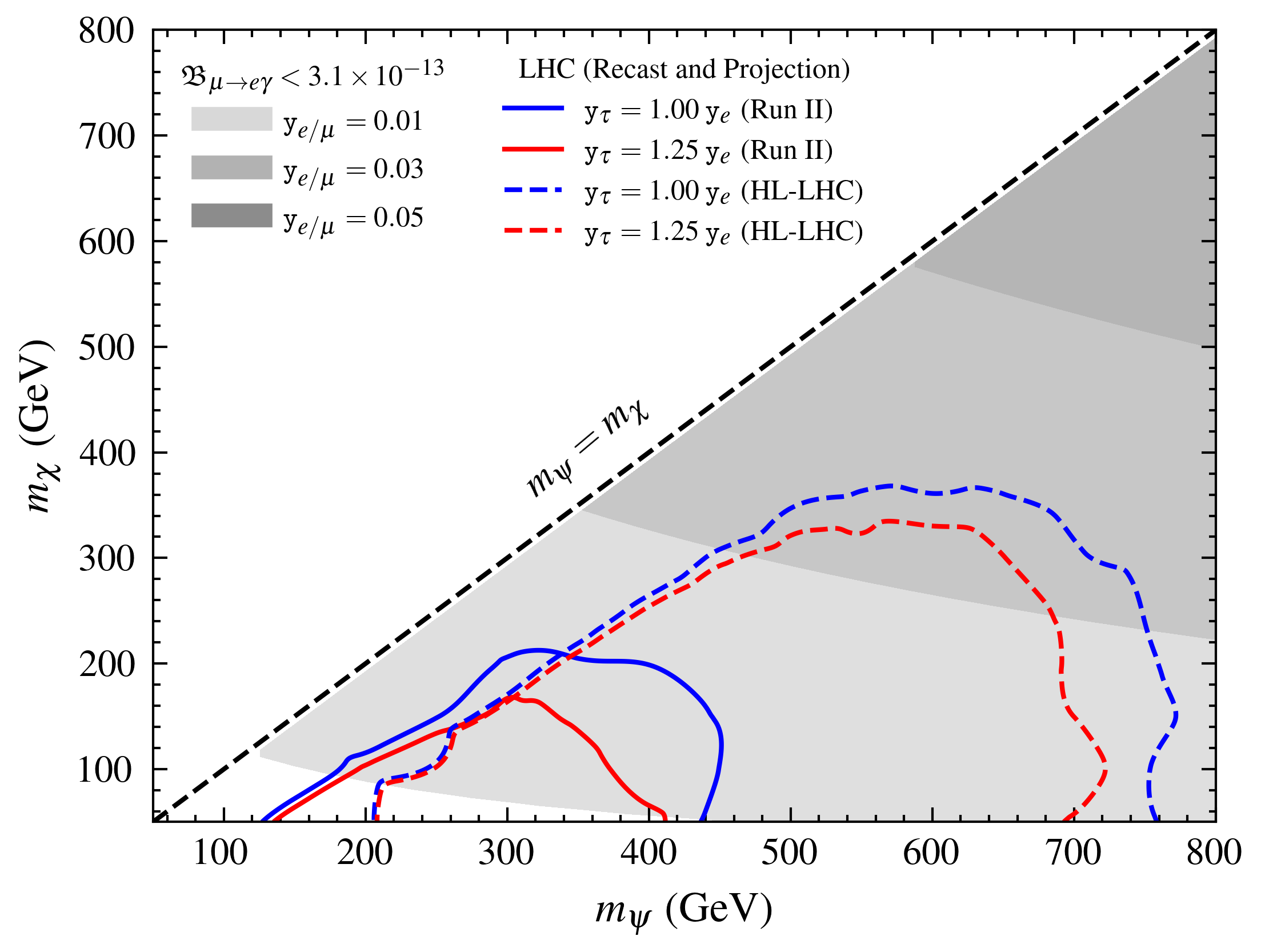}
\caption{Bounds on $m_{\psi}-m_{\chi}$ plane. The colored lines bound the closed region disallowed by \textit{Solid}: Recast of ATLAS dilepton + MET search (13 TeV, 139 fb$^{-1}$) \textit{Dashed}: Projection of dilepton + MET search at HL-LHC (14 TeV, 3000 fb$^{-1}$). The grey shaded region corresponds to the parameter space allowed from the LFV $\mu$ decay for different $\mathtt{y}_{e/\mu}$ ($\mathtt{y}_{\tau}$ arbitrary). The region above each shade is allowed indefinitely for that shade. $m_{\psi} < m_{\chi}$ is disallowed from on-shell production of $\psi$ and is separated by the dashed black line.}
\label{fig:recast}
\end{figure}

Most dark matter searches at the LHC focus on signatures predicted by the Minimal Supersymmetric Standard Model (MSSM), a popular extension of the Standard Model. In the MSSM, supersymmetry (SUSY) introduces new particles, the lightest of which, often a neutralino, serves as a dark matter candidate. These searches primarily look for missing transverse energy (MET) alongside other SUSY particles like squarks, sleptons or gluinos, which could decay into Standard Model particles and dark matter. However, the downside to this strategy is that this renders the analysis to cater only to a specific model. The common practice in such scenarios is to recast the existing LHC analyses in the context of the concerned BSM model. In our case, we recast the LHC di-lepton + MET signal studied at LHC experiments. The process is shown in Fig. \ref{fig:2l}. Regarding the di-tau signal, $\tau$ tagging is usually done concerning the hadronic decay mode of $\tau$ lepton ($\tau$ decay to pions and neutrino), which emerge as $\tau$ jets. The lepton decay mode of $\tau$ is disguised as lepton + MET signals and hence is difficult to segregate in processes in which missing particles are already present. However, the hadronic $\tau$ tagging turns out to be a strenuous task at the LHC as the light jets ($g, u, d, s, c$ jets), which are omnipotent at the hadron collider, can mimic the $\tau$ jets to some extent. This problem is aggravated by the fact that the $\tau$ jets emerge from EW processes, whereas the light jets dominantly appear from QCD backgrounds having large cross-sections. Hence, we restrict ourselves to the di-lepton signal only. We recast the ATLAS slepton pair decay to di-lepton and neutralinos (which appear as MET) at 13 TeV LHC at an integrated luminosity of 139 fb$^{-1}$ \cite{ATLAS:2019lff}. The model implementation is done using \texttt{FeynRules}. The MC events are generated in \texttt{MG5\_aMC} \cite{Alwall:2011uj} and the events are showered using \texttt{Pythia8} \cite{Sjostrand:2007gs}. The showered events are fed into \texttt{CheckMATE2} \cite{Dercks:2016npn} (build upon \texttt{Delphes3} \cite{deFavereau:2013fsa} and \texttt{Fastjet3} \cite{Cacciari:2005hq,Cacciari:2008gp,Cacciari:2011ma}). \texttt{CheckMATE2} uses CL$_{s}$ method \cite{Read:2002hq} to determine the 95\% C.L. exclusion limits. The events are generated at different $\left(m_{\psi}, m_{\chi}\right)$ benchmarks. The 95\% C.L. exclusion limit is shown in Fig. \ref{fig:recast}. Event selections are made using the following selection cuts: Opposite sign leptons, $p^{\ell \ell}_{T} > 25$ GeV, $M_{\ell \ell} > 25$ GeV and $N_{b} = 0$, where $p^{\ell \ell}_{T}$ is the vector sum of the $p_{T}$ of the leptons, $M_{\ell \ell}$ is the invariant mass of the leptons and $N_{b}$ is the number of b-jets. The signal regions are defined in Tab\,.~\ref{tab:sr1}.

\begin{table}[htb!]
\centering
\begin{tabular}{|c|c|}
\hline
Different flavor leptons, $n_{j} = 0$ & Same flavor leptons, $n_{j} = 0$ \\ \hline
$M_{\ell \ell} > 100$ GeV & $M_{\ell \ell} > 121.2$ GeV \\
$\slashed{E}_{T} > 110$ GeV & $\slashed{E}_{T} > 110$ GeV \\
$\slashed{E}^{sig}_{T} > 10 \sqrt{\rm GeV}$ & $\slashed{E}^{sig}_{T} > 10 \sqrt{\rm GeV}$ \\
$m_{T2} > 100$ GeV & $m_{T2} > 100$ GeV \\ \hline
Different flavor leptons, $n_{j} = 1$ & Same flavor leptons, $n_{j} = 1$ \\ \hline
$M_{\ell \ell} > 100$ GeV & $M_{\ell \ell} > 121.2$ GeV \\
$\slashed{E}_{T} > 110$ GeV & $\slashed{E}_{T} > 110$ GeV \\
$\slashed{E}^{sig}_{T} > 10 \sqrt{\rm GeV}$ & $\slashed{E}^{sig}_{T} > 10 \sqrt{\rm GeV}$ \\
$m_{T2} > 100$ GeV & $m_{T2} > 100$ GeV \\ \hline
\end{tabular}
\caption{Signal regions for ATLAS 13 TeV 139 fb$^{-1}$ recast. Here, $n_{j}$ is the number of light jets, $\slashed{E}_{T}$ is the MET, $\slashed{E}^{sig}_{T}$ is the MET significance defined as $\slashed{E}_{T}/\sqrt{H_{T}}$ where $H_{T}$ is the scalar sum of $p_{T}$ of visible particles and $m_{T2}$ is the stranverse mass \cite{Lester:1999tx,Barr:2003rg,Cheng:2008hk,Bai:2012gs}.}
\label{tab:sr1}
\end{table}
A similar analysis is repeated for HL-LHC 14 TeV 3000 fb$^{-1}$ using \texttt{CheckMATE2} projection card \texttt{dilepton\_HL}. The signal regions are defined to be same as \cite{ATLAS:2014zve}. The 95\% C.L. exclusion limit for the HL-LHC projection is shown in Fig. \ref{fig:recast}. We observe that the exclusion limits almost double to that of the LHC 13 TeV case.

\subsection{Search at future lepton colliders}
\begin{figure}[htb!]
\centering
\subfloat[]{\begin{tikzpicture}[baseline={(current bounding box.center)},style={scale=0.8, transform shape}]
\begin{feynman}
\vertex(a);
\vertex[above left = 1cm and 1cm  of a] (a1){\(q\)};
\vertex[below left = 1cm and 1cm  of a] (a2){\(\overline{q}\)};
\vertex[right =2cm  of a] (b);
\vertex[above right = 0.5cm and 0.5cm  of b] (b1);
\vertex[below right = 0.5cm and 0.5cm  of b] (b2);
\vertex[above right = 0.5cm and 0.5cm  of b1] (b3){\(\chi\)};
\vertex[below right = 0.5cm and 0.5cm  of b2] (b4){\(\chi^{*}\)};
\vertex[right = 0.5cm  of b1] (b11){\(\ell^{-}/\tau^{-}\)};
\vertex[right = 0.5cm  of b2] (b22){\(\ell^{+}/\tau^{+}\)};
\diagram*{
(a1) -- [line width=0.25mm,fermion, arrow size=0.7pt,style=black] (a),
(a) -- [line width=0.25mm,fermion, arrow size=0.7pt,style=black] (a2),
(a) -- [line width=0.25mm,boson, arrow size=0.7pt,style=gray!75,edge label'={\({\color{black}\rm\gamma/Z} \)}] (b),
(b) -- [line width=0.25mm,fermion, arrow size=0.7pt,style=gray!75,edge label={\({\color{black}\rm\psi^-} \)}] (b1),
(b1) -- [line width=0.25mm,charged scalar, arrow size=0.7pt,style=black] (b3),
(b1) -- [line width=0.25mm,fermion, arrow size=0.7pt,style=black] (b11),
(b2) -- [line width=0.25mm,fermion, arrow size=0.7pt,style=gray!75,edge label={\({\color{black}\rm\psi^+} \)}] (b),
(b22) -- [line width=0.25mm,fermion, arrow size=0.7pt,style=black] (b2),
(b4) -- [line width=0.25mm,charged scalar, arrow size=0.7pt,style=black] (b2)};
\end{feynman}
\end{tikzpicture}\label{fig:2l}}\quad
\subfloat[]{\begin{tikzpicture}[baseline={(current bounding box.center)},style={scale=0.8, transform shape}]
\begin{feynman}
\vertex(a);
\vertex[above left = 1cm and 1cm  of a] (a1){\(e^-\)};
\vertex[below left = 1cm and 1cm  of a] (a2){\(e^+\)};
\vertex[right =2cm  of a] (b);
\vertex[above right = 0.5cm and 0.5cm  of b] (b1);
\vertex[below right = 0.5cm and 0.5cm  of b] (b2);
\vertex[above right = 0.5cm and 0.5cm  of b1] (b3){\(\chi\)};
\vertex[below right = 0.5cm and 0.5cm  of b2] (b4){\(\chi^{*}\)};
\vertex[right = 0.5cm  of b1] (b11){\(\ell^{-}/\tau^{-}\)};
\vertex[right = 0.5cm  of b2] (b22){\(\ell^{+}/\tau^{+}\)};
\diagram*{
(a1) -- [line width=0.25mm,fermion, arrow size=0.7pt,style=black] (a),
(a) -- [line width=0.25mm,fermion, arrow size=0.7pt,style=black] (a2),
(a) -- [line width=0.25mm,boson, arrow size=0.7pt,style=gray!75,edge label'={\({\color{black}\rm\gamma/Z} \)}] (b),
(b) -- [line width=0.25mm,fermion, arrow size=0.7pt,style=gray!75,edge label={\({\color{black}\rm\psi^-}\)}] (b1),
(b1) -- [line width=0.25mm,charged scalar, arrow size=0.7pt,style=black] (b3),
(b1) -- [line width=0.25mm,fermion, arrow size=0.7pt,style=black] (b11),
(b2) -- [line width=0.25mm,fermion, arrow size=0.7pt,style=gray!75,edge label={\({\color{black}\rm\psi^+} \)}] (b),
(b22) -- [line width=0.25mm,fermion, arrow size=0.7pt,style=black] (b2),
(b4) -- [line width=0.25mm,charged scalar, arrow size=0.7pt,style=black] (b2)};
\end{feynman}
\end{tikzpicture}\label{fig:2l-a}}\quad
\subfloat[]{\begin{tikzpicture}[baseline={(current bounding box.center)},style={scale=0.8, transform shape}]
\begin{feynman}
\vertex(a);
\vertex[above left = 0.25cm and 1.5cm  of a] (a01){\(e^-\)};
\vertex[below = 2.0cm  of a] (b);
\vertex[below left = 0.25cm and 1.5cm  of b] (a02){\(e^+\)};
\vertex[right = 1cm  of a] (a1);
\vertex[right = 1cm  of b] (b1);
\vertex[ right = 0.5cm and 1cm  of a1] (a10){\(\ell^{-}/\tau^{-}\)};
\vertex[above right = 0.5cm and 1cm  of a1] (a11){\(\chi\)};
\vertex[ right = 0.5cm and 1cm  of b1] (b10){\(\ell^{+}/\tau^{+}\)};
\vertex[below right = 0.5cm and 1cm  of b1] (b11){\(\chi^{*}\)};
\diagram*{
(a01) -- [line width=0.25mm,fermion, arrow size=0.7pt,style=black] (a),
(b) -- [line width=0.25mm,charged scalar, arrow size=0.7pt,style=gray!75,edge label'={\({\color{black}\rm\chi} \)}] (a),
(b) -- [line width=0.25mm,fermion, arrow size=0.7pt,style=black] (a02),
(b1) -- [line width=0.25mm,fermion, arrow size=0.7pt,style=gray!75,edge label={\({\color{black}\rm\psi^+} \)}] (b),
(b10) -- [line width=0.25mm,fermion, arrow size=0.7pt,style=black,edge label'={\({\color{black}\rm} \)}] (b1),
(b11) -- [line width=0.25mm,charged scalar, arrow size=0.7pt,style=black,edge label'={\({\color{black}\rm} \)}] (b1),
(a) -- [line width=0.25mm,fermion, arrow size=0.7pt,style=gray!75,edge label={\({\color{black}\rm\psi^-} \)}] (a1),
(a1) -- [line width=0.25mm,fermion, arrow size=0.7pt,style=black,edge label'={\({\color{black}\rm} \)}] (a10),
(a1) -- [line width=0.25mm,charged scalar, arrow size=0.7pt,style=black,edge label'={\({\color{black}\rm} \)}] (a11)};
\end{feynman}
\end{tikzpicture}\label{fig:2l-b}}
\caption{Feynman diagrams \ref{fig:2l} (\ref{fig:2l-a}, \ref{fig:2l-b}) correspond to di-lepton/di-tau + $\slashed{E}_{T} (\slashed{E})$ signal at the LHC (ILC).}
\label{fig:mue}
\end{figure}
\begin{figure}[h!]
\centering
\includegraphics[width=0.45\linewidth]{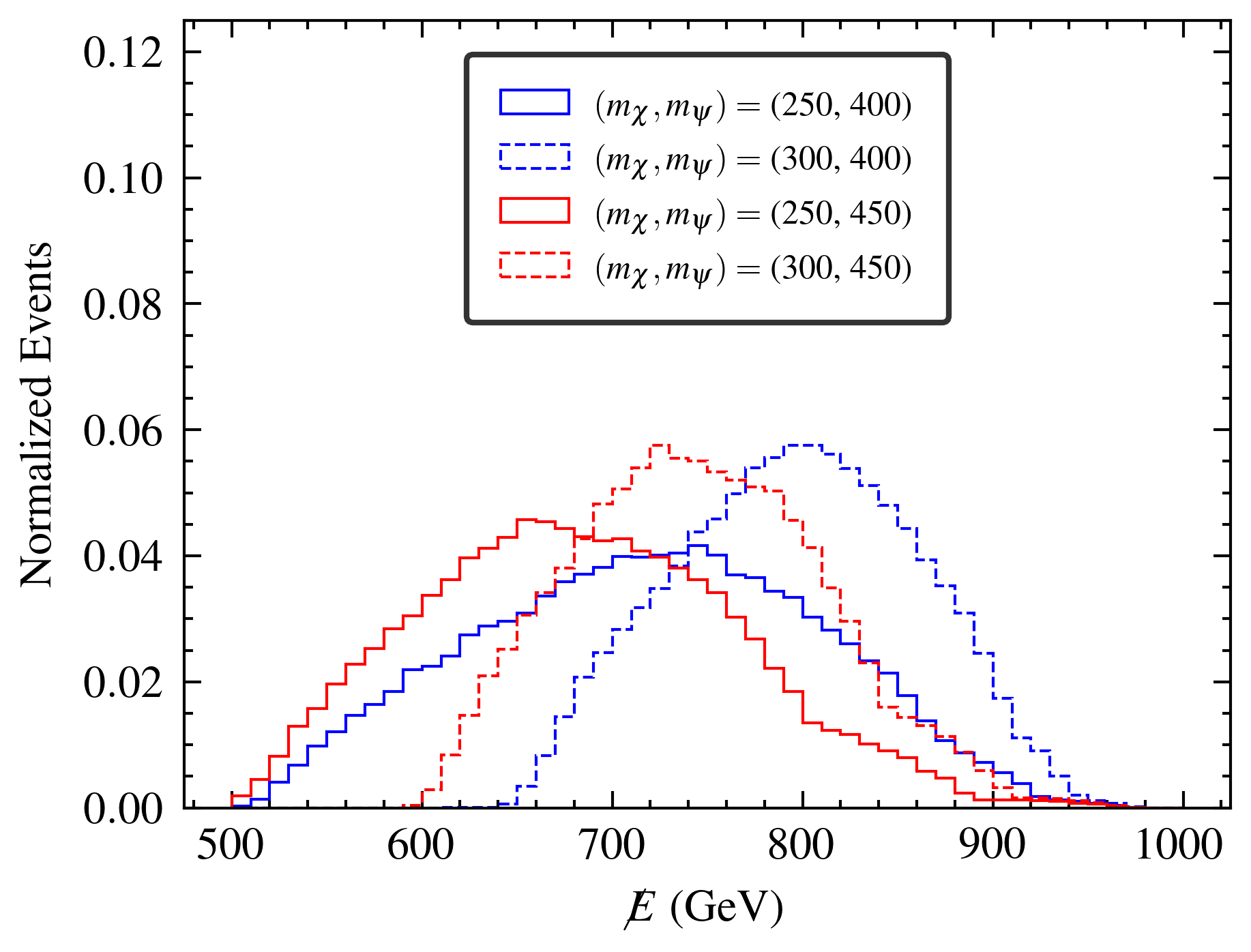}\quad
\includegraphics[width=0.45\linewidth]{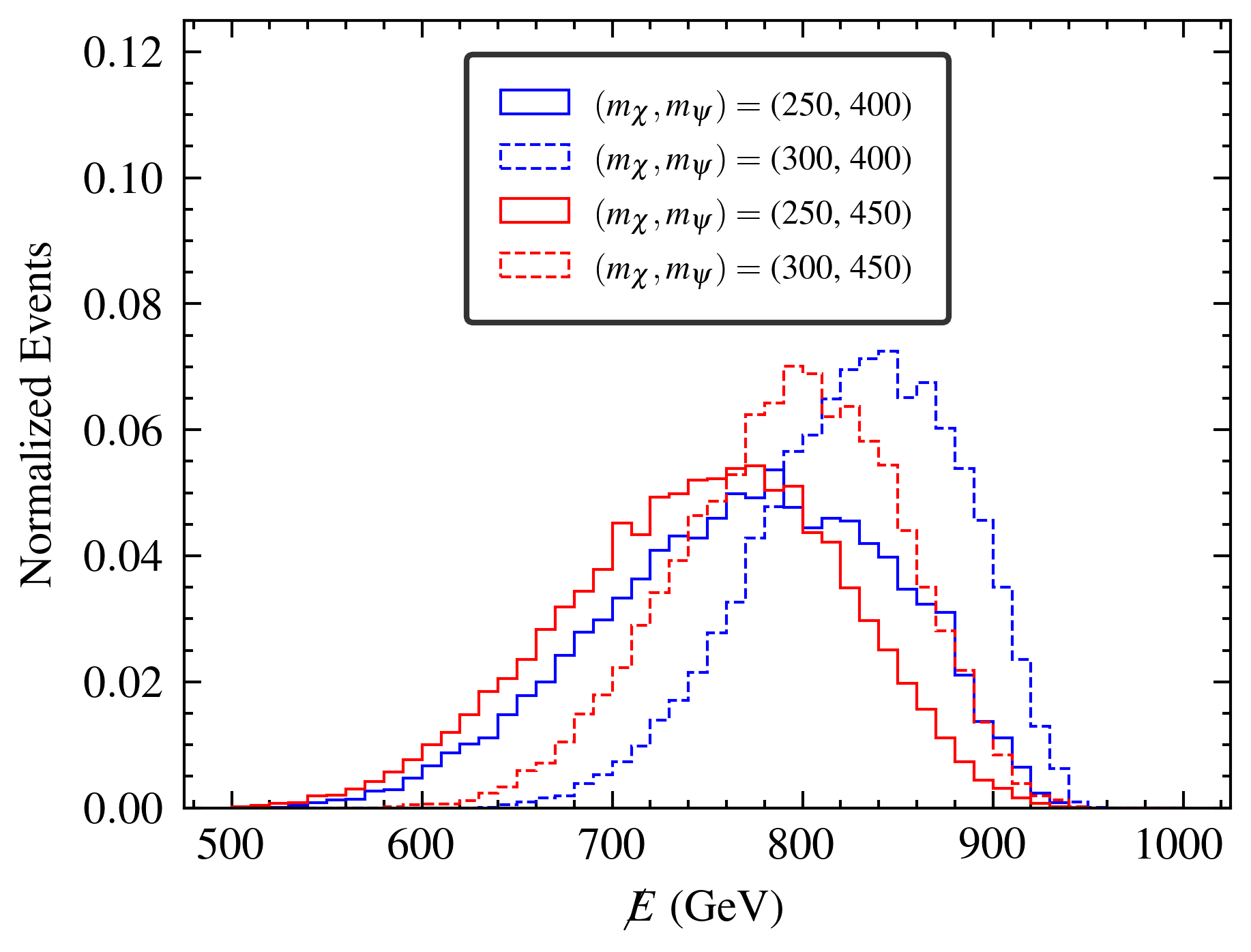}
\caption{Missing energy ($\slashed{E}$) distributions for di-lepton (left) and di-tau (right) signals for different $(m_{\psi}, m_{\chi})$ benchmarks.}
\label{fig:me}
\end{figure}

The dilepton signal processes corresponding to WIMP production at lepton colliders are shown in Fig. \ref{fig:mue}. The choice of lepton collider has a 3-fold justification. Firstly, the near-absence of hadronic activities at lepton colliders provides a cleaner environment for the study of missing energy signals. The reduced QCD backgrounds ensure lesser contamination of $\tau$ jets from light QCD jets. Secondly, the lepton portal connection of the WIMP opens up a t-channel possibility (as shown in Fig. \ref{fig:2l-b}), where the lepton portal coupling explicitly enters the production cross-section. However, at low centre-of-mass (CM) energies, the s-channel will dominate, and the effect of the t-channel will be subdued\footnote{Multi-TeV muon colliders can be better setups to observe significant contributions from the t-channel.}. Thirdly, since lepton colliders have a definite CM energy of the hard processes, this allows us to introduce the variable, missing energy ($\slashed{E}$), apart from MET. This variable is significant as this encodes information about the mass of the DM and the VLL. This is illustrated for di-lepton/di-tau + $\slashed{E}$ signal in Fig. \ref{fig:me}.

\begin{figure}[h!]
\centering
\includegraphics[width=0.45\linewidth]{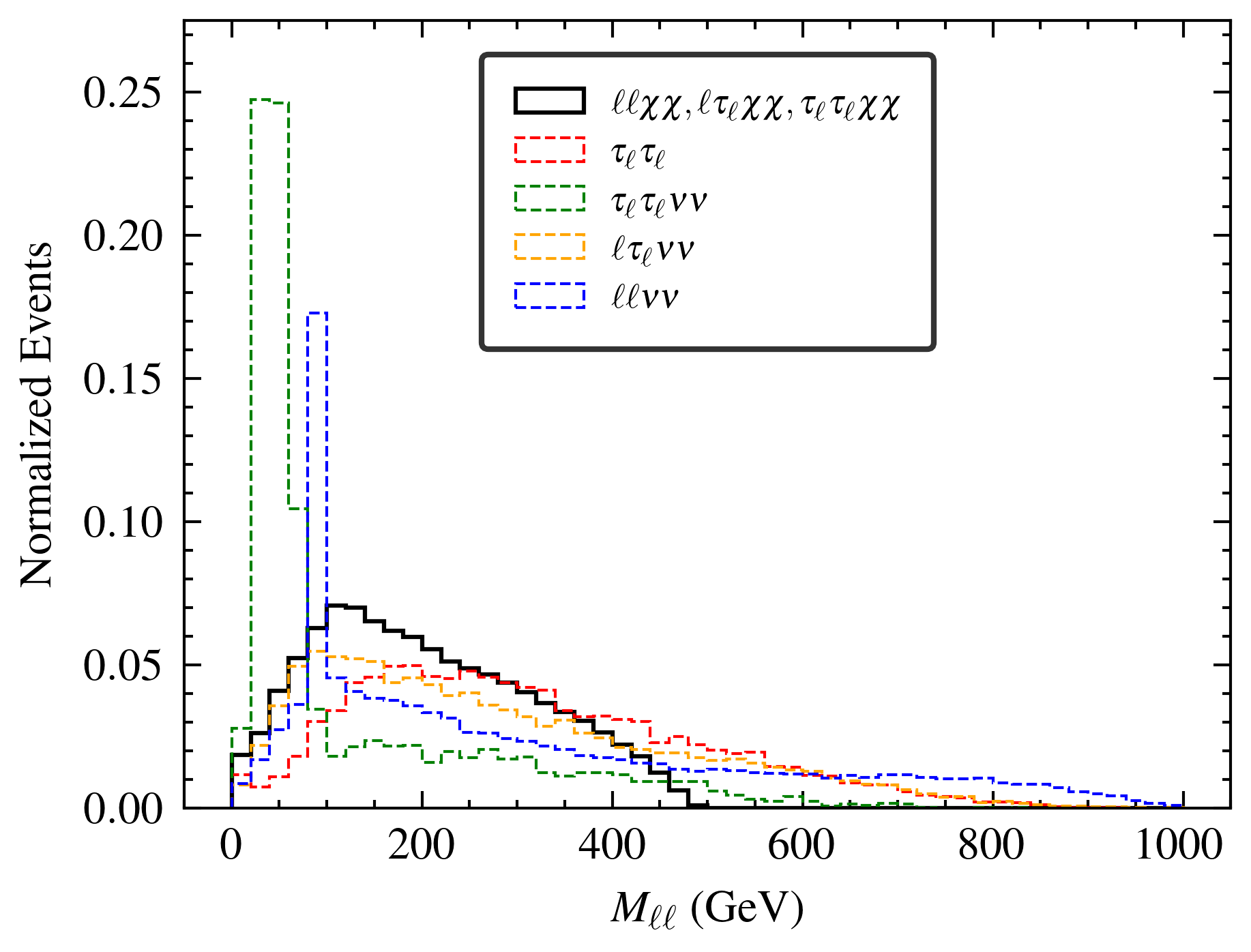}\quad
\includegraphics[width=0.45\linewidth]{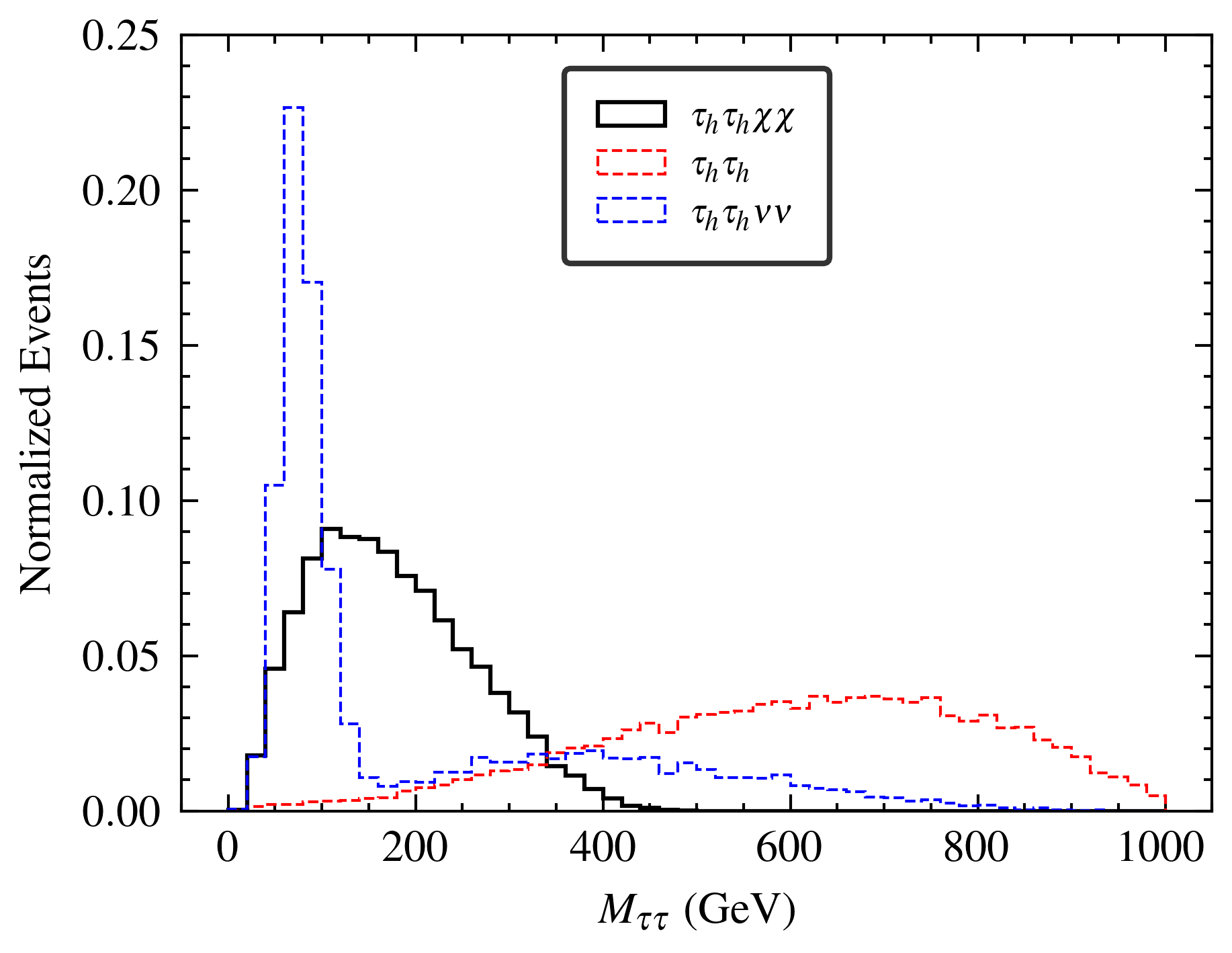}\\
\includegraphics[width=0.45\linewidth]{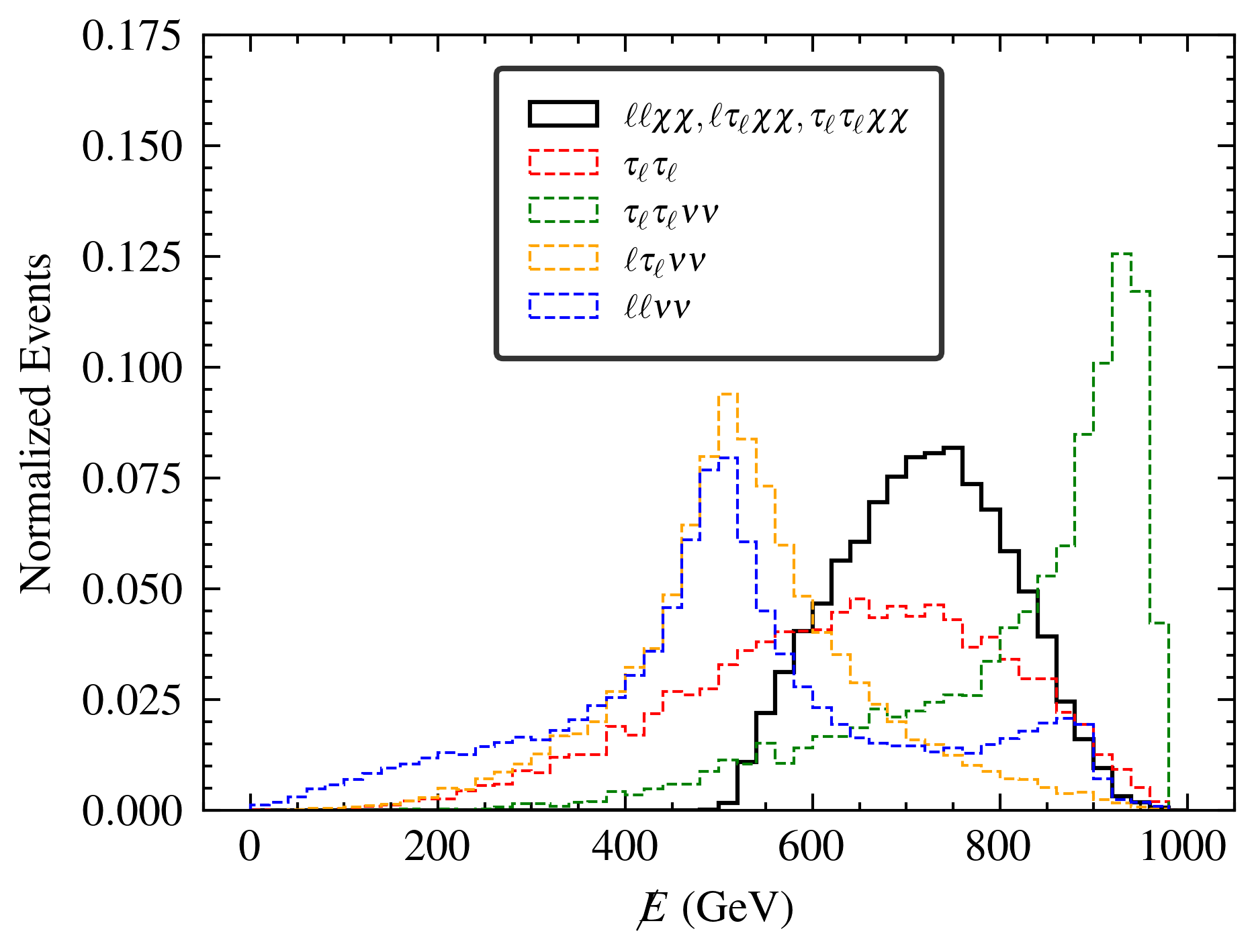}\quad
\includegraphics[width=0.45\linewidth]{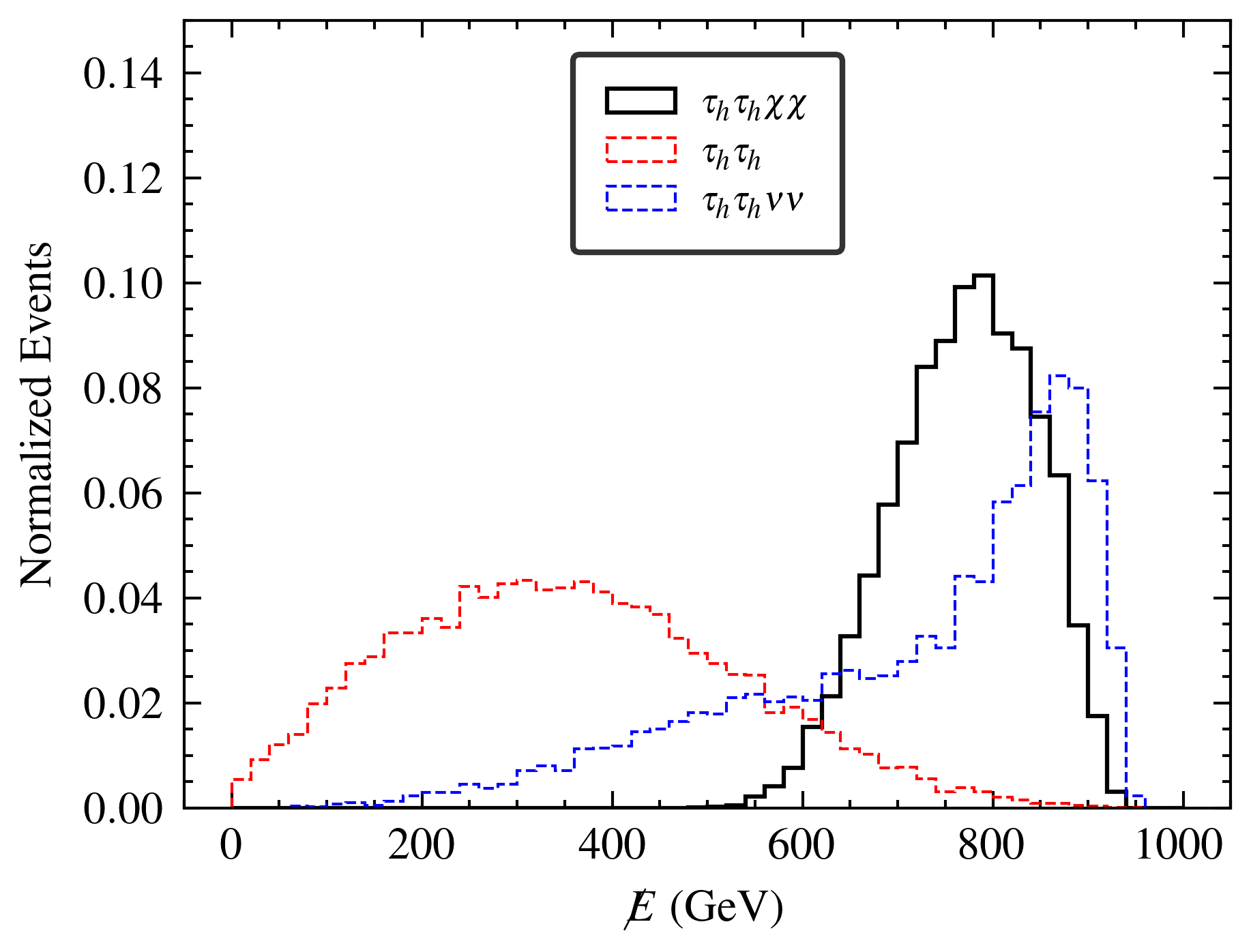}\\
\includegraphics[width=0.45\linewidth]{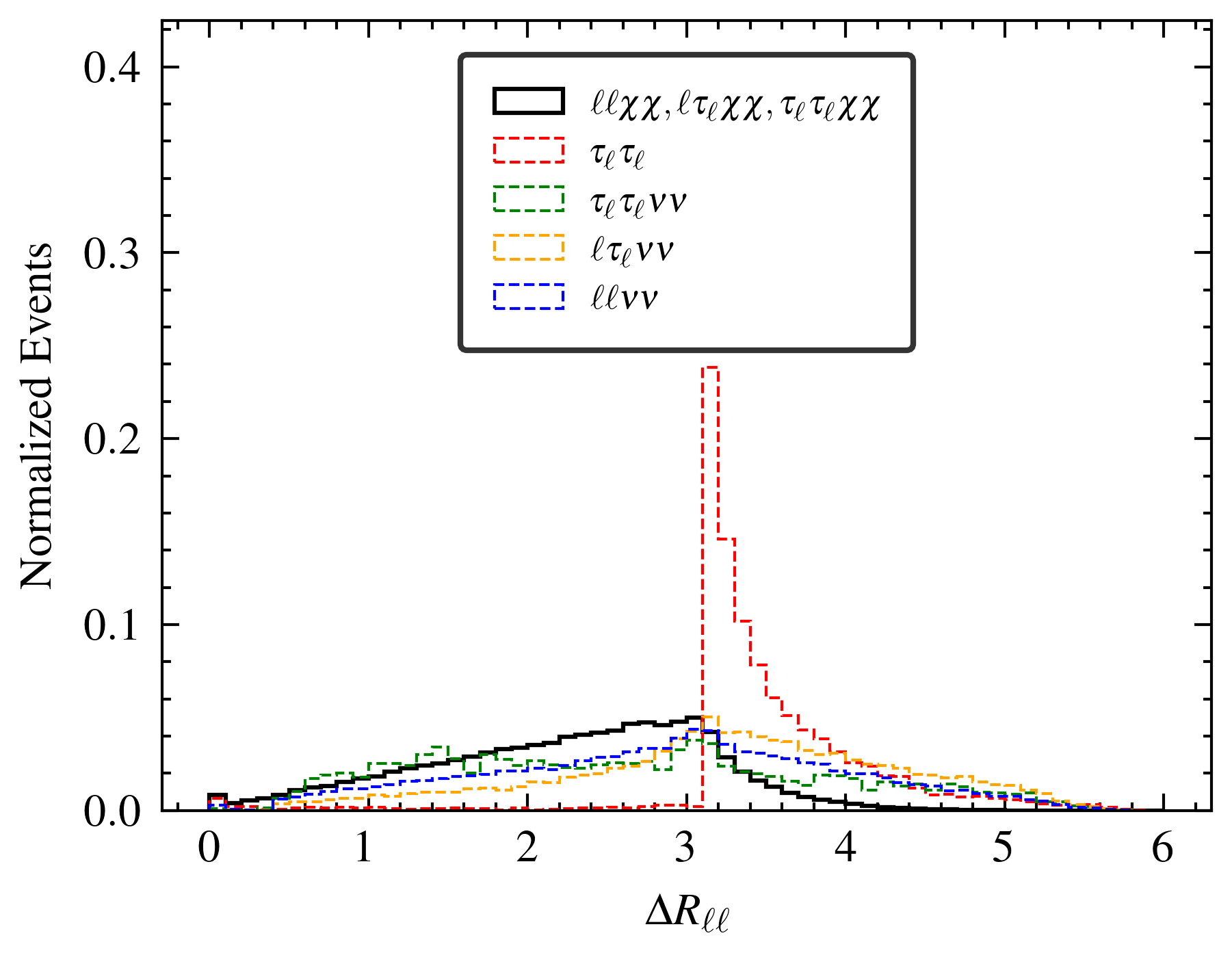}\quad
\includegraphics[width=0.45\linewidth]{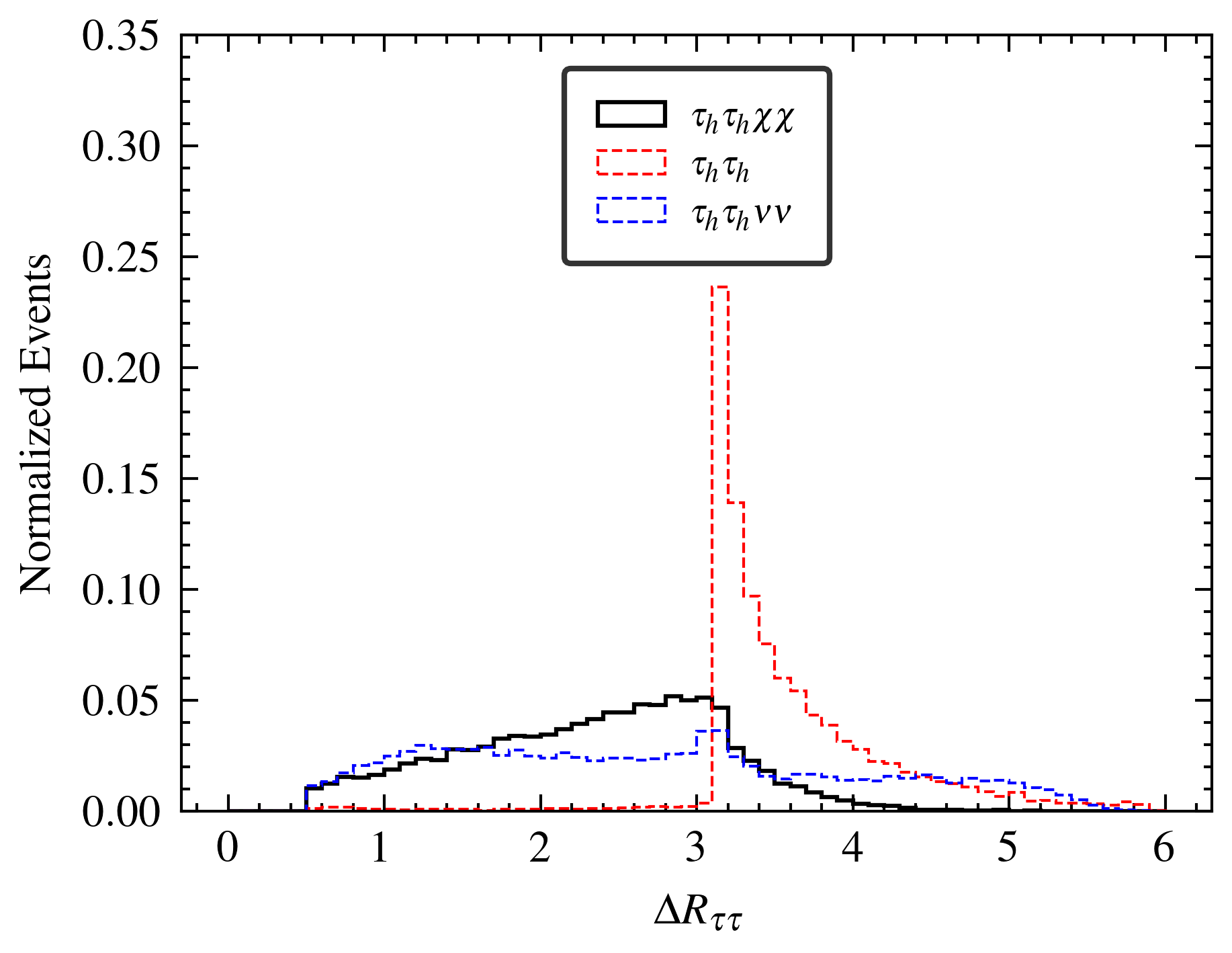}
\caption{Kinematic distributions corresponding to signal and background processes of di-lepton + $\slashed{E}$ (\textit{left}) and di-tau + $\slashed{E}$ (\textit{right}) signal at ILC 1 TeV. The signal correspond to the benchmark: $m_{\psi} = 400$ GeV, $m_{\chi} = 250$ GeV, $\mathtt{y}_{e} = \mathtt{y}_{\mu} = \mathtt{y}_{\tau} = 0.01$.}
\label{fig:d1}
\end{figure}

We perform the analysis at the ILC ($\sqrt{s} =$ 1 TeV). For the di-lepton signal, we choose two opposite sign leptons. We further choose events with no jets. The signal processes are $\ell \ell \chi \chi$, $\ell \tau_{\ell} \chi \chi$ and $\tau_{\ell} \tau_{\ell} \chi \chi$ where, $\tau_{\ell}$ is the leptonic $\tau$ decay products. The relevant SM backgrounds are $\ell \ell \nu \nu$, $\ell \tau_{\ell} \nu \nu$, $\tau_{\ell} \tau_{\ell} \nu \nu$ arising from hard processes like $WW,~ZZ,~\ell\ell Z,~\nu \nu Z,$ etc. and $\tau_{\ell} \tau_{\ell}$ (fully leptonic decay mode of $\tau$ pair production).
Similarly, for the di-tau signal, we choose two $\tau$ jets with no additional leptons or jets in the event. The tagging efficiency of $\tau$ jets is 60\%, and the efficiency of mistagging a light jet as $\tau$ jet is 1\%. The signal process is $\tau_{h} \tau_{h} \chi \chi$, where $\tau_{h}$ is the hadronic tau decay product ($\tau$ jet). The relevant SM backgrounds are $\tau_{h} \tau_{h} \nu \nu$ arising from hard processes like $WW,~ZZ,~\tau \tau Z,~\nu \nu Z,$ etc. and $\tau_{h} \tau_{h}$ (fully hadronic decay mode of $\tau$ pair production). Contribution to the background from light jet final states is found to be negligible and hence not considered. The relevant kinematic variable distributions for signal and background processes are shown in Fig. \ref{fig:d1}.

\begin{figure}[htb!]
\centering
\includegraphics[width=0.475\linewidth]{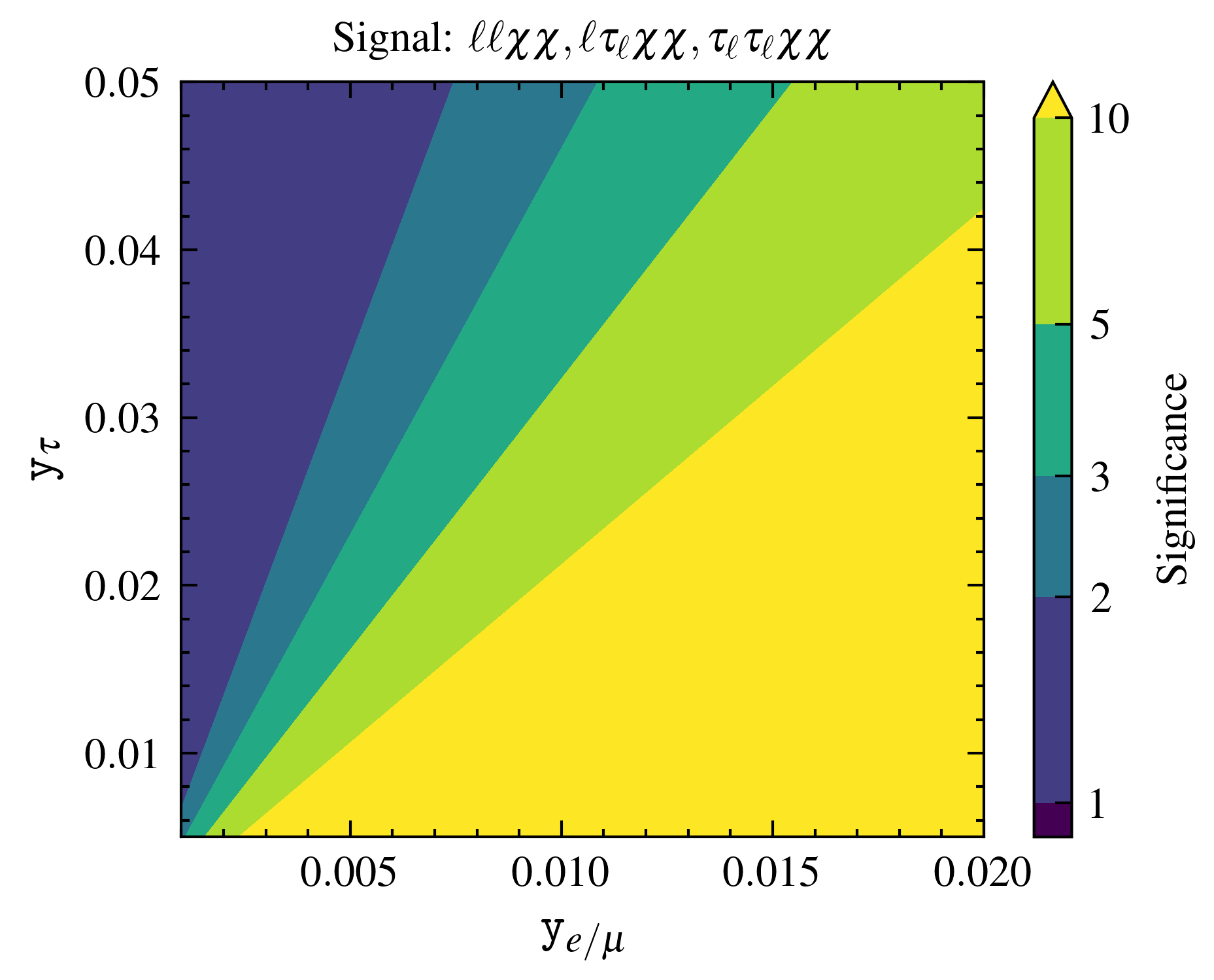}\quad
\includegraphics[width=0.475\linewidth]{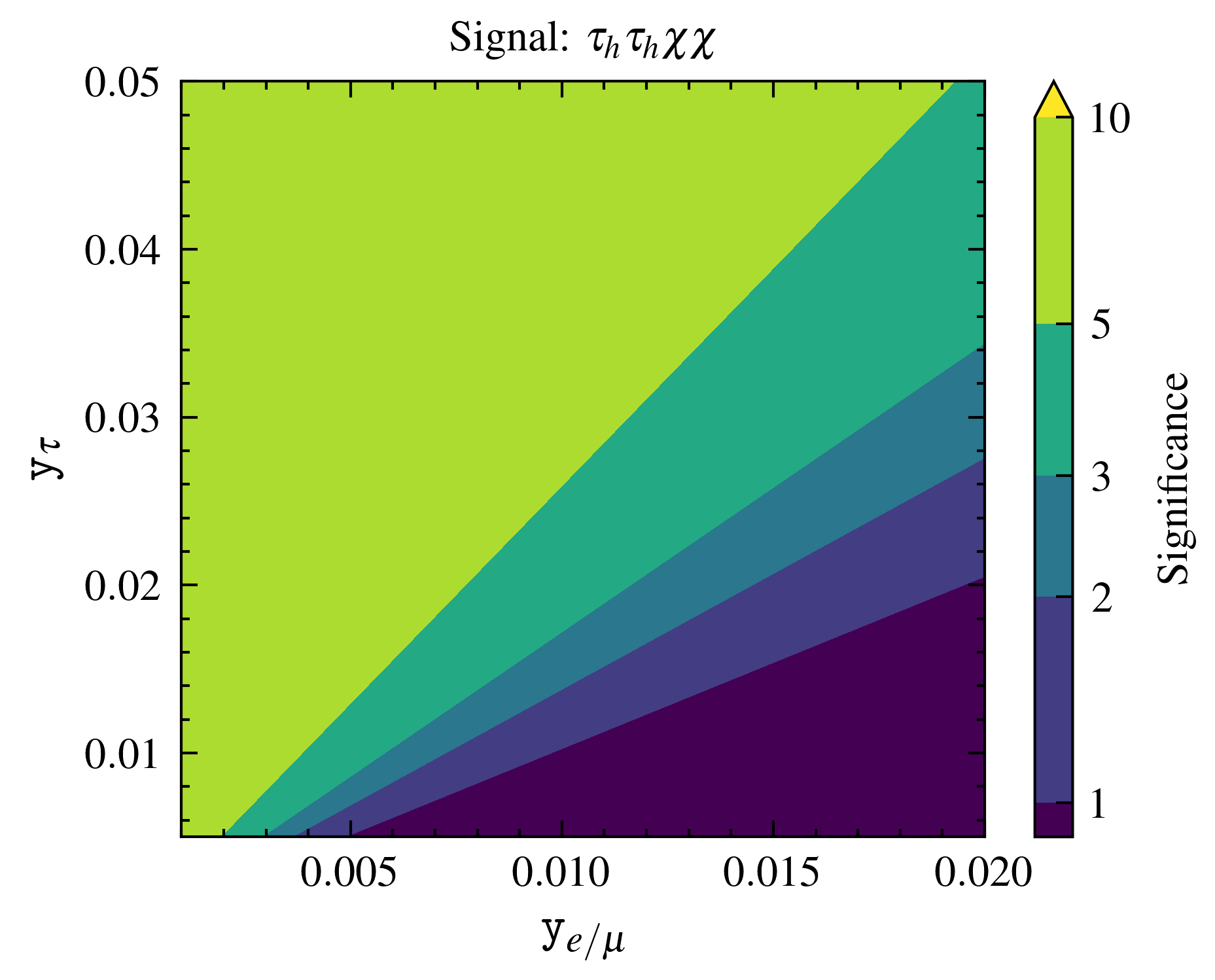}
\caption{Signal significance contours in $\mathtt{y}_{e/\mu}-\mathtt{y}_{\tau}$ plane. The integrated luminosity is taken to be $\mathfrak{L} = 100$ fb$^{-1}$. The signal correspond to the benchmark: $m_{\psi} = 400$ GeV, $m_{\chi} = 250$ GeV.}
\label{fig:sig}
\end{figure}

The signal and background cross-sections corresponding to the di-lepton and di-tau signal are shown in Tab\,.~\ref{tab:cuts}. For segregation of the signal from the background, we implement the following subsequent cuts, guided by the distributions in Fig. \ref{fig:d1}:
\begin{itemize}
\itemsep0em
\item $\mathcal{C}_{1}: M_{\ell \ell} / M_{\tau \tau} > 100$ GeV,
\item $\mathcal{C}_{2}: \slashed{E} > 500$ GeV,
\item $\mathcal{C}_{3}: \Delta R_{\ell \ell}/\Delta R_{\tau \tau} < 3$.
\end{itemize}
Here, $M_{\ell \ell}$ $(M_{\tau \tau})$ is the invariant mass of the lepton ($\tau$ jet) pairs. The variable $\Delta R_{\ell \ell}$ $(\Delta R_{\tau \tau})$ is the distance between the leptons ($\tau$ jets) on the detector $(\eta, \phi)$ plane. The cuts are chosen identically for both signal processes. The cross-sections prior and posterior to the cuts are tabulated in Tab\,.~\ref{tab:cuts}.
\begin{table}[htb!]
\centering
{\renewcommand{\arraystretch}{1.0}
\begin{tabular}{|c|c|c|c|c|}
\hline
Signal and Background Processes & $\sigma_{0}$ (in fb) & $\sigma_{1}$ (in fb) & $\sigma_{2}$ (in fb) & $\sigma_{3}$ (in fb) \\ \hline
Signal: $\ell \ell \chi \chi$, $\ell \tau_{\ell} \chi \chi$, $\tau_{\ell} \tau_{\ell} \chi \chi$ & 33.37 & 26.68 & 26.67 & 19.93 \\ \hline
$\tau_{\ell} \tau_{\ell}$ & 8.92 & 8.22 & 6.36 & 0.09 \\
$\tau_{\ell} \tau_{\ell} \nu \nu$ & 2.60 & 0.88 & 0.78 & 0.07 \\
$\ell \tau_{\ell} \nu \nu$ & 32.16 & 26.72 & 14.04 & 5.74 \\
$\ell \ell \nu \nu$ & 249.94 & 184.49 & 64.54 & 43.83 \\ \hline
Background (Total)  & 293.62 & 220.31 & 85.72 & 49.73 \\ \hline
\end{tabular}}
\vspace{0.25cm}

{\renewcommand{\arraystretch}{1.0}
\begin{tabular}{|c|c|c|c|c|}
\hline
Signal and Background Processes & $\sigma_{0}$ (in fb) & $\sigma_{1}$ (in fb) & $\sigma_{2}$ (in fb) & $\sigma_{3}$ (in fb) \\ \hline
Signal: $\tau_{h} \tau_{h} \chi \chi$ & 1.14 & 0.90 & 0.90 & 0.66 \\ \hline
$\tau_{h} \tau_{h}$ & 16.21 & 16.07 & 3.29 & 0.03 \\
$\tau_{h} \tau_{h} \nu \nu$ & 5.20 & 2.50 & 1.75 & 0.58 \\ \hline
Background (Total) & 21.41 & 18.57 & 5.04 & 0.61 \\ \hline
\end{tabular}}
\caption{Cross sections following subsequent cuts for di-lepton signal, including leptonic $\tau$ decays (\textit{top}) and di-tau (hadronic) signal (\textit{bottom}). $\sigma_0$, $\sigma_1$, $\sigma_2$ and $\sigma_3$ are the cross section post sequential cuts, $\mathcal{C}_{1}$, $\mathcal{C}_{2}$ and $\mathcal{C}_{3}$ respectively. The signal correspond to the benchmark: $m_{\psi} = 400$ GeV, $m_{\chi} = 250$ GeV, $\mathtt{y}_{e} = \mathtt{y}_{\mu} = \mathtt{y}_{\tau} = 0.01$.}
\label{tab:cuts}
\end{table}
Choosing an invariant mass cut, $M_{\ell \ell} / M_{\tau \tau} > 100$ GeV wipes out backgrounds where the leptons/$\tau$ jets are products of $Z$ decay. The most important variable is the missing energy of the event. The presence of massive DM results in missing energies to peak at higher values compared to most SM backgrounds. A missing energy cut of $\slashed{E} > 500$ GeV significantly reduces the major backgrounds in the case of both signals while keeping the signal numbers almost unaltered. Finally, $\tau$ pair backgrounds can be significantly reduced by choosing $\Delta R_{\ell \ell}/\Delta R_{\tau \tau}$ to be less than 3. Here, the signal correspond to the benchmark: $m_{\psi} = 400$ GeV, $m_{\chi} = 250$ GeV, $\mathtt{y}_{e} = \mathtt{y}_{\mu} = \mathtt{y}_{\tau} = 0.01$. Tuning these lepton portal couplings alters the branching ratios of $\psi$ decay, which significantly affects the signal cross sections. The signal significance contours plotted on $\mathtt{y}_{e/\mu}-\mathtt{y}_{\tau}$ plane are shown for both the signal processes in Fig\,.~\ref{fig:sig}. The signal significance is defined as:
\begin{equation}
{\rm Significance} = \frac{\sigma_{\rm s}}{\sqrt{\sigma_{\rm b}}} \times \sqrt{\mathfrak{L}}\,.
\end{equation}
Here, $\sigma_{\rm s}$ and $\sigma_{\rm b}$ are the signal and background cross sections post cutflow. $\mathfrak{L}$ is the integrated luminosity. The couplings $\mathtt{y}_{e/\mu}$ are strongly constrained from $\mu \to e \gamma$ LFV decay, however the same doesn't apply for $\mathtt{y}_{\tau}$. Hence, for certain parameter space regions di-tau may appear to be better choice as signal despite lower cross section. This has been illustrated in the plots. The parameter space is unaffected by future sensitivities of $\mu\to e \gamma$ branching measurements at MEG II.
\section{Summary and Conclusion}
\label{sec6}
In this article, we have studied a minimal extension of the Standard Model (SM) containing a real and a complex scalar DM, stable under $\mathcal{Z}_2\otimes\mathcal{Z}_3$ symmetry. Additionally, we introduced a Dirac vector-like lepton (VLL) with hypercharge $-1$, which transforms under $\mathcal{Z}_3$ symmetry in a similar manner to the complex scalar DM. This VLL connects the DM and the lepton sector, hence becoming subjected to constraints from lepton-flavor violation on the DM relic density allowed parameter space.
In section \ref{sec3}, we discussed the constraints from lepton flavor anomalies and lepton flavor violating decays. Although lepton flavor-conserving processes, such as muon \(g-2\) and electron \(g-2\), are less restrictive in our analysis region (\(\mathtt{y}_{\ell} \lesssim 0.1\)), the lepton flavor-violating (LFV) decay mode \(\mu \to e \gamma\) significantly constrains the parameter space. In our analysis framework, the constraints from Higgs boson decays to di-leptons are less stringent than those from LFV processes and can therefore be disregarded.

In section \ref{sec4}, we provide a comprehensive discussion of the DM phenomenology. In the WIMP-pFIMP scenario, the cBEQ has been written in the standard manner, but the only difference arises from the pFIMP interaction terms, most importantly the conversion terms, which is key to thermalization of FIMP, i.e. pFIMP scenario.
The total DM relic density, attributed to both WIMPs and pFIMPs, is calculated by solving the cBEQ. This has been discussed in the two subsections that focus on the effective spin-independent WIMP/pFIMP-nucleon scattering cross-section and the production of SM particles through WIMP/pFIMP annihilation.
The limits from LFV are also superimposed, resulting in the exclusion of most points below $m_{\rm DM} \sim 100~\rm GeV$.  In the case of WIMP, some points can still be probed at future DD experiments, PandaX-xT or DARWIN (200 t y), etc.
Concurrently, the direct detection of pFIMP is only possible via the WIMP loop and most of the points are consistent with current DD observations, and we look forward to future detection results. The indirect detection limits could also be applicable to the WIMP annihilation to the electron-positron (tree), bottom pair (tree) and photon pair (box-loop), and also semi-annihilation (tree) to the Higgs.
So, the observed and projection limits from Fermi-LAT, AMS$-02$, H.E.S.S, and CTA (projection) also put an exclusion on the relic density and DD-allowed parameter space. The WIMP self-annihilation to electron-positron only excludes the low mass regime, while the Higgs resonance regime is excluded by the annihilation to bottom pair. Additionally, around the Higgs mass, certain points are excluded for large Higgs portal coupling \((\lambda_{\chi\rm H} \gtrsim 0.1)\), which is relevant for semi-annihilation of WIMP.
However, the annihilation into two photons remains unaffected at current experimental sensitivity. For pFIMPs, annihilation into bottom quark pairs happens only through the WIMP loop process, making them mostly unaffected by Fermi-LAT constraints, except near the Higgs resonance.
In section \ref{sec5}, we explored how our model could be tested in collider experiments. We recast the results from existing LHC studies on di-lepton + MET events in the context of our model, allowing us to determine the exclusion limits for the masses of the WIMP and VLL. We also estimate how these limits might improve at the HL-LHC run, finding that the exclusion limit roughly doubles.
Additionally, we proposed a DM search strategy for future lepton colliders, analyzing di-lepton/di-tau + missing energy events at the $1~\rm TeV$ run of the ILC. Finally, we compare the two signal processes and highlight the parameter space on the $\mathtt{y}_{e/\mu}-\mathtt{y}_{\tau}$ plane where each signal is likely to be most significant.

With a minimal extension to the SM that includes a real scalar, a complex scalar, and a VLL, we can successfully explain the DM relic density and LFV constraints. At the same time, this model remains consistent with current limits from direct, indirect, and collider experiments, while allowing for the future detection of WIMPs and pFIMPs in these searches. Not only that, such lepton portal DM models will also leave a signature at LFV decays and can be probed at future low-energy experiments as well. This nice feature occurs due to the presence of the Higgs portal, which is related to direct and indirect searches for DM, as well as lepton portal interactions relevant to LFV decay and collider searches for DM. Through our analysis, we found that both portal interactions significantly contribute to the DM relic density.
\acknowledgments
DP extends his gratitude to Prof. Subhaditya Bhattacharya for his valuable discussions and insightful comments throughout the draft preparation.

\newpage
\bibliographystyle{JHEP}
\bibliography{Hdm}

\providecommand{\href}[2]{#2}\begingroup\raggedright\begin{thebibliography}{100}

\bibitem{Clowe:2006eq}
D.~Clowe, M.~Bradac, A.H.~Gonzalez, M.~Markevitch, S.W.~Randall, C.~Jones
  et~al., \emph{{A direct empirical proof of the existence of dark matter}},
  \href{https://doi.org/10.1086/508162}{\emph{Astrophys. J. Lett.} {\bfseries
  648} (2006) L109} [\href{https://arxiv.org/abs/astro-ph/0608407}{{\ttfamily
  astro-ph/0608407}}].

\bibitem{Komatsu:2014ioa}
{\scshape WMAP Science Team} collaboration, \emph{{Results from the Wilkinson
  Microwave Anisotropy Probe}},
  \href{https://doi.org/10.1093/ptep/ptu083}{\emph{PTEP} {\bfseries 2014}
  (2014) 06B102} [\href{https://arxiv.org/abs/1404.5415}{{\ttfamily
  1404.5415}}].

\bibitem{Sofue:2000jx}
Y.~Sofue and V.~Rubin, \emph{{Rotation curves of spiral galaxies}},
  \href{https://doi.org/10.1146/annurev.astro.39.1.137}{\emph{Ann. Rev. Astron.
  Astrophys.} {\bfseries 39} (2001) 137}
  [\href{https://arxiv.org/abs/astro-ph/0010594}{{\ttfamily
  astro-ph/0010594}}].

\bibitem{Hayashi:2006kw}
E.~Hayashi and S.D.M.~White, \emph{{How Rare is the Bullet Cluster?}},
  \href{https://doi.org/10.1111/j.1745-3933.2006.00184.x}{\emph{Mon. Not. Roy.
  Astron. Soc.} {\bfseries 370} (2006) L38}
  [\href{https://arxiv.org/abs/astro-ph/0604443}{{\ttfamily
  astro-ph/0604443}}].

\bibitem{Zwicky:1933gu}
F.~Zwicky, \emph{{Die Rotverschiebung von extragalaktischen Nebeln}},
  \href{https://doi.org/10.1007/s10714-008-0707-4}{\emph{Helv. Phys. Acta}
  {\bfseries 6} (1933) 110}.

\bibitem{Zwicky:1937zza}
F.~Zwicky, \emph{{On the Masses of Nebulae and of Clusters of Nebulae}},
  \href{https://doi.org/10.1086/143864}{\emph{Astrophys. J.} {\bfseries 86}
  (1937) 217}.

\bibitem{Trimble:1987ee}
V.~Trimble, \emph{{Existence and Nature of Dark Matter in the Universe}},
  \href{https://doi.org/10.1146/annurev.aa.25.090187.002233}{\emph{Ann. Rev.
  Astron. Astrophys.} {\bfseries 25} (1987) 425}.

\bibitem{Griest:1989wd}
K.~Griest and M.~Kamionkowski, \emph{{Unitarity Limits on the Mass and Radius
  of Dark Matter Particles}},
  \href{https://doi.org/10.1103/PhysRevLett.64.615}{\emph{Phys. Rev. Lett.}
  {\bfseries 64} (1990) 615}.

\bibitem{Jungman:1995df}
G.~Jungman, M.~Kamionkowski and K.~Griest, \emph{{Supersymmetric dark matter}},
  \href{https://doi.org/10.1016/0370-1573(95)00058-5}{\emph{Phys. Rept.}
  {\bfseries 267} (1996) 195}
  [\href{https://arxiv.org/abs/hep-ph/9506380}{{\ttfamily hep-ph/9506380}}].

\bibitem{Badziak:2017uto}
M.~Badziak, M.~Olechowski and P.~Szczerbiak, \emph{{Spin-dependent constraints
  on blind spots for thermal singlino-higgsino dark matter with(out) light
  singlets}}, \href{https://doi.org/10.1007/JHEP07(2017)050}{\emph{JHEP}
  {\bfseries 07} (2017) 050}
  [\href{https://arxiv.org/abs/1705.00227}{{\ttfamily 1705.00227}}].

\bibitem{Badziak:2015exr}
M.~Badziak, M.~Olechowski and P.~Szczerbiak, \emph{{Blind spots for neutralino
  dark matter in the NMSSM}},
  \href{https://doi.org/10.1007/JHEP03(2016)179}{\emph{JHEP} {\bfseries 03}
  (2016) 179} [\href{https://arxiv.org/abs/1512.02472}{{\ttfamily
  1512.02472}}].

\bibitem{Dey:2019lyr}
A.~Dey, J.~Lahiri and B.~Mukhopadhyaya, \emph{{LHC signals of a heavy doublet
  Higgs as dark matter portal: cut-based approach and improvement with gradient
  boosting and neural networks}},
  \href{https://doi.org/10.1007/JHEP09(2019)004}{\emph{JHEP} {\bfseries 09}
  (2019) 004} [\href{https://arxiv.org/abs/1905.02242}{{\ttfamily
  1905.02242}}].

\bibitem{PhysRevD.43.3191}
K.~Griest and D.~Seckel, \emph{Three exceptions in the calculation of relic
  abundances}, \href{https://doi.org/10.1103/PhysRevD.43.3191}{\emph{Phys. Rev.
  D} {\bfseries 43} (1991) 3191}.

\bibitem{Edsjo:1997bg}
J.~Edsjo and P.~Gondolo, \emph{{Neutralino relic density including
  coannihilations}},
  \href{https://doi.org/10.1103/PhysRevD.56.1879}{\emph{Phys. Rev. D}
  {\bfseries 56} (1997) 1879}
  [\href{https://arxiv.org/abs/hep-ph/9704361}{{\ttfamily hep-ph/9704361}}].

\bibitem{DAgnolo:2017dbv}
R.T.~D'Agnolo, D.~Pappadopulo and J.T.~Ruderman, \emph{{Fourth Exception in the
  Calculation of Relic Abundances}},
  \href{https://doi.org/10.1103/PhysRevLett.119.061102}{\emph{Phys. Rev. Lett.}
  {\bfseries 119} (2017) 061102}
  [\href{https://arxiv.org/abs/1705.08450}{{\ttfamily 1705.08450}}].

\bibitem{DiazSaez:2024nrq}
B.~D\'\i{}az~S\'aez, J.~Lahiri and K.~M\"ohling, \emph{{Coscattering in the
  extended singlet-scalar Higgs portal}},
  \href{https://doi.org/10.1088/1475-7516/2024/10/001}{\emph{JCAP} {\bfseries
  10} (2024) 001} [\href{https://arxiv.org/abs/2404.19057}{{\ttfamily
  2404.19057}}].

\bibitem{Hall:2009bx}
L.J.~Hall, K.~Jedamzik, J.~March-Russell and S.M.~West, \emph{{Freeze-In
  Production of FIMP Dark Matter}},
  \href{https://doi.org/10.1007/JHEP03(2010)080}{\emph{JHEP} {\bfseries 03}
  (2010) 080} [\href{https://arxiv.org/abs/0911.1120}{{\ttfamily 0911.1120}}].

\bibitem{Elahi:2014fsa}
F.~Elahi, C.~Kolda and J.~Unwin, \emph{{UltraViolet Freeze-in}},
  \href{https://doi.org/10.1007/JHEP03(2015)048}{\emph{JHEP} {\bfseries 03}
  (2015) 048} [\href{https://arxiv.org/abs/1410.6157}{{\ttfamily 1410.6157}}].

\bibitem{Blennow:2013jba}
M.~Blennow, E.~Fernandez-Martinez and B.~Zaldivar, \emph{{Freeze-in through
  portals}}, \href{https://doi.org/10.1088/1475-7516/2014/01/003}{\emph{JCAP}
  {\bfseries 01} (2014) 003} [\href{https://arxiv.org/abs/1309.7348}{{\ttfamily
  1309.7348}}].

\bibitem{Bhattacharya:2022vxm}
S.~Bhattacharya, J.~Lahiri and D.~Pradhan, \emph{{Detection possibility of a
  pseudo-FIMP in the presence of a thermal WIMP}},
  \href{https://doi.org/10.1103/PhysRevD.109.095031}{\emph{Phys. Rev. D}
  {\bfseries 109} (2024) 095031}
  [\href{https://arxiv.org/abs/2212.14846}{{\ttfamily 2212.14846}}].

\bibitem{Bhattacharya:2022dco}
S.~Bhattacharya, D.~Pradhan and J.~Lahiri, \emph{{Dynamics of pseudofeebly
  interacting massive particles in presence of thermal dark matter}},
  \href{https://doi.org/10.1103/PhysRevD.108.L111702}{\emph{Phys. Rev. D}
  {\bfseries 108} (2023) L111702}
  [\href{https://arxiv.org/abs/2212.07622}{{\ttfamily 2212.07622}}].

\bibitem{Bai:2014osa}
Y.~Bai and J.~Berger, \emph{{Lepton Portal Dark Matter}},
  \href{https://doi.org/10.1007/JHEP08(2014)153}{\emph{JHEP} {\bfseries 08}
  (2014) 153} [\href{https://arxiv.org/abs/1402.6696}{{\ttfamily 1402.6696}}].

\bibitem{Kawamura:2020qxo}
J.~Kawamura, S.~Okawa and Y.~Omura, \emph{{Current status and muon $g-2$
  explanation of lepton portal dark matter}},
  \href{https://doi.org/10.1007/JHEP08(2020)042}{\emph{JHEP} {\bfseries 08}
  (2020) 042} [\href{https://arxiv.org/abs/2002.12534}{{\ttfamily
  2002.12534}}].

\bibitem{PandaX:2024pjr}
{\scshape PandaX} collaboration, \emph{{Search for lepton portal dark matter in
  the PandaX-4T experiment}},
  \href{https://arxiv.org/abs/2408.14730}{{\ttfamily 2408.14730}}.

\bibitem{DiazSaez:2022nhp}
B.~D\'\i{}az~S\'aez and K.~Ghorbani, \emph{{Z $_{3}$ scalar dark matter with
  strong positron fluxes}},
  \href{https://doi.org/10.1088/1475-7516/2023/02/002}{\emph{JCAP} {\bfseries
  02} (2023) 002} [\href{https://arxiv.org/abs/2203.09282}{{\ttfamily
  2203.09282}}].

\bibitem{Asadi:2023csb}
P.~Asadi, A.~Radick and T.-T.~Yu, \emph{{Interplay of freeze-in and freeze-out:
  Lepton-flavored dark matter and muon colliders}},
  \href{https://doi.org/10.1103/PhysRevD.110.035022}{\emph{Phys. Rev. D}
  {\bfseries 110} (2024) 035022}
  [\href{https://arxiv.org/abs/2312.03826}{{\ttfamily 2312.03826}}].

\bibitem{Mandal:2018czf}
R.~Mandal, \emph{{Fermionic dark matter in leptoquark portal}},
  \href{https://doi.org/10.1140/epjc/s10052-018-6192-3}{\emph{Eur. Phys. J. C}
  {\bfseries 78} (2018) 726}
  [\href{https://arxiv.org/abs/1808.07844}{{\ttfamily 1808.07844}}].

\bibitem{Bhattacharya:2013hva}
S.~Bhattacharya, A.~Drozd, B.~Grzadkowski and J.~Wudka, \emph{{Two-Component
  Dark Matter}}, \href{https://doi.org/10.1007/JHEP10(2013)158}{\emph{JHEP}
  {\bfseries 10} (2013) 158} [\href{https://arxiv.org/abs/1309.2986}{{\ttfamily
  1309.2986}}].

\bibitem{Bhattacharya:2016ysw}
S.~Bhattacharya, P.~Poulose and P.~Ghosh, \emph{{Multipartite Interacting
  Scalar Dark Matter in the light of updated LUX data}},
  \href{https://doi.org/10.1088/1475-7516/2017/04/043}{\emph{JCAP} {\bfseries
  04} (2017) 043} [\href{https://arxiv.org/abs/1607.08461}{{\ttfamily
  1607.08461}}].

\bibitem{DuttaBanik:2016jzv}
A.~Dutta~Banik, M.~Pandey, D.~Majumdar and A.~Biswas, \emph{{Two component
  WIMP\textendash{}FImP dark matter model with singlet fermion, scalar and
  pseudo scalar}},
  \href{https://doi.org/10.1140/epjc/s10052-017-5221-y}{\emph{Eur. Phys. J. C}
  {\bfseries 77} (2017) 657}
  [\href{https://arxiv.org/abs/1612.08621}{{\ttfamily 1612.08621}}].

\bibitem{Bhattacharya:2021rwh}
S.~Bhattacharya, S.~Chakraborti and D.~Pradhan, \emph{{Electroweak symmetry
  breaking and WIMP-FIMP dark matter}},
  \href{https://doi.org/10.1007/JHEP07(2022)091}{\emph{JHEP} {\bfseries 07}
  (2022) 091} [\href{https://arxiv.org/abs/2110.06985}{{\ttfamily
  2110.06985}}].

\bibitem{Bhattacharya:2024nla}
S.~Bhattacharya, L.~Kolay and D.~Pradhan, \emph{{Multiparticle scalar dark
  matter with $\mathbb{Z}_N$ symmetry}},
  \href{https://arxiv.org/abs/2410.16275}{{\ttfamily 2410.16275}}.

\bibitem{Athron:2021iuf}
P.~Athron, C.~Bal\'azs, D.H.J.~Jacob, W.~Kotlarski, D.~St\"ockinger and
  H.~St\"ockinger-Kim, \emph{{New physics explanations of a$_{\mu}$ in light of
  the FNAL muon $g-2$ measurement}},
  \href{https://doi.org/10.1007/JHEP09(2021)080}{\emph{JHEP} {\bfseries 09}
  (2021) 080} [\href{https://arxiv.org/abs/2104.03691}{{\ttfamily
  2104.03691}}].

\bibitem{Belanger:2012zr}
G.~Belanger, K.~Kannike, A.~Pukhov and M.~Raidal, \emph{{$Z_3$ Scalar Singlet
  Dark Matter}},
  \href{https://doi.org/10.1088/1475-7516/2013/01/022}{\emph{JCAP} {\bfseries
  01} (2013) 022} [\href{https://arxiv.org/abs/1211.1014}{{\ttfamily
  1211.1014}}].

\bibitem{Planck:2018vyg}
{\scshape Planck} collaboration, \emph{{Planck 2018 results. VI. Cosmological
  parameters}},
  \href{https://doi.org/10.1051/0004-6361/201833910}{\emph{Astron. Astrophys.}
  {\bfseries 641} (2020) A6}
  [\href{https://arxiv.org/abs/1807.06209}{{\ttfamily 1807.06209}}].

\bibitem{L3:2001xsz}
{\scshape L3} collaboration, \emph{{Search for heavy neutral and charged
  leptons in $e^{+} e^{-}$ annihilation at LEP}},
  \href{https://doi.org/10.1016/S0370-2693(01)01005-X}{\emph{Phys. Lett. B}
  {\bfseries 517} (2001) 75}
  [\href{https://arxiv.org/abs/hep-ex/0107015}{{\ttfamily hep-ex/0107015}}].

\bibitem{ALEPH:2002nwp}
{\scshape ALEPH} collaboration, \emph{{Absolute lower limits on the masses of
  selectrons and sneutrinos in the MSSM}},
  \href{https://doi.org/10.1016/S0370-2693(02)02471-1}{\emph{Phys. Lett. B}
  {\bfseries 544} (2002) 73}
  [\href{https://arxiv.org/abs/hep-ex/0207056}{{\ttfamily hep-ex/0207056}}].

\bibitem{OPAL:2003nhx}
{\scshape OPAL} collaboration, \emph{{Search for anomalous production of
  dilepton events with missing transverse momentum in e+ e- collisions at
  s**(1/2) = 183-Gev to 209-GeV}},
  \href{https://doi.org/10.1140/epjc/s2003-01466-y}{\emph{Eur. Phys. J. C}
  {\bfseries 32} (2004) 453}
  [\href{https://arxiv.org/abs/hep-ex/0309014}{{\ttfamily hep-ex/0309014}}].

\bibitem{DELPHI:2003uqw}
{\scshape DELPHI} collaboration, \emph{{Searches for supersymmetric particles
  in e+ e- collisions up to 208-GeV and interpretation of the results within
  the MSSM}}, \href{https://doi.org/10.1140/epjc/s2003-01355-5}{\emph{Eur.
  Phys. J. C} {\bfseries 31} (2003) 421}
  [\href{https://arxiv.org/abs/hep-ex/0311019}{{\ttfamily hep-ex/0311019}}].

\bibitem{L3:2003fyi}
{\scshape L3} collaboration, \emph{{Search for scalar leptons and scalar quarks
  at LEP}}, \href{https://doi.org/10.1016/j.physletb.2003.10.010}{\emph{Phys.
  Lett. B} {\bfseries 580} (2004) 37}
  [\href{https://arxiv.org/abs/hep-ex/0310007}{{\ttfamily hep-ex/0310007}}].

\bibitem{Athron:2018ipf}
P.~Athron, J.M.~Cornell, F.~Kahlhoefer, J.~Mckay, P.~Scott and S.~Wild,
  \emph{{Impact of vacuum stability, perturbativity and XENON1T on global fits
  of $\mathbb {Z}_2$ and $\mathbb {Z}_3$ scalar singlet dark matter}},
  \href{https://doi.org/10.1140/epjc/s10052-018-6314-y}{\emph{Eur. Phys. J. C}
  {\bfseries 78} (2018) 830}
  [\href{https://arxiv.org/abs/1806.11281}{{\ttfamily 1806.11281}}].

\bibitem{Choi:2021yps}
S.-M.~Choi, J.~Kim, P.~Ko and J.~Li, \emph{{A multi-component SIMP model with
  U(1)$_{X}$\textrightarrow{} Z$_{2}$ \texttimes{} Z$_{3}$}},
  \href{https://doi.org/10.1007/JHEP09(2021)028}{\emph{JHEP} {\bfseries 09}
  (2021) 028} [\href{https://arxiv.org/abs/2103.05956}{{\ttfamily
  2103.05956}}].

\bibitem{ATLAS:2023tkt}
{\scshape ATLAS} collaboration, \emph{{Combination of searches for invisible
  decays of the Higgs boson using 139 fb\ensuremath{-}1 of proton-proton
  collision data at s=13 TeV collected with the ATLAS experiment}},
  \href{https://doi.org/10.1016/j.physletb.2023.137963}{\emph{Phys. Lett. B}
  {\bfseries 842} (2023) 137963}
  [\href{https://arxiv.org/abs/2301.10731}{{\ttfamily 2301.10731}}].

\bibitem{CMS:2023sdw}
{\scshape CMS} collaboration, \emph{{A search for decays of the Higgs boson to
  invisible particles in events with a top-antitop quark pair or a vector boson
  in proton-proton collisions at $\sqrt{s} = 13\,\text {Te}\hspace{-.08em}\text
  {V} $}}, \href{https://doi.org/10.1140/epjc/s10052-023-11952-7}{\emph{Eur.
  Phys. J. C} {\bfseries 83} (2023) 933}
  [\href{https://arxiv.org/abs/2303.01214}{{\ttfamily 2303.01214}}].

\bibitem{CMS:2022ley}
{\scshape CMS} collaboration, \emph{{Measurement of the Higgs boson width and
  evidence of its off-shell contributions to ZZ production}},
  \href{https://doi.org/10.1038/s41567-022-01682-0}{\emph{Nature Phys.}
  {\bfseries 18} (2022) 1329}
  [\href{https://arxiv.org/abs/2202.06923}{{\ttfamily 2202.06923}}].

\bibitem{LHCHiggsCrossSectionWorkingGroup:2016ypw}
{\scshape LHC Higgs Cross Section Working Group} collaboration, \emph{{Handbook
  of LHC Higgs Cross Sections: 4. Deciphering the Nature of the Higgs Sector}},
   \href{https://arxiv.org/abs/1610.07922}{{\ttfamily 1610.07922}}.

\bibitem{CMS:2021qbc}
{\scshape CMS} collaboration, \emph{{ Precision measurement of the Z invisible
  width with the CMS experiment in pp collisions at $\sqrt{s}=13~\mathrm{TeV}$
  }}, .

\bibitem{CMS:2022ett}
{\scshape CMS} collaboration, \emph{{Precision measurement of the Z boson
  invisible width in pp collisions at s=13 TeV}},
  \href{https://doi.org/10.1016/j.physletb.2022.137563}{\emph{Phys. Lett. B}
  {\bfseries 842} (2023) 137563}
  [\href{https://arxiv.org/abs/2206.07110}{{\ttfamily 2206.07110}}].

\bibitem{ALEPH:2005ab}
{\scshape ALEPH, DELPHI, L3, OPAL, SLD, LEP Electroweak Working Group, SLD
  Electroweak Group, SLD Heavy Flavour Group} collaboration, \emph{{Precision
  electroweak measurements on the $Z$ resonance}},
  \href{https://doi.org/10.1016/j.physrep.2005.12.006}{\emph{Phys. Rept.}
  {\bfseries 427} (2006) 257}
  [\href{https://arxiv.org/abs/hep-ex/0509008}{{\ttfamily hep-ex/0509008}}].

\bibitem{Chacko:2001xd}
Z.~Chacko and G.D.~Kribs, \emph{{Constraints on lepton flavor violation in the
  MSSM from the muon anomalous magnetic moment measurement}},
  \href{https://doi.org/10.1103/PhysRevD.64.075015}{\emph{Phys. Rev. D}
  {\bfseries 64} (2001) 075015}
  [\href{https://arxiv.org/abs/hep-ph/0104317}{{\ttfamily hep-ph/0104317}}].

\bibitem{Lindner:2016bgg}
M.~Lindner, M.~Platscher and F.S.~Queiroz, \emph{{A Call for New Physics : The
  Muon Anomalous Magnetic Moment and Lepton Flavor Violation}},
  \href{https://doi.org/10.1016/j.physrep.2017.12.001}{\emph{Phys. Rept.}
  {\bfseries 731} (2018) 1} [\href{https://arxiv.org/abs/1610.06587}{{\ttfamily
  1610.06587}}].

\bibitem{Barducci:2018esg}
D.~Barducci, A.~Deandrea, S.~Moretti, L.~Panizzi and H.~Prager,
  \emph{{Characterizing dark matter interacting with extra charged leptons}},
  \href{https://doi.org/10.1103/PhysRevD.97.075006}{\emph{Phys. Rev. D}
  {\bfseries 97} (2018) 075006}
  [\href{https://arxiv.org/abs/1801.02707}{{\ttfamily 1801.02707}}].

\bibitem{Aoyama:2019ryr}
T.~Aoyama, T.~Kinoshita and M.~Nio, \emph{{Theory of the Anomalous Magnetic
  Moment of the Electron}},
  \href{https://doi.org/10.3390/atoms7010028}{\emph{Atoms} {\bfseries 7} (2019)
  28}.

\bibitem{Acaroglu:2022hrm}
H.~Acaro\u{g}lu, P.~Agrawal and M.~Blanke, \emph{{Lepton-flavoured scalar dark
  matter in Dark Minimal Flavour Violation}},
  \href{https://doi.org/10.1007/JHEP05(2023)106}{\emph{JHEP} {\bfseries 05}
  (2023) 106} [\href{https://arxiv.org/abs/2211.03809}{{\ttfamily
  2211.03809}}].

\bibitem{Leveille:1977rc}
J.P.~Leveille, \emph{{The Second Order Weak Correction to (G-2) of the Muon in
  Arbitrary Gauge Models}},
  \href{https://doi.org/10.1016/0550-3213(78)90051-2}{\emph{Nucl. Phys. B}
  {\bfseries 137} (1978) 63}.

\bibitem{Acaroglu:2023cza}
H.~Acaro\u{g}lu, M.~Blanke and M.~Tabet, \emph{{Opening the Higgs portal to
  lepton-flavoured dark matter}},
  \href{https://doi.org/10.1007/JHEP11(2023)079}{\emph{JHEP} {\bfseries 11}
  (2023) 079} [\href{https://arxiv.org/abs/2309.10700}{{\ttfamily
  2309.10700}}].

\bibitem{DAlise:2022ypp}
A.~D'Alise et~al., \emph{{Standard model anomalies: lepton flavour
  non-universality, g \ensuremath{-} 2 and W-mass}},
  \href{https://doi.org/10.1007/JHEP08(2022)125}{\emph{JHEP} {\bfseries 08}
  (2022) 125} [\href{https://arxiv.org/abs/2204.03686}{{\ttfamily
  2204.03686}}].

\bibitem{Schwartz:2014sze}
M.D.~Schwartz, \emph{{Quantum Field Theory and the Standard Model}}, Cambridge
  University Press (3, 2014).

\bibitem{Peskin:1995ev}
M.E.~Peskin and D.V.~Schroeder, \emph{{An Introduction to quantum field
  theory}}, Addison-Wesley, Reading, USA (1995).

\bibitem{pal2014introductory}
P.~Pal, \emph{An Introductory Course of Particle Physics}, Taylor $\&$ Francis
  (2014).

\bibitem{Fan:2022eto}
X.~Fan, T.G.~Myers, B.A.D.~Sukra and G.~Gabrielse, \emph{{Measurement of the
  Electron Magnetic Moment}},
  \href{https://doi.org/10.1103/PhysRevLett.130.071801}{\emph{Phys. Rev. Lett.}
  {\bfseries 130} (2023) 071801}
  [\href{https://arxiv.org/abs/2209.13084}{{\ttfamily 2209.13084}}].

\bibitem{Aoyama:2012qma}
T.~Aoyama, M.~Hayakawa, T.~Kinoshita and M.~Nio, \emph{{Quantum electrodynamics
  calculation of lepton anomalous magnetic moments: Numerical approach to the
  perturbation theory of QED}},
  \href{https://doi.org/10.1093/ptep/pts030}{\emph{PTEP} {\bfseries 2012}
  (2012) 01A107}.

\bibitem{Morel:2020dww}
L.~Morel, Z.~Yao, P.~Clad\'e and S.~Guellati-Kh\'elifa, \emph{{Determination of
  the fine-structure constant with an accuracy of 81 parts per trillion}},
  \href{https://doi.org/10.1038/s41586-020-2964-7}{\emph{Nature} {\bfseries
  588} (2020) 61}.

\bibitem{Parker:2018vye}
R.H.~Parker, C.~Yu, W.~Zhong, B.~Estey and H.~M\"uller, \emph{{Measurement of
  the fine-structure constant as a test of the Standard Model}},
  \href{https://doi.org/10.1126/science.aap7706}{\emph{Science} {\bfseries 360}
  (2018) 191} [\href{https://arxiv.org/abs/1812.04130}{{\ttfamily
  1812.04130}}].

\bibitem{Muong-2:2021ojo}
{\scshape Muon g-2} collaboration, \emph{{Measurement of the Positive Muon
  Anomalous Magnetic Moment to 0.46 ppm}},
  \href{https://doi.org/10.1103/PhysRevLett.126.141801}{\emph{Phys. Rev. Lett.}
  {\bfseries 126} (2021) 141801}
  [\href{https://arxiv.org/abs/2104.03281}{{\ttfamily 2104.03281}}].

\bibitem{Muong-2:2023cdq}
{\scshape Muon g-2} collaboration, \emph{{Measurement of the Positive Muon
  Anomalous Magnetic Moment to 0.20~ppm}},
  \href{https://doi.org/10.1103/PhysRevLett.131.161802}{\emph{Phys. Rev. Lett.}
  {\bfseries 131} (2023) 161802}
  [\href{https://arxiv.org/abs/2308.06230}{{\ttfamily 2308.06230}}].

\bibitem{Datta:2023iln}
A.~Datta, D.~Marfatia and L.~Mukherjee,
  \emph{{B\textrightarrow{}K\ensuremath{\nu}\ensuremath{\nu}\textasciimacron{},
  MiniBooNE and muon g-2 anomalies from a dark sector}},
  \href{https://doi.org/10.1103/PhysRevD.109.L031701}{\emph{Phys. Rev. D}
  {\bfseries 109} (2024) L031701}
  [\href{https://arxiv.org/abs/2310.15136}{{\ttfamily 2310.15136}}].

\bibitem{Volkotrub:2888478}
{\scshape ATLAS} collaboration, \emph{{Tau anomalous magnetic moment
  measurements at ATLAS and CMS}}, .

\bibitem{Verducci:2023cgx}
M.~Verducci, C.~Roda, V.~Cavasinni and N.~Vignaroli, \emph{{Study of the
  measurement of the \ensuremath{\tau} lepton anomalous magnetic moment in high
  energy lead-lead collisions at the LHC}},
  \href{https://doi.org/10.1103/PhysRevD.110.052001}{\emph{Phys. Rev. D}
  {\bfseries 110} (2024) 052001}
  [\href{https://arxiv.org/abs/2307.15160}{{\ttfamily 2307.15160}}].

\bibitem{MEGII:2018kmf}
{\scshape MEG II} collaboration, \emph{{The design of the MEG II experiment}},
  \href{https://doi.org/10.1140/epjc/s10052-018-5845-6}{\emph{Eur. Phys. J. C}
  {\bfseries 78} (2018) 380}
  [\href{https://arxiv.org/abs/1801.04688}{{\ttfamily 1801.04688}}].

\bibitem{MEGII:2023ltw}
{\scshape MEG II} collaboration, \emph{{A search for $\upmu ^+ \rightarrow
  \textrm{e}^+ \upgamma $ with the first dataset of the MEG~II experiment}},
  \href{https://doi.org/10.1140/epjc/s10052-024-12416-2}{\emph{Eur. Phys. J. C}
  {\bfseries 84} (2024) 216}
  [\href{https://arxiv.org/abs/2310.12614}{{\ttfamily 2310.12614}}].

\bibitem{BaBar:2009hkt}
{\scshape BaBar} collaboration, \emph{{Searches for Lepton Flavor Violation in
  the Decays tau+- ---\ensuremath{>} e+- gamma and tau+- ---\ensuremath{>} mu+-
  gamma}}, \href{https://doi.org/10.1103/PhysRevLett.104.021802}{\emph{Phys.
  Rev. Lett.} {\bfseries 104} (2010) 021802}
  [\href{https://arxiv.org/abs/0908.2381}{{\ttfamily 0908.2381}}].

\bibitem{Belle:2021ysv}
{\scshape Belle} collaboration, \emph{{Search for lepton-flavor-violating
  tau-lepton decays to $\ell\gamma$ at Belle}},
  \href{https://doi.org/10.1007/JHEP10(2021)019}{\emph{JHEP} {\bfseries 10}
  (2021) 19} [\href{https://arxiv.org/abs/2103.12994}{{\ttfamily 2103.12994}}].

\bibitem{CMS:2022urr}
{\scshape CMS} collaboration, \emph{{Search for the Higgs boson decay to a pair
  of electrons in proton-proton collisions at s=13TeV}},
  \href{https://doi.org/10.1016/j.physletb.2023.137783}{\emph{Phys. Lett. B}
  {\bfseries 846} (2023) 137783}
  [\href{https://arxiv.org/abs/2208.00265}{{\ttfamily 2208.00265}}].

\bibitem{ATLAS:2022vkf}
{\scshape ATLAS} collaboration, \emph{{A detailed map of Higgs boson
  interactions by the ATLAS experiment ten years after the discovery}},
  \href{https://doi.org/10.1038/s41586-022-04893-w}{\emph{Nature} {\bfseries
  607} (2022) 52} [\href{https://arxiv.org/abs/2207.00092}{{\ttfamily
  2207.00092}}].

\bibitem{CMS:2023pte}
{\scshape CMS} collaboration, \emph{{Search for the lepton-flavor violating
  decay of the Higgs boson and additional Higgs bosons in the e$\mu$ final
  state in proton-proton collisions at $\sqrt{s}$ = 13 TeV}},
  \href{https://doi.org/10.1103/PhysRevD.108.072004}{\emph{Phys. Rev. D}
  {\bfseries 108} (2023) 072004}
  [\href{https://arxiv.org/abs/2305.18106}{{\ttfamily 2305.18106}}].

\bibitem{CMS:2021rsq}
{\scshape CMS} collaboration, \emph{{Search for lepton-flavor violating decays
  of the Higgs boson in the $\mu\tau$ and e$\tau$ final states in proton-proton
  collisions at $\sqrt{s}$ = 13 TeV}},
  \href{https://doi.org/10.1103/PhysRevD.104.032013}{\emph{Phys. Rev. D}
  {\bfseries 104} (2021) 032013}
  [\href{https://arxiv.org/abs/2105.03007}{{\ttfamily 2105.03007}}].

\bibitem{ATLAS:2023mvd}
{\scshape ATLAS} collaboration, \emph{{Searches for lepton-flavour-violating
  decays of the Higgs boson into $e\tau$ and $\mu\tau$ in $\sqrt{s}=13$ TeV
  $pp$ collisions with the ATLAS detector}},
  \href{https://doi.org/10.1007/JHEP07(2023)166}{\emph{JHEP} {\bfseries 07}
  (2023) 166} [\href{https://arxiv.org/abs/2302.05225}{{\ttfamily
  2302.05225}}].

\bibitem{Hong:2020qxc}
T.T.~Hong, H.T.~Hung, H.H.~Phuong, L.T.T.~Phuong and L.T.~Hue,
  \emph{{Lepton-flavor-violating decays of the SM-like Higgs boson
  $h\rightarrow e_ie_j$, and $e_i \rightarrow e_j\, \gamma $ in a flipped 3-3-1
  model}}, \href{https://doi.org/10.1093/ptep/ptaa026}{\emph{PTEP} {\bfseries
  2020} (2020) 043B03} [\href{https://arxiv.org/abs/2002.06826}{{\ttfamily
  2002.06826}}].

\bibitem{Fajfer:2021cxa}
S.~Fajfer, J.F.~Kamenik and M.~Tammaro, \emph{{Interplay of New Physics effects
  in (g \ensuremath{-} 2)$_{\ell}$ and h \textrightarrow{}
  \ensuremath{\ell}$^{+}$\ensuremath{\ell}$^{-}$ \textemdash{} lessons from
  SMEFT}}, \href{https://doi.org/10.1007/JHEP06(2021)099}{\emph{JHEP}
  {\bfseries 06} (2021) 099}
  [\href{https://arxiv.org/abs/2103.10859}{{\ttfamily 2103.10859}}].

\bibitem{Stebbins:2019xjr}
A.~Stebbins and G.~Krnjaic, \emph{{New Limits on Charged Dark Matter from
  Large-Scale Coherent Magnetic Fields}},
  \href{https://doi.org/10.1088/1475-7516/2019/12/003}{\emph{JCAP} {\bfseries
  12} (2019) 003} [\href{https://arxiv.org/abs/1908.05275}{{\ttfamily
  1908.05275}}].

\bibitem{Munoz:2018pzp}
J.B.~Mu\~noz and A.~Loeb, \emph{{A small amount of mini-charged dark matter
  could cool the baryons in the early Universe}},
  \href{https://doi.org/10.1038/s41586-018-0151-x}{\emph{Nature} {\bfseries
  557} (2018) 684} [\href{https://arxiv.org/abs/1802.10094}{{\ttfamily
  1802.10094}}].

\bibitem{DeRujula:1989fe}
A.~De~Rujula, S.L.~Glashow and U.~Sarid, \emph{{CHARGED DARK MATTER}},
  \href{https://doi.org/10.1016/0550-3213(90)90227-5}{\emph{Nucl. Phys. B}
  {\bfseries 333} (1990) 173}.

\bibitem{Agrawal:2016quu}
P.~Agrawal, F.-Y.~Cyr-Racine, L.~Randall and J.~Scholtz, \emph{{Make Dark
  Matter Charged Again}},
  \href{https://doi.org/10.1088/1475-7516/2017/05/022}{\emph{JCAP} {\bfseries
  05} (2017) 022} [\href{https://arxiv.org/abs/1610.04611}{{\ttfamily
  1610.04611}}].

\bibitem{Kadota:2016tqq}
K.~Kadota, T.~Sekiguchi and H.~Tashiro, \emph{{A new constraint on millicharged
  dark matter from galaxy clusters}},
  \href{https://arxiv.org/abs/1602.04009}{{\ttfamily 1602.04009}}.

\bibitem{Davidson:2000hf}
S.~Davidson, S.~Hannestad and G.~Raffelt, \emph{{Updated bounds on millicharged
  particles}}, \href{https://doi.org/10.1088/1126-6708/2000/05/003}{\emph{JHEP}
  {\bfseries 05} (2000) 003}
  [\href{https://arxiv.org/abs/hep-ph/0001179}{{\ttfamily hep-ph/0001179}}].

\bibitem{Iles:2024zka}
E.~Iles, S.~Heeba and K.~Schutz, \emph{{Direct Detection of the Millicharged
  Background}},  \href{https://arxiv.org/abs/2407.21096}{{\ttfamily
  2407.21096}}.

\bibitem{Berlin:2024lwe}
A.~Berlin and D.~Hooper, \emph{{High-Energy Neutrinos From Millicharged Dark
  Matter Annihilation in the Sun}},
  \href{https://arxiv.org/abs/2407.04768}{{\ttfamily 2407.04768}}.

\bibitem{Fiorillo:2024upk}
D.F.G.~Fiorillo and E.~Vitagliano, \emph{{Self-interacting dark sectors in
  supernovae are fluid}},  \href{https://arxiv.org/abs/2404.07714}{{\ttfamily
  2404.07714}}.

\bibitem{Alguero:2023zol}
G.~Alguero, G.~Belanger, F.~Boudjema, S.~Chakraborti, A.~Goudelis, S.~Kraml
  et~al., \emph{{micrOMEGAs 6.0: N-component dark matter}},
  \href{https://doi.org/10.1016/j.cpc.2024.109133}{\emph{Comput. Phys. Commun.}
  {\bfseries 299} (2024) 109133}
  [\href{https://arxiv.org/abs/2312.14894}{{\ttfamily 2312.14894}}].

\bibitem{Christensen:2008py}
N.D.~Christensen and C.~Duhr, \emph{{FeynRules - Feynman rules made easy}},
  \href{https://doi.org/10.1016/j.cpc.2009.02.018}{\emph{Comput. Phys. Commun.}
  {\bfseries 180} (2009) 1614}
  [\href{https://arxiv.org/abs/0806.4194}{{\ttfamily 0806.4194}}].

\bibitem{Alloul:2013bka}
A.~Alloul, N.D.~Christensen, C.~Degrande, C.~Duhr and B.~Fuks, \emph{{FeynRules
  2.0 - A complete toolbox for tree-level phenomenology}},
  \href{https://doi.org/10.1016/j.cpc.2014.04.012}{\emph{Comput. Phys. Commun.}
  {\bfseries 185} (2014) 2250}
  [\href{https://arxiv.org/abs/1310.1921}{{\ttfamily 1310.1921}}].

\bibitem{XENON:2018voc}
{\scshape XENON} collaboration, \emph{{Dark Matter Search Results from a One
  Ton-Year Exposure of XENON1T}},
  \href{https://doi.org/10.1103/PhysRevLett.121.111302}{\emph{Phys. Rev. Lett.}
  {\bfseries 121} (2018) 111302}
  [\href{https://arxiv.org/abs/1805.12562}{{\ttfamily 1805.12562}}].

\bibitem{XENON:2023cxc}
{\scshape XENON} collaboration, \emph{{First Dark Matter Search with Nuclear
  Recoils from the XENONnT Experiment}},
  \href{https://doi.org/10.1103/PhysRevLett.131.041003}{\emph{Phys. Rev. Lett.}
  {\bfseries 131} (2023) 041003}
  [\href{https://arxiv.org/abs/2303.14729}{{\ttfamily 2303.14729}}].

\bibitem{LZ:2022lsv}
{\scshape LZ} collaboration, \emph{{First Dark Matter Search Results from the
  LUX-ZEPLIN (LZ) Experiment}},
  \href{https://doi.org/10.1103/PhysRevLett.131.041002}{\emph{Phys. Rev. Lett.}
  {\bfseries 131} (2023) 041002}
  [\href{https://arxiv.org/abs/2207.03764}{{\ttfamily 2207.03764}}].

\bibitem{PandaX:2024oxq}
{\scshape PandaX} collaboration, \emph{{PandaX-xT: a Multi-ten-tonne Liquid
  Xenon Observatory at the China Jinping Underground Laboratory}},
  \href{https://arxiv.org/abs/2402.03596}{{\ttfamily 2402.03596}}.

\bibitem{Baudis:2023pzu}
L.~Baudis, \emph{{Dual-phase xenon time projection chambers for~rare-event
  searches}}, \href{https://doi.org/10.1098/rsta.2023.0083}{\emph{Phil. Trans.
  Roy. Soc. Lond. A} {\bfseries 382} (2023) 20230083}
  [\href{https://arxiv.org/abs/2311.05320}{{\ttfamily 2311.05320}}].

\bibitem{Ibarra:2013zia}
A.~Ibarra, A.S.~Lamperstorfer and J.~Silk, \emph{{Dark matter annihilations and
  decays after the AMS-02 positron measurements}},
  \href{https://doi.org/10.1103/PhysRevD.89.063539}{\emph{Phys. Rev. D}
  {\bfseries 89} (2014) 063539}
  [\href{https://arxiv.org/abs/1309.2570}{{\ttfamily 1309.2570}}].

\bibitem{Weng:2020zmr}
{\scshape AMS} collaboration, \emph{{Towards Understanding the Origin of
  Cosmic-Ray Positrons}}, \href{https://doi.org/10.22323/1.358.0091}{\emph{PoS}
  {\bfseries ICRC2019} (2020) 091}.

\bibitem{Essig:2013goa}
R.~Essig, E.~Kuflik, S.D.~McDermott, T.~Volansky and K.M.~Zurek,
  \emph{{Constraining Light Dark Matter with Diffuse X-Ray and Gamma-Ray
  Observations}}, \href{https://doi.org/10.1007/JHEP11(2013)193}{\emph{JHEP}
  {\bfseries 11} (2013) 193} [\href{https://arxiv.org/abs/1309.4091}{{\ttfamily
  1309.4091}}].

\bibitem{Fermi-LAT:2015kyq}
{\scshape Fermi-LAT} collaboration, \emph{{Updated search for spectral lines
  from Galactic dark matter interactions with pass 8 data from the Fermi Large
  Area Telescope}},
  \href{https://doi.org/10.1103/PhysRevD.91.122002}{\emph{Phys. Rev. D}
  {\bfseries 91} (2015) 122002}
  [\href{https://arxiv.org/abs/1506.00013}{{\ttfamily 1506.00013}}].

\bibitem{Fermi-LAT:2016uux}
{\scshape Fermi-LAT, DES} collaboration, \emph{{Searching for Dark Matter
  Annihilation in Recently Discovered Milky Way Satellites with Fermi-LAT}},
  \href{https://doi.org/10.3847/1538-4357/834/2/110}{\emph{Astrophys. J.}
  {\bfseries 834} (2017) 110}
  [\href{https://arxiv.org/abs/1611.03184}{{\ttfamily 1611.03184}}].

\bibitem{Meurer:2009ir}
{\scshape Fermi-LAT} collaboration, \emph{{Dark Matter Searches with the Fermi
  Large Area Telescope}}, {\emph{AIP Conf. Proc.} {\bfseries 719} (2009) 1085}
  [\href{https://arxiv.org/abs/0904.2348}{{\ttfamily 0904.2348}}].

\bibitem{Fermi-LAT:2015att}
{\scshape Fermi-LAT} collaboration, \emph{{Searching for Dark Matter
  Annihilation from Milky Way Dwarf Spheroidal Galaxies with Six Years of Fermi
  Large Area Telescope Data}},
  \href{https://doi.org/10.1103/PhysRevLett.115.231301}{\emph{Phys. Rev. Lett.}
  {\bfseries 115} (2015) 231301}
  [\href{https://arxiv.org/abs/1503.02641}{{\ttfamily 1503.02641}}].

\bibitem{HESS:2016mib}
{\scshape H.E.S.S.} collaboration, \emph{{Search for dark matter annihilations
  towards the inner Galactic halo from 10 years of observations with H.E.S.S}},
  \href{https://doi.org/10.1103/PhysRevLett.117.111301}{\emph{Phys. Rev. Lett.}
  {\bfseries 117} (2016) 111301}
  [\href{https://arxiv.org/abs/1607.08142}{{\ttfamily 1607.08142}}].

\bibitem{Silverwood:2014yza}
H.~Silverwood, C.~Weniger, P.~Scott and G.~Bertone, \emph{{A realistic
  assessment of the CTA sensitivity to dark matter annihilation}},
  \href{https://doi.org/10.1088/1475-7516/2015/03/055}{\emph{JCAP} {\bfseries
  03} (2015) 055} [\href{https://arxiv.org/abs/1408.4131}{{\ttfamily
  1408.4131}}].

\bibitem{ATLAS:2019lff}
{\scshape ATLAS} collaboration, \emph{{Search for electroweak production of
  charginos and sleptons decaying into final states with two leptons and
  missing transverse momentum in $\sqrt{s}=13$ TeV $pp$ collisions using the
  ATLAS detector}},
  \href{https://doi.org/10.1140/epjc/s10052-019-7594-6}{\emph{Eur. Phys. J. C}
  {\bfseries 80} (2020) 123}
  [\href{https://arxiv.org/abs/1908.08215}{{\ttfamily 1908.08215}}].

\bibitem{Alwall:2011uj}
J.~Alwall, M.~Herquet, F.~Maltoni, O.~Mattelaer and T.~Stelzer, \emph{{MadGraph
  5 : Going Beyond}},
  \href{https://doi.org/10.1007/JHEP06(2011)128}{\emph{JHEP} {\bfseries 06}
  (2011) 128} [\href{https://arxiv.org/abs/1106.0522}{{\ttfamily 1106.0522}}].

\bibitem{Sjostrand:2007gs}
T.~Sjostrand, S.~Mrenna and P.Z.~Skands, \emph{{A Brief Introduction to PYTHIA
  8.1}}, \href{https://doi.org/10.1016/j.cpc.2008.01.036}{\emph{Comput. Phys.
  Commun.} {\bfseries 178} (2008) 852}
  [\href{https://arxiv.org/abs/0710.3820}{{\ttfamily 0710.3820}}].

\bibitem{Dercks:2016npn}
D.~Dercks, N.~Desai, J.S.~Kim, K.~Rolbiecki, J.~Tattersall and T.~Weber,
  \emph{{CheckMATE 2: From the model to the limit}},
  \href{https://doi.org/10.1016/j.cpc.2017.08.021}{\emph{Comput. Phys. Commun.}
  {\bfseries 221} (2017) 383}
  [\href{https://arxiv.org/abs/1611.09856}{{\ttfamily 1611.09856}}].

\bibitem{deFavereau:2013fsa}
{\scshape DELPHES 3} collaboration, \emph{{DELPHES 3, A modular framework for
  fast simulation of a generic collider experiment}},
  \href{https://doi.org/10.1007/JHEP02(2014)057}{\emph{JHEP} {\bfseries 02}
  (2014) 057} [\href{https://arxiv.org/abs/1307.6346}{{\ttfamily 1307.6346}}].

\bibitem{Cacciari:2005hq}
M.~Cacciari and G.P.~Salam, \emph{{Dispelling the $N^{3}$ myth for the $k_t$
  jet-finder}},
  \href{https://doi.org/10.1016/j.physletb.2006.08.037}{\emph{Phys. Lett. B}
  {\bfseries 641} (2006) 57}
  [\href{https://arxiv.org/abs/hep-ph/0512210}{{\ttfamily hep-ph/0512210}}].

\bibitem{Cacciari:2008gp}
M.~Cacciari, G.P.~Salam and G.~Soyez, \emph{{The anti-$k_t$ jet clustering
  algorithm}}, \href{https://doi.org/10.1088/1126-6708/2008/04/063}{\emph{JHEP}
  {\bfseries 04} (2008) 063} [\href{https://arxiv.org/abs/0802.1189}{{\ttfamily
  0802.1189}}].

\bibitem{Cacciari:2011ma}
M.~Cacciari, G.P.~Salam and G.~Soyez, \emph{{FastJet User Manual}},
  \href{https://doi.org/10.1140/epjc/s10052-012-1896-2}{\emph{Eur. Phys. J. C}
  {\bfseries 72} (2012) 1896}
  [\href{https://arxiv.org/abs/1111.6097}{{\ttfamily 1111.6097}}].

\bibitem{Read:2002hq}
A.L.~Read, \emph{{Presentation of search results: The $CL_s$ technique}},
  \href{https://doi.org/10.1088/0954-3899/28/10/313}{\emph{J. Phys. G}
  {\bfseries 28} (2002) 2693}.

\bibitem{Lester:1999tx}
C.G.~Lester and D.J.~Summers, \emph{{Measuring masses of semiinvisibly decaying
  particles pair produced at hadron colliders}},
  \href{https://doi.org/10.1016/S0370-2693(99)00945-4}{\emph{Phys. Lett. B}
  {\bfseries 463} (1999) 99}
  [\href{https://arxiv.org/abs/hep-ph/9906349}{{\ttfamily hep-ph/9906349}}].

\bibitem{Barr:2003rg}
A.~Barr, C.~Lester and P.~Stephens, \emph{{m(T2): The Truth behind the
  glamour}}, \href{https://doi.org/10.1088/0954-3899/29/10/304}{\emph{J. Phys.
  G} {\bfseries 29} (2003) 2343}
  [\href{https://arxiv.org/abs/hep-ph/0304226}{{\ttfamily hep-ph/0304226}}].

\bibitem{Cheng:2008hk}
H.-C.~Cheng and Z.~Han, \emph{{Minimal Kinematic Constraints and m(T2)}},
  \href{https://doi.org/10.1088/1126-6708/2008/12/063}{\emph{JHEP} {\bfseries
  12} (2008) 063} [\href{https://arxiv.org/abs/0810.5178}{{\ttfamily
  0810.5178}}].

\bibitem{Bai:2012gs}
Y.~Bai, H.-C.~Cheng, J.~Gallicchio and J.~Gu, \emph{{Stop the Top Background of
  the Stop Search}}, \href{https://doi.org/10.1007/JHEP07(2012)110}{\emph{JHEP}
  {\bfseries 07} (2012) 110} [\href{https://arxiv.org/abs/1203.4813}{{\ttfamily
  1203.4813}}].

\bibitem{ATLAS:2014zve}
{\scshape ATLAS} collaboration, \emph{{Search for direct production of
  charginos, neutralinos and sleptons in final states with two leptons and
  missing transverse momentum in $pp$ collisions at $\sqrt{s} =$ 8 TeV with the
  ATLAS detector}}, \href{https://doi.org/10.1007/JHEP05(2014)071}{\emph{JHEP}
  {\bfseries 05} (2014) 071} [\href{https://arxiv.org/abs/1403.5294}{{\ttfamily
  1403.5294}}].

\end{thebibliography}\endgroup
\end{document}